\documentclass[review]{elsarticle}
\usepackage[left=3.5cm, right=3.5cm, top=3cm]{geometry}

\usepackage{titlesec}
\titleformat{\section}
{\normalfont\Large\bfseries}%
{\thesection}{1em}{}
\titlespacing*{\section}
{0pt}{1.5\baselineskip}{1.5\baselineskip}

\titleformat{\subsection}
{\normalfont\large\bfseries}%
{\thesubsection}{1em}{}
\titlespacing*{\subsection}
{0pt}{1.5\baselineskip}{1.5\baselineskip}

\titleformat{\subsubsection}
{\normalfont\bfseries}%
{\thesubsubsection}{1em}{}
\titlespacing*{\subsubsection}
{0pt}{1.5\baselineskip}{1.\baselineskip}

\usepackage{multicol}
\usepackage[utf8]{inputenc}
\usepackage{siunitx}
\usepackage{comment}
\usepackage[toc,page,title]{appendix}

\usepackage{dsfont}
\usepackage{soul}
\usepackage{enumerate}
\usepackage[shortlabels]{enumitem}

\usepackage{amsmath,amsfonts,amsthm,amssymb}
\usepackage{empheq}
\usepackage{stmaryrd}
\usepackage[ruled,vlined]{algorithm2e}
\usepackage{mathalfa, eufrak, yfonts, mathrsfs}
\DeclareMathOperator*{\argmin}{argmin}
\DeclareMathOperator*{\argmax}{argmax}

\newtheorem{lemma}{Lemma}[subsection]

\newtheorem{theorem}{Theorem}[subsection]

\newtheorem{corollary}{Corollary}[subsection]

\theoremstyle{definition}

\theoremstyle{definition}
\newtheorem*{definition*}{Definition}

\newtheorem{proposition}{Proposition}[subsection]

\newtheorem{example}{Example}[subsection]

\newcommand{\supp}[1]{\sloppy\text{supp$(#1)$}}

\renewenvironment{proof}{{\bfseries Proof.}}{\qed}

\usepackage{amssymb,amsmath,amsfonts,amsthm} 
\usepackage{stmaryrd}
\usepackage[ruled,vlined]{algorithm2e}

\theoremstyle{definition}
\newtheorem*{remark}{Remark}





\newtheorem*{example*}{\small{Example}}
\usepackage{graphicx}
\graphicspath{ {./img/} }
\usepackage{xcolor}
\usepackage{comment}

\usepackage[makeroom]{cancel}
\usepackage{pgfplots}
\usepackage{tikz}
\usetikzlibrary{decorations.pathreplacing}
\usetikzlibrary{positioning}
\usetikzlibrary{arrows}
\usetikzlibrary{arrows.meta}
\usetikzlibrary{shapes.geometric}
\usetikzlibrary{shapes.misc}
\tikzset{cross/.style={cross out, draw=red, minimum size=40*(#1-\pgflinewidth), inner sep=0pt, outer sep=0pt},
	cross/.default={1pt}}

\makeatletter
\newcommand{\osymbol}[1]{\mathbin{\mathpalette\make@circled#1}}
\newcommand{\make@circled}[2]{
	\ooalign{$\m@th#1\smallbigcirc{#1}$\cr\hidewidth$\m@th#1#2$\hidewidth\cr}
}
\newcommand{\smallbigcirc}[1]{
	\vcenter{\hbox{\scalebox{1}{$\m@th#1\bigcirc$}}}
}
\makeatother

\usepackage{pdfpages}
\usepackage{graphicx}
\usepackage{lineno,hyperref}
\modulolinenumbers[5]

\usepackage[export]{adjustbox}
\usepackage{subcaption}

\usepackage{parskip}
\usepackage[utf8]{inputenc}

\begin{document}

\begin{frontmatter}
	
	\title{
		Efficient M\"obius Transformations and their applications to 
		Dempster-Shafer Theory: Clarification and implementation\tnoteref{mytitlenote}}
	\tnotetext[mytitlenote]{This work was carried out and co-funded in the framework of the Labex MS2T and the Hauts-de-France region of France. It was supported by the French Government, through the program ``Investments for the future'' managed by the National Agency for Research (Reference ANR-11-IDEX-0004-02).}
	\author{Maxime Chaveroche$^*$, Franck Davoine, V\'eronique Cherfaoui}
	\ead{name.surname@hds.utc.fr}
	\address{Sorbonne University Alliance, Universit\'e de technologie de Compi\`egne, CNRS, Heudiasyc,\\
		CS 60319 - 60203 Compi\`egne Cedex, France}
	\cortext[mycorrespondingauthor]{Corresponding author}
	\begin{abstract}
		Dempster-Shafer Theory (DST) generalizes Bayesian probability theory, offering useful additional information, but suffers from a high computational burden. A lot of work has been done to reduce the complexity of computations used in information fusion with Dempster's rule. The main approaches exploit either the structure of Boolean lattices or the information contained in belief sources. Each has its merits depending on the situation.
		In this paper, we propose sequences of graphs for the computation of the zeta and M\"obius transformations that optimally exploit both the structure of distributive semilattices and the information contained in belief sources. We call them the \textit{Efficient M\"obius Transformations} (EMT).
		We show that the complexity of the EMT is always inferior to the complexity of algorithms that consider the whole lattice, such as the \textit{Fast M\"obius Transform} (FMT) for all DST transformations. We then explain how to use them to fuse two belief sources.
		More generally, our EMTs apply to any function in any finite distributive lattice, focusing on a meet-closed or join-closed subset.	
		This article extends our work published at the international conference on \textit{Scalable Uncertainty Management} (SUM) \cite{me}. It clarifies it, brings some minor corrections and provides implementation details such as data structures and algorithms applied to DST.
	\end{abstract}
\end{frontmatter}



\section{Introduction}

Dempster-Shafer Theory (DST) \cite{shafer76} is an elegant formalism that generalizes Bayesian probability theory. It is more expressive by giving the possibility for a source to represent its belief in the state of a variable not only by assigning credit directly to a possible state (strong evidence) but also by assigning credit to any subset (weaker evidence) of the set $\Omega$ of all possible states. This assignment of credit is called a \textit{mass function} and provides meta-information to quantify the level of uncertainty about one's believes considering the way one established them, which is critical for decision making.

Nevertheless, this information comes with a cost: considering $2^{|\Omega|}$ potential values instead of only $|\Omega|$ can lead to computationally and spatially expensive algorithms. They can become difficult to use for more than a dozen possible states (e.g. 20 states in $\Omega$ generate more than a million subsets), although we may need to consider large frames of discernment (e.g. for classification or identification tasks). Moreover, these algorithms not being tractable anymore beyond a few dozen states means their performances greatly degrade before that, which further limits their application to real-time applications. To tackle this issue, a lot of work has been done to reduce the complexity of transformations used to combine belief sources with Dempster's rule \cite{dempster68}. We distinguish between two approaches that we call \textit{powerset-based} and \textit{evidence-based}.

The \textit{powerset-based} approach concerns all algorithms based on the structure of the powerset $2^{\Omega}$ of the frame of discernment $\Omega$. They have a complexity dependent on $|\Omega|$. Early works \cite{barnett81,gordon85,shenoy86,shafer87} proposed optimizations by restricting the structure of evidence to only singletons and their negation, which greatly restrains the expressiveness of DST. Later, a family of optimal algorithms working in the general case, i.e. the ones based on the \textit{Fast M\"obius Transform} (FMT) \cite{FMT}, was discovered. Their complexity is $O(|\Omega|. 2^{|\Omega|})$ in time and $O(2^{|\Omega|})$ in space. It has become the de facto standard for the computation of every transformation in DST. Consequently, efforts were made to reduce the size of $\Omega$ to benefit from the optimal algorithms of the FMT. More specifically, \cite{wilson2000} refers to the process of conditioning by the \textit{combined core} (intersection of the unions of all \textit{focal sets} of each belief source) and \textit{lossless coarsening} (merging of elements of $\Omega$ which always appear together in focal sets). Also, Monte Carlo methods \cite{wilson2000} have been proposed but depend on a number of trials that must be large and grows with $|\Omega |$, in addition to not being exact.

The \textit{evidence-based} approach concerns all algorithms that aim to reduce the computations to the only subsets that contain information (\textit{evidence}), called \textit{focal sets}, which are usually far less numerous than $2^{|\Omega|}$. This approach, also refered to as the \textit{obvious} one, implicitly originates from the seminal work of Shafer \cite{shafer76} and is often more efficient than the powerset-based one since it only depends on information contained in sources in a quadratic way. Doing so, it allows for the exploitation of the full potential of DST by enabling us to choose any frame of discernment, without concern about its size. Moreover, the evidence-based approach benefits directly from the use of approximation methods, some of which are very efficient \cite{sarabi-jamab16}. Therefore, this approach seems superior to the FMT in most use cases, above all when $|\Omega|$ is large, where an algorithm with exponential complexity is just intractable.

It is also possible to easily find evidence-based algorithms computing all DST transformation, except for the conjunctive and disjunctive decompositions for which we recently proposed a method \cite{me_gretsi, chaveroche2021efficient}. 

However, since these algorithms rely only on the information contained in sources, they do not exploit the structure of the powerset to reduce the complexity, leading to situations in which the FMT can be more efficient if almost every subset contains information, i.e. if the number of focal sets tends towards $2^{|\Omega|}$ \cite{wilson2000}, all the most when no approximation method is employed.

In this paper, we fuse these two approaches into one, proposing new sequences of graphs, in the same fashion as the FMT, that are always more efficient than the FMT and can in addition benefit from evidence-based optimizations. We call them the \textit{Efficient M\"obius Transformations} (EMT). More generally, our approach applies to any function defined on a finite distributive lattice.

Outside the scope of DST, \cite{bjorklundtrimmed} is related to our approach in the sense that we both try to remove redundancy in the computation of the zeta and M\"obius transforms on the subset lattice $2^\Omega$. However, they only consider the redundancy of computing the image of a subset that is known to be null beforehand. To do so, they only visit sets that are accessible from the focal sets of lowest rank by successive unions with each element of $\Omega$. Here, we demonstrate that it is possible to avoid far more computations by reducing them to specific sets so that each image is only computed once. These sets are the focal points described in \cite{me_gretsi, chaveroche2021efficient}. The study of their properties will be carried out in depth in an upcoming article \cite{me_journal}. Besides, our method is more general since it applies to any finite distributive lattice.

Furthermore, an important result of our work resides in the optimal computation of the zeta and M\"obius transforms in any intersection-closed family $F$ of sets from $2^\Omega$, i.e. with a complexity $O(|\Omega|.|F|)$. Indeed, in the work of \cite{bjorklund2016fast} on the optimal computation of these transforms in any finite lattice $L$, they embedded $L$ into the Boolean lattice $2^\Omega$, obtaining an intersection-closed family $F$ as its equivalent, and found a meta-procedure building a circuit of size $O(|\Omega|.|F|)$ computing the zeta and M\"obius transforms. However, they did not managed to build this circuit in less than $O(|\Omega|.2^{|\Omega|})$. Given $F$, our Theorem \ref{mob_opti_F} in this paper directly computes this circuit with a complexity that can be as low as $O(|\Omega|.|F|)$ in some instances, while being much simpler.

This paper is organized as follows: Section \ref{SUM:preliminaries} will present the elements on which our method is built. Section \ref{emt} will present our EMT. Section \ref{discussion} will discuss their complexity and their usage both in general and in the case of DST. Finally, we will conclude this article with section \ref{SUM:conclusion}.

\section{Background of our method}\label{SUM:preliminaries}

Let $(P, \leq)$ be a finite\footnote{The following definitions hold for lower semifinite partially ordered sets as well, i.e. partially ordered sets such that the number of elements of $P$ lower in the sense of $\leq$ than another element of $P$ is finite. But for the sake of simplicity, we will only talk of finite partially ordered sets.} set partially ordered by $\leq$. 

\subsection{Zeta transform} 
The zeta transform $g: P \rightarrow \mathbb{R}$ of a function $f: P \rightarrow \mathbb{R}$ is defined as follows:
\begin{align}\label{zeta_trans}
\forall y \in P,\quad g(y) = \sum_{x \leq y} f(x)
\end{align}
This can be extended to the multiplication as the \textit{multiplicative zeta transform}:
\begin{align*}
\forall y \in P,\quad g(y) = \prod_{x \leq y} f(x)
\end{align*}
\begin{example}\label{SUM:b_from_m}
	In DST, the implicability function $b$ is defined as the zeta transform of the mass function $m$ in $(2^\Omega, \subseteq)$, i.e.:
	\begin{align*}
	\forall B \in 2^\Omega,\quad b(B) = \sum_{A \subseteq B} m(A)
	\end{align*}
\end{example}
\begin{example}\label{SUM:b_from_v}
	In DST, the implicability function $b$ is also the inverse of the multiplicative zeta transform of the disjunctive weight function $v$ in $(2^\Omega, \subseteq)$, i.e.:
	\begin{align*}
	\forall B \in 2^\Omega,\quad b(B) = \prod_{A \subseteq B} v(A)^{-1}
	\end{align*}
\end{example}
\begin{example}\label{SUM:q_from_m}
	In DST, the commonality function $q$ is defined as the zeta transform of the mass function $m$ in $(2^\Omega, \supseteq)$, i.e.:
	\begin{align*}
	\forall B \in 2^\Omega,\quad q(B) = \sum_{A \supseteq B} m(A)
	\end{align*}
\end{example}
\begin{example}\label{SUM:q_from_w}
	In DST, the commonality function $q$ is also the inverse of the multiplicative zeta transform of the conjunctive weight function $w$ in $(2^\Omega, \supseteq)$, i.e.:
	\begin{align*}
	\forall B \in 2^\Omega,\quad q(B) = \prod_{A \supseteq B} w(A)^{-1}
	\end{align*}
\end{example}

\subsection{M\"obius transform} The M\"obius transform of $g$ is $f$. It is defined as follows:
\begin{align}\label{SUM:mob_trans}
\forall y \in P,\quad f(y) = \sum_{x \leq y} g(x) . \mu_{P,\leq}(x,y)
\end{align}
where $\mu_{P, \leq}$ is the M\"obius function of $(P, \leq)$ (See \cite{rota}). There is also a multiplicative version with the same properties that can be seen as the exponential of the M\"obius transform of $\log \circ ~g$:
\begin{align*}
\forall y \in P,\quad f(y) = \prod_{x \leq y} g(x)^{\mu_{P,\leq}(x,y)}
\end{align*}

\begin{example}\label{SUM:m_from_b}
	In DST, the mass function $m$ is the M\"obius transform of the implicability function $b$ in $(2^\Omega, \subseteq)$, i.e.:
	\begin{align*}
	\forall B \in 2^\Omega,\quad m(B) = \sum_{A \subseteq B} b(A).\mu_{2^\Omega, \subseteq}(A,B)
	\end{align*}
	where for any $A,B \in 2^\Omega$, the M\"obius function evaluates to $\mu_{2^\Omega, \subseteq}(A,B) = (-1)^{|B|-|A|}$, as recalled in \cite{FMT}.
\end{example}

\begin{example}\label{SUM:v_from_b}
	In DST, the disjunctive weight function $v$ is the inverse of the multiplicative M\"obius transform of the implicability function $b$ in $(2^\Omega, \subseteq)$, i.e.:
	\begin{align*}
	\forall B \in 2^\Omega,\quad v(B) = \prod_{A \subseteq B} b(A)^{-\mu_{2^\Omega, \subseteq}(A,B)}
	\end{align*}
\end{example}

\begin{example}\label{SUM:m_from_q}
	In DST, the mass function $m$ is the M\"obius transform of the commonality function $q$ in $(2^\Omega, \supseteq)$, i.e.:
	\begin{align*}
	\forall B \in 2^\Omega,\quad m(B) = \sum_{A \supseteq B} q(A).\mu_{2^\Omega, \supseteq}(A,B)
	\end{align*}
	where for any $A,B \in 2^\Omega$, the M\"obius function also evaluates to $\mu_{2^\Omega, \supseteq}(A,B) = (-1)^{|B|-|A|}$.
\end{example}

\begin{example}\label{SUM:w_from_q}
	In DST, the conjunctive weight function $w$ is the inverse of the multiplicative M\"obius transform of the commonality function $q$ in $(2^\Omega, \subseteq)$, i.e.:
	\begin{align*}
	\forall B \in 2^\Omega,\quad w(B) = \prod_{A \supseteq B} q(A)^{-\mu_{2^\Omega, \supseteq}(A,B)}
	\end{align*}
\end{example}

\subsection{Sequence of graphs and computation of the zeta transform}\label{char_FMT}

\def\h{1.25}

To yield $g(y)$ for some $y\in P$, we must sum all values $f(x)$ such that $x \leq y$. Our objective is to do it in the minimum number of operations, i.e. to minimize the number of terms to sum. If we only compute $g(y)$ alone, we have to pick every element in $\{ x \in P ~/~ x \leq y \}$ and sum their associated values through $f$. However, if we compute $g(y)$ for all elements $y\in P$ at once, we can organize and mix these computations so that partial sums are reused for more than one value through $g$. Indeed, for any element $y \in P$, if there is an element $z\in P$ such that $y \leq z$, we have $g(z) = g(y) + \sum_{\substack{x \leq z\\x \not\leq y}} f(x)$.
So, we want to recursively build partial sums so that we can get the full sum on each $g(y)$ by only summing the values on some elements from $\{ x \in P ~/~ x \leq y \}$. In other words, we would like to define an ordered sequence of transformations computing $g$ from $f$. 

Let us adopt the formalism of graph theory. Let $G_\leq$ be a directed acyclic graph in which the set of its nodes matches $P$ and each arrow is directed by $\leq$. Let $E_\leq$ be the set of its arrows. We have $E_\leq = \{ (x, y) \in P^2 ~/~ x \leq y \}$ and $G_\leq = (P, E_\leq)$. Thus, computing $g(y)$ alone is equivalent to visiting the node $y$ from all nodes $x$ of $G_\leq$ such that there is an arrow $(x,y) \in E_\leq$. Each ``visit'' to node $y$ from a node $x$ corresponds to the computation of the operation $f(x) + \cdot$, where $\cdot$ represents the current state of the sum associated with $y$. Thus, $G_\leq$, combined with the binary operator $f(\cdot) + \cdot$, describes the transformation of 0 into $g$. More concisely, we will equivalently initialize our algorithm with values through $f$ instead of 0 and exploit the combination of $G_<$ and +. We will say that the transformation \textit{$(G_<, f, +)$ computes the zeta transform of $f$ in $(P, \leq)$}.
In the end, we want to minimize the number of ``visits'' to be made to all $y$, i.e. we want to minimize the total number of arrows to follow to compute every $g(y)$. Therefore, the question is: Is there an ordered sequence of graphs that can compute $g$ from $f$ with less arrows in total than $(G_<, f, +)$ ?

Let $I_{P}$ be the set containing all identity arrows of $G_\leq$, i.e. $I_{P} = \{ (x, y) \in P^2 ~/~ x = y \}$.
Consider that all elements $y\in P$ are initialized with $f(y)$. We are interested in finding a sequence of graphs that is equivalent to the arrows of $G_<$.
Let $(H_i)_{i\in\llbracket 1, n\rrbracket}$ be a sequence of $n$ directed acyclic graphs $H_i = (P, E_i)$. We will note $((H_i)_{i\in\llbracket 1, n\rrbracket}, f, +)$ the computation that transforms $f$ into $h_1$ through the arrows of $E_1$, then transforms $h_1$ into $h_2$ through the arrows of $E_2$, and so on until the transformation of $h_{n-1}$ into $h_n$ through the arrows of $E_n$. 
We ignore all identity arrows in these computations.
So, this sequence of graphs requires us to consider $|E_1| + |E_2|+ \cdots + |E_n|$ arrows, but transforms $f$ into $h_n$ in $|E_1\backslash I_{P}| + |E_2 \backslash I_{P}|+ \cdots + |E_n \backslash I_{P}|$ operations.
\begin{proposition}\label{zeta_opti_FMT}
	Let $\Omega = \{ \omega_1, \omega_2, \dots, \omega_n\}$. One particular sequence of interest is $(H_i)_{i\in\llbracket 1, n\rrbracket}$, where $H_i = (2^\Omega, E_i)$ and:
	$$E_i = \{ (X,Y) \in 2^\Omega \times 2^\Omega /~ Y = X \cup \{\omega_i\} \}.$$
	This sequence computes the same zeta transformations as $G_\subset = (2^\Omega, E_\subset)$, where $E_\subset = \{ (X,Y) \in 2^\Omega \times 2^\Omega ~/~ X \subset Y \}$.
	
\end{proposition}

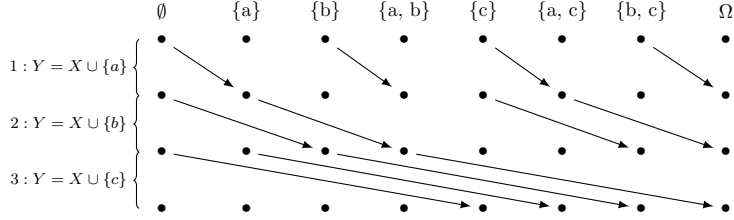
\begin{figure}[t]
	\centering
	\hspace{-0.4cm}
	\begin{tikzpicture}[scale=0.75, every node/.style={transform shape},
	node/.style={draw, dot,minimum size=0.2cm, inner sep=0pt},
	det/.style={draw, diamond,minimum size=1.1cm, inner sep=0pt},
	rect/.style={draw, rectangle,minimum size=1.1cm, inner sep=2pt}
	]
	
	\node (none) at (-5, 0) {$\emptyset$};
	\node (a) at (-3.5, 0) {\{a\}};
	\node (b) at (-2.1, 0) {\{b\}};
	\node (ab) at (-0.7, 0) {\{a, b\}};
	\node (c) at (0.7, 0) {\{c\}};
	\node (ac) at (2.1, 0) {\{a, c\}};
	\node (bc) at (3.5, 0) {\{b, c\}};
	\node (abc) at (5, 0) {$\Omega$};
	
	\node (none1) at (-5, -0.5) {$\bullet$};
	\node (a1) at (-3.5, -0.5) {$\bullet$};
	\node (b1) at (-2.1, -0.5) {$\bullet$};
	\node (ab1) at (-0.7, -0.5) {$\bullet$};
	\node (c1) at (0.7, -0.5) {$\bullet$};
	\node (ac1) at (2.1, -0.5) {$\bullet$};
	\node (bc1) at (3.5, -0.5) {$\bullet$};
	\node (abc1) at (5, -0.5) {$\bullet$};
	
	\draw [decorate,decoration={brace,amplitude=2pt},yshift=0pt]
	(-5.4,-1.5) -- (-5.4,-0.5) node [black,midway,xshift=-1.25cm] {\footnotesize
		$1: Y = X \cup \{a\}$};
	
	\node (none2) at (-5, -1.5) {$\bullet$};
	\node (a2) at (-3.5, -1.5) {$\bullet$};
	\node (b2) at (-2.1, -1.5) {$\bullet$};
	\node (ab2) at (-0.7, -1.5) {$\bullet$};
	\node (c2) at (0.7, -1.5) {$\bullet$};
	\node (ac2) at (2.1, -1.5) {$\bullet$};
	\node (bc2) at (3.5, -1.5) {$\bullet$};
	\node (abc2) at (5, -1.5) {$\bullet$};
	
	\draw [decorate,decoration={brace,amplitude=2pt},yshift=0pt]
	(-5.4,-2.5) -- (-5.4,-1.5) node [black,midway,xshift=-1.25cm] {\footnotesize
		$2: Y = X \cup \{b\}$};
	
	\node (none3) at (-5, -2.5) {$\bullet$};
	\node (a3) at (-3.5, -2.5) {$\bullet$};
	\node (b3) at (-2.1, -2.5) {$\bullet$};
	\node (ab3) at (-0.7, -2.5) {$\bullet$};
	\node (c3) at (0.7, -2.5) {$\bullet$};
	\node (ac3) at (2.1, -2.5) {$\bullet$};
	\node (bc3) at (3.5, -2.5) {$\bullet$};
	\node (abc3) at (5, -2.5) {$\bullet$};
	
	\draw [decorate,decoration={brace,amplitude=2pt},yshift=0pt]
	(-5.4,-3.5) -- (-5.4,-2.5) node [black,midway,xshift=-1.25cm] {\footnotesize
		$3: Y = X \cup \{c\}$};
	
	\node (none4) at (-5, -3.5) {$\bullet$};
	\node (a4) at (-3.5, -3.5) {$\bullet$};
	\node (b4) at (-2.1, -3.5) {$\bullet$};
	\node (ab4) at (-0.7, -3.5) {$\bullet$};
	\node (c4) at (0.7, -3.5) {$\bullet$};
	\node (ac4) at (2.1, -3.5) {$\bullet$};
	\node (bc4) at (3.5, -3.5) {$\bullet$};
	\node (abc4) at (5, -3.5) {$\bullet$};
	
	\draw[->,>=latex] (none1) to (a2);
	\draw[->,>=latex] (b1) to (ab2);
	\draw[->,>=latex] (c1) to (ac2);
	\draw[->,>=latex] (bc1) to (abc2);
	
	\draw[->,>=latex] (none2) to (b3);
	\draw[->,>=latex] (a2) to (ab3);
	\draw[->,>=latex] (c2) to (bc3);
	\draw[->,>=latex] (ac2) to (abc3);
	
	\draw[->,>=latex] (none3) to (c4);
	\draw[->,>=latex] (a3) to (ac4);
	\draw[->,>=latex] (b3) to (bc4);
	\draw[->,>=latex] (ab3) to (abc4);
	
	\end{tikzpicture}
	\caption{\small{Illustration representing the paths generated by the arrows contained in the sequence $(H_i)_{i\in\llbracket 1, 3\rrbracket}$, where $H_i = (2^\Omega, E_i)$ and $E_i = \{ (X,Y) \in 2^\Omega \times 2^\Omega /~ Y = X \cup \{\omega_i\} \}$ and $\Omega = \{ a,b,c \}$. This sequence computes the same zeta transformations as $G_\subset = (2^\Omega, E_\subset)$, where $E_\subset = \{ (X,Y) \in 2^\Omega \times 2^\Omega ~/~ X \subset Y \}$. A dot represents the node of its column. Its row $i$ corresponds to both the tail of a potential arrow in $H_i$ and the head of a potential arrow in $H_{i-1}$. The last row corresponds to the heads of all potential arrows that could be in $H_3$. The arrows represents the actual arrows in each graph $H_i$. Identity arrows are ignored in computations and not displayed here for the sake of clarity. If they were, there would be vertical arrows in every column, linking each dot in row $i$ to the next dot of same node in row $i+1$. This representation is derived from the one used in \cite{FMT}.}}
	\label{fig:FMT_sub}
\end{figure}

\begin{example}\label{ex:num_FMT}
	Let us say that $\Omega = \{a, b, c\}$. Crossing ignored arrows (i.e. identity arrows), we have:
	\begin{itemize}
		\item $E_1 = \{ (X,Y) \in 2^\Omega \times 2^\Omega /~ Y = X \cup \{ a \} \} = \{$
		\begin{align*}
			 &\quad(\emptyset, \{ a \}),~ \xcancel{(\{a\}, \{a\})},~ (\{b\}, \{a,b\}),~ \xcancel{(\{a,b\}, \{a,b\})},~ \\
			 &\quad(\{c\}, \{a,c\}),~ \xcancel{(\{a,c\}), \{a,c\})},~ (\{b,c\}, \Omega),~ \xcancel{(\Omega, \Omega)}\\
			 &\}
		\end{align*}
		\item $E_2 = \{ (X,Y) \in 2^\Omega \times 2^\Omega /~ Y = X \cup \{ b \} \} = \{$
		\begin{align*}
			&\quad(\emptyset, \{ b \}),~ (\{a\}, \{a,b\}),~ \xcancel{(\{b\}, \{b\})},~ \xcancel{(\{a,b\}, \{a,b\})},~ \\
			&\quad(\{c\}, \{b,c\}),~ (\{a,c\}), \Omega),~ \xcancel{(\{b,c\}, \{b,c\})},~ \xcancel{(\Omega, \Omega)}\\
			&\}
		\end{align*}
		\item $E_3 = \{ (X,Y) \in 2^\Omega \times 2^\Omega /~ Y = X \cup \{ c \} \} = \{$
		\begin{align*}
			&\quad(\emptyset, \{ c \}),~ (\{a\}, \{a,c\}),~ (\{b\}, \{b,c\}),~ (\{a,b\}, \Omega),~ \\
			&\quad\xcancel{(\{c\}, \{c\})},~ \xcancel{(\{a,c\}), \{a,c\})},~ \xcancel{(\{b,c\}, \{b,c\})},~ \xcancel{(\Omega, \Omega)}\\
			&\}
		\end{align*}
	\end{itemize}
	Fig. \ref{fig:FMT_sub} illustrates this sequence. Check that, after execution of $((H_i)_{i\in\llbracket 1, n\rrbracket}, f, +)$, each element $y$ of $2^\Omega$ is associated with the sum $\sum_{x \subseteq y} f(x)$. For instance, let us take a look at $\Omega$. Initially, each element of $2^\Omega$ is associated with its value through $f$. Then, at step 1, we can see that the value on $\Omega$ is summed with $f(\{ b,c \})$. At step 2, it is summed with $h_1(\{ a,c \})$, which is equal to $f(\{c\}) + f(\{ a,c \})$, following step 1. Finally, at step 3, it is summed with $h_2(\{a,b\})$, which is equal to $h_1(\{a\}) + h_1(\{a,b\})$ following step 2, which is itself equal to $f(\emptyset) + f(\{a\}) + f(\{b\}) + f(\{a,b\})$, following step 1. Gathering all these terms, we get that $h_3(\Omega) = f(\Omega) + f(\{ b,c \}) + f(\{c\}) + f(\{ a,c \}) + f(\emptyset) + f(\{a\}) + f(\{b\}) + f(\{a,b\})$.
\end{example}

\begin{figure}[t]
	\centering
	\hspace{0.4cm}
	\begin{tikzpicture}[scale=0.75, every node/.style={transform shape},
	node/.style={draw, dot,minimum size=0.2cm, inner sep=0pt},
	det/.style={draw, diamond,minimum size=1.1cm, inner sep=0pt},
	rect/.style={draw, rectangle,minimum size=1.1cm, inner sep=2pt}
	]
	
	\node (none) at (-5, 0) {$\emptyset$};
	\node (a) at (-3.5, 0) {\{a\}};
	\node (b) at (-2.1, 0) {\{b\}};
	\node (ab) at (-0.7, 0) {\{a, b\}};
	\node (c) at (0.7, 0) {\{c\}};
	\node (ac) at (2.1, 0) {\{a, c\}};
	\node (bc) at (3.5, 0) {\{b, c\}};
	\node (abc) at (5, 0) {$\Omega$};
	
	\node (none1) at (-5, -0.5) {$\bullet$};
	\node (a1) at (-3.5, -0.5) {$\bullet$};
	\node (b1) at (-2.1, -0.5) {$\bullet$};
	\node (ab1) at (-0.7, -0.5) {$\bullet$};
	\node (c1) at (0.7, -0.5) {$\bullet$};
	\node (ac1) at (2.1, -0.5) {$\bullet$};
	\node (bc1) at (3.5, -0.5) {$\bullet$};
	\node (abc1) at (5, -0.5) {$\bullet$};
	
	\draw [decorate,decoration={brace,amplitude=2pt},yshift=0pt]
	(-5.4,-1.5) -- (-5.4,-0.5) node [black,midway,xshift=-1.25cm] {\footnotesize
		$1: X = Y \cup \{a\}$};
	
	\node (none2) at (-5, -1.5) {$\bullet$};
	\node (a2) at (-3.5, -1.5) {$\bullet$};
	\node (b2) at (-2.1, -1.5) {$\bullet$};
	\node (ab2) at (-0.7, -1.5) {$\bullet$};
	\node (c2) at (0.7, -1.5) {$\bullet$};
	\node (ac2) at (2.1, -1.5) {$\bullet$};
	\node (bc2) at (3.5, -1.5) {$\bullet$};
	\node (abc2) at (5, -1.5) {$\bullet$};
	
	\draw [decorate,decoration={brace,amplitude=2pt},yshift=0pt]
	(-5.4,-2.5) -- (-5.4,-1.5) node [black,midway,xshift=-1.25cm] {\footnotesize
		$2: X = Y \cup \{b\}$};
	
	\node (none3) at (-5, -2.5) {$\bullet$};
	\node (a3) at (-3.5, -2.5) {$\bullet$};
	\node (b3) at (-2.1, -2.5) {$\bullet$};
	\node (ab3) at (-0.7, -2.5) {$\bullet$};
	\node (c3) at (0.7, -2.5) {$\bullet$};
	\node (ac3) at (2.1, -2.5) {$\bullet$};
	\node (bc3) at (3.5, -2.5) {$\bullet$};
	\node (abc3) at (5, -2.5) {$\bullet$};
	
	\draw [decorate,decoration={brace,amplitude=2pt},yshift=0pt]
	(-5.4,-3.5) -- (-5.4,-2.5) node [black,midway,xshift=-1.25cm] {\footnotesize
		$3: X = Y \cup \{c\}$};
	
	\node (none4) at (-5, -3.5) {$\bullet$};
	\node (a4) at (-3.5, -3.5) {$\bullet$};
	\node (b4) at (-2.1, -3.5) {$\bullet$};
	\node (ab4) at (-0.7, -3.5) {$\bullet$};
	\node (c4) at (0.7, -3.5) {$\bullet$};
	\node (ac4) at (2.1, -3.5) {$\bullet$};
	\node (bc4) at (3.5, -3.5) {$\bullet$};
	\node (abc4) at (5, -3.5) {$\bullet$};
	
	\draw[->,>=latex] (a1) to (none2);
	\draw[->,>=latex] (ab1) to (b2);
	\draw[->,>=latex] (ac1) to (c2);
	\draw[->,>=latex] (abc1) to (bc2);
	
	\draw[->,>=latex] (b2) to (none3);
	\draw[->,>=latex] (ab2) to (a3);
	\draw[->,>=latex] (bc2) to (c3);
	\draw[->,>=latex] (abc2) to (ac3);
	
	\draw[->,>=latex] (c3) to (none4);
	\draw[->,>=latex] (ac3) to (a4);
	\draw[->,>=latex] (bc3) to (b4);
	\draw[->,>=latex] (abc3) to (ab4);
	
	\end{tikzpicture}
	\caption{\small{Illustration representing the paths generated by the arrows contained in the sequence $(H_i)_{i\in\llbracket 1, 3\rrbracket}$, where $H_i = (2^\Omega, E_i)$ and $E_i = \{ (X,Y) \in 2^\Omega \times 2^\Omega /~ X = Y \cup \{\omega_i\} \}$ and $\Omega = \{ a,b,c \}$. This sequence computes the same zeta transformations as $G_\supset = (2^\Omega, E_\supset)$, where $E_\supset = \{ (X,Y) \in 2^\Omega \times 2^\Omega ~/~ X \supset Y \}$. 
			}}
	\label{fig:FMT_sup}
\end{figure}
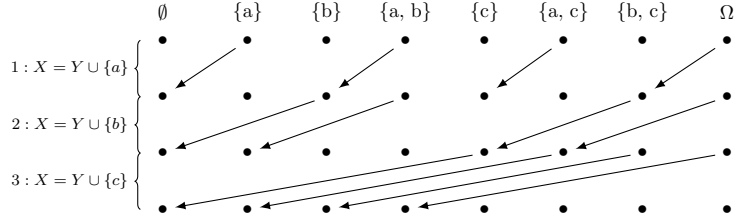

\begin{proposition}\label{zeta_opti_FMT_dual}
	The dual of this particular sequence in $(2^\Omega, \supseteq)$ is $(H_i)_{i\in\llbracket 1, n\rrbracket}$, where $H_i = (2^\Omega, E_i)$ and:
	$$E_i = \{ (X,Y) \in 2^\Omega \times 2^\Omega /~ X = Y \cup \{\omega_i\} \}.$$
	This sequence computes the same zeta transformations as $G_\supset = (2^\Omega, E_\supset)$, where $E_\supset = \{ (X,Y) \in 2^\Omega \times 2^\Omega ~/~ X \supset Y \}$.
\end{proposition}
\begin{example}
 Fig. \ref{fig:FMT_sup} illustrates the dual sequence for zeta transforms in $(2^\Omega, \supseteq)$ and $\Omega = \{a,b,c\}$.
\end{example}

\begin{remark}
	These two sequences of graphs are the foundation of the \textit{Fast M\"obius Transform} (FMT) algorithms. Their execution is $O(n.2^n)$ in time and $O(2^n)$ in space.
	As we can see, the FMT presented here proposes two transformations $((H_i)_{i\in\llbracket 1, n\rrbracket}, f, +)$ that computes the same transformation as respectively $(G_\subset, f, +)$ and $(G_\supset, f, +)$ for any function $f:2^\Omega \rightarrow \mathbb{R}$. The authors proved them to be the optimal transformations for any set $\Omega$, i.e. the one that uses the fewest arrows, independently of the function $f$ to be considered. This means that they do not take into account the neutral values of $f$ for the operator +, i.e. where $f$ evaluates to 0, contrary to our approach. This is why our method is able to feature a lower complexity than this \textit{optimal} FMT.
\end{remark}

More generally, Theorem 3 of \cite{FMT} defines a necessary and sufficient condition to verify that a transformation $((H_i)_{i\in\llbracket 1, n\rrbracket}, f, +)$ computes (ignoring identity arrows) the same transformation as $(G_<, f, +)$. 
It is stated in our terms as follows:
\begin{theorem}\label{theorem:FMT}
Let $(H_i)_{i\in\llbracket 1, n\rrbracket}$ be a sequence of directed acyclic graphs $H_i = (P, E_i)$. Let us pose $A_i = E_i \cup I_P$. The transformation $((H_i)_{i\in\llbracket 1, n\rrbracket}, f, +)$ computes (ignoring identity arrows) the same transformation as $(G_<, f, +)$
if and only if each set of arrows satisfies $E_i \subseteq E_\leq$ and every arrow $e \in E_\leq$ can be decomposed as a unique path $(e_1, e_2, \dots, e_{n}) \in A_1 \times A_2 \times \dots \times A_{n}$, where the tail of $e_1$ is the tail of $e$ and the head of $e_n$ is the head of $e$. Recall that a path is a sequence of arrows in which $\forall i\in \llbracket 1, n-1\rrbracket$, the head of $e_i$ is the tail of $e_{i+1}$.
\end{theorem}
\begin{example}
	Let us prove Proposition \ref{zeta_opti_FMT}. We had $\Omega = \{ \omega_1, \omega_2, \dots, \omega_n\}$. The sequence of graphs $(H_i)_{i\in\llbracket 1, n\rrbracket}$ computes the same zeta transformations as $G_\subset = (2^\Omega, E_\subset)$, where $E_\subset = \{ (X,Y) \in 2^\Omega \times 2^\Omega ~/~ X \subset Y \}$ if:
	$$E_i = \{ (X,Y) \in 2^\Omega \times 2^\Omega /~ Y = X \cup \{\omega_i\} \},$$
	where $i \in \{1, \dots, n\}$ and $H_i = (2^\Omega, E_i)$. 
	
	\begin{proof}
		Obviously, for any $i \in \llbracket 1, n \rrbracket$, we have $E_i \subseteq E_\subseteq$.
		In addition, each set $X \subseteq \Omega$ is composed by definition of at most $n$ elements from $\Omega$. Thus, it is possible to reach in at most $n$ steps from $X$ any set $Y \subseteq \Omega$ such that $X \subseteq Y$ with identity arrows and arrows where the head is just the tail with exactly one other element from $\Omega$, i.e. with identity arrows or arrows of $(H_i)_{i\in\llbracket 1, n\rrbracket}$. Moreover, since each step corresponds to a distinct element from $\Omega$, there is exactly one path from $X$ to $Y$: at each step corresponding to the elements in $Y\backslash X$, follow the arrow that adds an element to the set reached in the previous step. Only identity arrows can be followed in the steps corresponding to the elements in $X\cap Y$. Otherwise, there would be an element missing or in excess relatively to $Y$ at the end of step $n$, which means that $Y$ could not be reached. Therefore, according to Theorem \ref{theorem:FMT}, the transformation $((H_i)_{i\in\llbracket 1, n\rrbracket}, f, +)$ computes the same transformation as $(G_\subset, f, +)$ for any function $f:2^\Omega \rightarrow \mathbb{R}$.
	\end{proof}
\end{example}

In addition, for a sequence of graphs computing zeta transforms in $(P, \leq)$, reversing its paths yields a sequence of graphs computing zeta transforms in $(P, \geq)$. Hence Proposition \ref{zeta_opti_FMT_dual}.

\subsection{Sequence of graphs and computation of the M\"obius transform}\label{char_mob}

Now, consider that we want to find a sequence of graphs that undoes the previous computation. We want to transform $g$ into $f$, i.e. the M\"obius transform of $g$ in $(P, \leq)$. For this, notice that for any step $i$ in the transformation $((H_i)_{i\in\llbracket 1, n\rrbracket}, f, +)$, we have for each node $y\in P$,
\begin{align*}
	h_i(y) = h_{i-1}(y) + \sum_{(x,y) \in E_i\backslash I_P} h_{i-1}(x)	\quad\Leftrightarrow\quad h_{i-1}(y) = h_i(y) - \sum_{(x,y) \in E_i\backslash I_P} h_{i-1}(x).
\end{align*}
So, as long as we know all $h_{i-1}(x)$ for all arrows $(x,y) \in E_i\backslash I_P$ at each step $i$ and for all $y\in P$, we can simply reverse the order of the sequence $(H_i)_{i\in\llbracket 1, n\rrbracket}$ and use the operator - instead of +. If this is verified, then $((H_{n-i+1})_{i\in\llbracket 1, n\rrbracket}, g, -)$ computes the M\"obius transform of $g$ in $(P, \leq)$. This condition can be translated as follows: for every arrow $(x, y) \in E_i\backslash I_P$, we have $h_i(x) = h_{i-1}(x)$. This condition is equivalent to stating that for every arrow $(x, y) \in E_i\backslash I_P$, there is no arrow $(w, x)$ in $E_i\backslash I_P$.

\begin{theorem}\label{theorem:mobius}
	Let $(H_i)_{i\in\llbracket 1, n\rrbracket}$ be a sequence of directed acyclic graphs $H_i = (P, E_i)$. Let $h_n$ be the function resulting from the transformation $((H_i)_{i\in\llbracket 1, n\rrbracket}, f, +)$, ignoring identity arrows. If for every arrow $(x, y) \in E_i\backslash I_P$, there is no arrow $(w, x)$ in $E_i\backslash I_P$, then $((H_{n-i+1})_{i\in\llbracket 1, n\rrbracket}, h_n, -)$ yields the initial function $f$.
\end{theorem}

Thus, if $((H_i)_{i\in\llbracket 1, n\rrbracket}, f, +)$ computes the zeta transform $g$ of $f$ in $(P, \leq)$ and Theorem \ref{theorem:mobius} is satisfied, then \textit{$((H_{n-i+1})_{i\in\llbracket 1, n\rrbracket}, g, -)$ computes the M\"obius transform $f$ of $g$ in $(P, \leq)$}.

\subsubsection{Application to the powerset lattice $2^\Omega$ (FMT)}\label{mob_opti_FMT}
Consider again the sequence $(H_i)_{i\in\llbracket 1, n\rrbracket}$, from the application of section \ref{zeta_opti_FMT}, that computes the zeta transform of $f$ in $(2^\Omega, \subseteq)$. If there is an arrow $(X, Y) \in E_i\backslash I_{2^\Omega}$, then $\omega_i \not\in X$. This means that there is no set $W$ in $2^\Omega$ such that $W \cup \{ \omega_i \} = X$, and so no arrow $(W, X)$ in $E_i\backslash I_{2^\Omega}$. Thus, according to Theorem \ref{theorem:mobius}, $((H_{n-i+1})_{i\in\llbracket 1, n\rrbracket}, h_n, -)$ computes the M\"obius transformation of $((H_i)_{i\in\llbracket 1, n\rrbracket}, f, +)$, i.e. the function $f$. Furthermore, given that $\Omega$ is a set and that each graph $H_i$ concerns an element $\omega_i$, independently from the others, any indexing (order) in the sequence $(H_i)_{i\in\llbracket 1, n\rrbracket}$ computes the zeta transformation. This implies that any order in $((H_{n-i+1})_{i\in\llbracket 1, n\rrbracket}, h_n, -)$ computes the M\"obius transformation of $((H_i)_{i\in\llbracket 1, n\rrbracket}, f, +)$. In particular, $((H_{i})_{i\in\llbracket 1, n\rrbracket}, h_n, -)$ computes the same transformation as $((H_{n-i+1})_{i\in\llbracket 1, n\rrbracket}, h_n, -)$.

\begin{example}\label{ex:num_FMT_mobius}
	Let us say that $\Omega = \{a, b, c\}$. We want to compute the M\"obius transform $f$ of $g$ in $(2^\Omega, \subseteq)$. Each arrow set $E_i$ has already been computed in Example \ref{ex:num_FMT}. Fig. \ref{fig:FMT_sub} illustrates any transformation based on $(H_{i})_{i\in\llbracket 1, n\rrbracket}$, including $((H_{i})_{i\in\llbracket 1, n\rrbracket}, h_n, -)$. 
	Check that, after execution of $((H_{i})_{i\in\llbracket 1, n\rrbracket}, h_n, -)$, each element $y$ of $2^\Omega$ is associated with $f(y)$. For instance, let us take a look again at $\Omega$. Initially, each element of $2^\Omega$ is associated with its value through $g$. Then, at step 1, we can see that to the value on $\Omega$ is subtracted $g(\{ b,c \})$. At step 2, to $h_1(\Omega)$ is subtracted $h_1(\{ a,c \})$, which is equal to $g(\{ a,c \}) - g(\{c\})$, following step 1. Finally, at step 3, to $h_2(\Omega)$ is subtracted $h_2(\{a,b\})$, which is equal to $h_1(\{a,b\}) - h_1(\{a\})$ following step 2, which is itself equal to $g(\{a,b\}) - g(\{b\}) - (g(\{a\}) - g(\emptyset))$, following step 1. Gathering all these terms, we get that 
	\begin{align*}
		h_3(\Omega) &= g(\Omega) - g(\{ b,c \}) - \left[~g(\{ a,c \}) - g(\{c\})~ \right] - \left[~g(\{a,b\}) - g(\{b\}) - \left[~g(\{a\}) - g(\emptyset)~\right]~\right]\\
		&= g(\Omega) - g(\{ b,c \}) - g(\{ a,c \}) + g(\{c\}) - g(\{a,b\}) + g(\{b\}) + g(\{a\}) - g(\emptyset)\\
		&= \sum_{X \subseteq \Omega} g(x) . (-1)^{|\Omega|-|X|}.
	\end{align*}
	As recalled in \cite{FMT}, the function that associates to each couple $(X,Y) \in 2^\Omega \times 2^\Omega$ the value $(-1)^{|Y|-|X|}$ is the M\"obius function $\mu$ in $(2^\Omega, \subseteq)$. So, according to Eq. \ref{SUM:mob_trans}, we have $h_3(\Omega) = f(\Omega)$.

\end{example}

\subsection{Order theory}

Let $(P, \leq)$ be a set partially ordered by the relation $\leq$.

\subsubsection{Meet / join} Let $S$ be a subset of $P$. If it is unique, the greatest element of $P$ that is less than all the elements of $S$ is called the meet (or infimum) of $S$. It is noted $\bigwedge S$. If $S = \{x, y\}$, we may also note it with the binary operator $\wedge$ such that $x \wedge y$. If it is unique, the least element of $P$ that is greater than all the elements of $S$ is called the join (or supremum) of $S$. It is noted $\bigvee S$. If $S = \{x, y\}$, we may also note it with the binary operator $\vee$ such that $x \vee y$.

\begin{example}
	In $(2^\Omega, \subseteq)$, the meet operator $\wedge$ is the intersection operator $\cap$, while the join operator $\vee$ is the union operator $\cup$.
\end{example}

\subsubsection{\textit{Lattice / semi-lattice}} If any non-empty subset of $P$ has a join, we say that $P$ is an upper semilattice. If any non-empty subset of $P$ has a meet, we say that $P$ is a lower semilattice. When $P$ is both, we say that $P$ is a lattice.

\begin{example}
	In $(2^\Omega, \subseteq)$, any non-empty subset $S$ has an intersection and a union in $2^\Omega$. They respectively represent the common elements of the sets in $S$ and the cumulative elements of all the sets in $S$. Thus, $2^\Omega$ is a lattice.
\end{example}

\subsubsection{Irreducible elements} In any partially ordered set, there are bottom elements such that they cannot be described as the join of two lesser elements. Such irreducible elements are called the \textit{join-irreducible elements of $P$} if they are not equal to the global minimum of $P$. We will note $^\vee\mathcal{I}(P)$ the set of all join-irreducible elements of $P$. Since none of them is $\bigwedge P$, this means that the join of any two join-irreducible elements yields a non-join-irreducible element of $P$. In fact, if $P$ is an upper semilattice (or a lattice), it is known that the join of all possible non-empty subset of $^\vee\mathcal{I}(P)$ yields all elements of $P$, except $\bigwedge P$. Formally, we write that all join-irreducible element $i$ verifies $i \neq \bigwedge P$ and for all elements $x,y \in P$, if $x < i$ and $y < i$, then $x \vee y < i$.

Dually, in any partially ordered set, there are top elements such that they cannot be described as the meet of two greater elements. Such irreducible elements are called the \textit{meet-irreducible elements of $P$} if they are not equal to the global maximum of $P$. We will note $^\wedge\mathcal{I}(P)$ the set of all meet-irreducible elements of $P$. Since none of them is $\bigvee P$, this means that the meet of any two meet-irreducible elements yields a non-meet-irreducible element of $P$. In fact, if $P$ is a lower semilattice (or a lattice), it is known that the meet of all possible non-empty subset of $^\wedge\mathcal{I}(P)$ yields all elements of $P$, except $\bigvee P$. Formally, we write that all meet-irreducible element $i$ verifies $i \neq \bigvee P$ and for all elements $x,y \in P$, if $x > i$ and $y > i$, then $x \wedge y > i$.

\begin{example}
	In $(2^\Omega, \subseteq)$, the join-irreducible elements are the singletons $\{ \omega \}$, where $\omega \in \Omega$. The meet-irreducible elements are their complement $\overline{\{ \omega \}} = \Omega\backslash\{ \omega \}$, where $\omega \in \Omega$.
\end{example}

\subsubsection{Distributive lattice} A distributive lattice $L$ is a lattice that satisfies the distributive law:
\begin{align}\label{distrib_law}
	\forall x, y, z \in L,\quad (x \wedge y) \vee (x \wedge z) = x \wedge (y \vee z)
\end{align}
Since $(x \wedge z) \vee z = z$, this condition is equivalent to:
\begin{align}\label{distrib_law_dual}
\forall x, y, z \in L,\quad (z \vee y) \wedge (z \vee x) = z \vee (y \wedge x)
\end{align}

\begin{example}
	In the powerset lattice $2^\Omega$, it holds for any sets $A,B,C \in 2^\Omega$ that $(A \cap B) \cup (A \cap C) = A \cap (B \cup C)$. Thus, the lattice $2^\Omega$ is a distributive lattice.
\end{example}

\subsubsection{Sublattice} A sublattice $S$ is simply a subset of a lattice $L$ that is itself a lattice, with the same meet and join operations as $L$. This means that for any two elements $x,y \in S$, we have both $x \vee y \in S$ and $x \wedge y \in S$, where $\vee$ and $\wedge$ are the join and meet operators of $L$.

\subsubsection{Upset / down set} An upset (or upward closed set) $S$ is a subset of $P$ such that all elements in $P$ greater than at least one element of $S$ is in $S$. The upper closure of an element $x \in P$ is noted $\uparrow x$ or $x^{\uparrow P}$ (when the encompassing set has to be specified). It is equal to $\{ y\in P ~/~ x \leq y \}$.

Dually, a down set (or downward closed set) $S$ is a subset of $P$ such that all elements in $P$ less than at least one element of $S$ is in $S$. The lower closure of an element $x \in P$ is noted $\downarrow x$ or $x^{\downarrow P}$ (when the encompassing set has to be specified). It is equal to $\{ y\in P ~/~ x \geq y \}$.

\subsection{Support elements and focal points}
Let $f: P \rightarrow \mathbb{R}$. 

\subsubsection{Support of a function in $P$} The support $\supp{f}$ of a function $f: P \rightarrow \mathbb{R}$ is defined as $\supp{f} = \{ x \in P ~/~ f(x) \neq 0 \}$. 

\begin{example}
	In DST, the set containing the focal sets of a mass function $m$ is $\supp{m}$.
\end{example}

\subsubsection{Focal points}

We note $^\vee\supp{f}$ the smallest join-closed subset of $P$ containing $\supp{f}$, i.e.:
\begin{align*}
^\vee\supp{f} = \left\lbrace \bigvee S ~/~ S \subseteq \supp{f},~S \neq \emptyset \right\rbrace
\end{align*}
We note $^\wedge\supp{f}$ the smallest meet-closed subset of $P$ containing $\supp{f}$, i.e.:
\begin{align*}
^\wedge\supp{f} = \left\lbrace \bigwedge S ~/~ S \subseteq \supp{f},~S \neq \emptyset \right\rbrace
\end{align*}
The set containing the \textit{focal points} $\mathring{\mathcal{F}}$ of a mass function $m$ from \cite{me_gretsi, chaveroche2021efficient} for the conjunctive weight function is $^\wedge\supp{m}$. For the disjunctive weight function, we use their dual focal points, defined by $^\vee\supp{m}$.

It has been proven in \cite{me_gretsi, chaveroche2021efficient} that the image of $2^\Omega$ through the conjunctive weight function can be computed without redundancies by only considering the focal points $^\wedge\supp{m}$ in the definition of the multiplicative M\"obius transform of the commonality function. The image of all set in $2^\Omega \backslash ^\wedge\supp{m}$ through the conjunctive weight function is 1. In the same way, the image of any set in $2^\Omega \backslash ^\wedge\supp{m}$ through the commonality function is only a duplicate of the image of a focal point in $^\wedge\supp{m}$. Its image can be recovered by searching for its smallest focal point superset in $^\wedge\supp{m}$. The same can be stated for the disjunctive weight function regarding the implicability function and $^\vee\supp{m}$.

In fact, as generalized in \cite{me_journal}, for any function $f:P\rightarrow \mathbb{R}$, its focal points $^\wedge\supp{f}$ are sufficient to define its zeta and M\"obius transforms in $(P, \geq)$, and $^\vee\supp{f}$ is sufficient to define its zeta and M\"obius transforms in $(P, \leq)$.

However, considering the case where $P$ is a finite lattice, naive algorithms that only consider $^o\supp{f}$, where $o \in \{ \vee, \wedge\}$ have upper bound complexities in $O(|^o\supp{f}|^2)$, which may be worse than the optimal complexity $O(|^\vee\mathcal{I}(P)|.|P|)$ of a procedure that considers the whole lattice $P$. In this paper, we propose computing schemes for $g(S)$, where $^o\supp{f} \subseteq S \subseteq P$ and $g$ is the zeta transform of $f$ in $(P, \leq)$, with complexities always less than $O(|^\vee\mathcal{I}(P)|.|P|)$, provided that $P$ is a finite distributive lattice. We also provide schemes for computing the M\"obius transform $f(S)$ of $g$ in $(P, \leq)$.

\section{Our Efficient M\"obius Transformations}\label{emt}

In this section, we consider a function $f: P \rightarrow \mathbb{R}$ where $P$ is a finite distributive lattice (e.g. the powerset lattice $2^\Omega$). We present here the sequences of graphs that can be exploited to compute our so-called \textit{Efficient M\"obius Transformations}. Theorem \ref{mob_opti_L} describes a way of computing the zeta and M\"obius transforms of a function based on the smallest sublattice $^{\mathcal{L}}\supp{f}$ of $P$ containing both $^\wedge\supp{f}$ and $^\vee\supp{f}$, which is defined in Proposition \ref{supp_lattice}. Theorem \ref{mob_opti_F} goes beyond this optimization by computing these transforms based only on $^o\supp{f}$, where $o \in \{ \vee, \wedge\}$. Nevertheless, this second approach requires the direct computation of $^o\supp{f}$, which has an upper bound complexity of $O(|\supp{f}|.|^o\supp{f}|)$, which may be more than $O(|^\vee\mathcal{I}(P)|.|P|)$ if $|\supp{f}| \gg |^\vee\mathcal{I}(P)|$.

\subsection{Preliminary results}

In this subsection, we provide some propositions that are useful for proving our main results (presented in the next subsection).

\begin{lemma}[\textit{Safe join}]\label{safe_join}
	Let us consider a finite distributive lattice $L$. For all join-irreducible element $i \in {^{\vee}\mathcal{I}(L)}$ and for all elements $x,y \in {L}$ that are not greater than or equal to $i$, i.e. $i \not\leq x$ and $i \not\leq y$, we have that their join is not greater or equal to $i$ either, i.e. $i \not\leq x \vee y$.
\end{lemma}
\begin{proof}
	By definition of a join-irreducible element, we know that $\forall i \in {^{\vee}\mathcal{I}(L)}$ and for all $a,b \in {L}$, if $a < i$ and $b < i$, then $a\vee b <i$. Moreover, for all $x,y \in {L}$ such that $i \not\leq x$ and $i \not\leq y$, we have equivalently $i \wedge x < i$ and $i \wedge y < i$. Thus, we get that $(i \wedge x) \vee (i \wedge y) < i$. Since $L$ satisfies the distributive law Eq. (\ref{distrib_law}), this implies that $(i \wedge x) \vee (i \wedge y) = i \wedge (x \vee y) < i$, which means that $i \not\leq x\vee y$.
\end{proof}

\begin{lemma}[\textit{Safe meet}]\label{safe_meet}
	Let us consider a finite distributive lattice $L$. For all meet-irreducible element $i \in {^{\wedge}\mathcal{I}(L)}$ and for all elements $x,y \in {L}$ that are not less than or equal to $i$, i.e. $i \not\geq x$ and $i \not\geq y$, we have that their meet is not less or equal to $i$ either, i.e. $i \not\geq x \wedge y$.
\end{lemma}
\begin{proof}
	By definition of a meet-irreducible element, we know that $\forall i \in {^{\wedge}\mathcal{I}(L)}$ and for all $a,b \in {L}$, if $a > i$ and $b > i$, then $a\wedge b >i$. Moreover, for all $x,y \in {L}$ such that $i \not\geq x$ and $i \not\geq y$, we have equivalently $i \vee x > i$ and $i \vee y > i$. Thus, we get that $(i \vee x) \wedge (i \vee y) > i$. Since $L$ satisfies the distributive law Eq. (\ref{distrib_law_dual}), this implies that $(i \vee x) \wedge (i \vee y) = i \vee (x \wedge y) < i$, which means that $i \not\geq x\vee y$.
\end{proof}

\begin{proposition}[\textit{Iota elements of a subset of $P$}]\label{iota_elements}
	For any subset $S \subseteq P$, we note $\iota(S)$ the set containing the join-irreducible elements of the smallest sublattice $L_S$ of $P$ containing $S$, i.e. $\iota(S) = {^\vee \mathcal{I}(L_S)}$. These so-called \textit{iota elements of $S$} can be obtained through the following equality:
	$$\iota(S) = \left\lbrace \bigwedge i^{\uparrow S}
	~/~ i\in {^\vee\mathcal{I}(P)},~ i^{\uparrow S} \neq \emptyset
	\right\rbrace,$$
	where $i^{\uparrow S}$ is the upper closure of $i$ in $S$, i.e. $\{ s\in S ~/~ i \leq s\}$.
\end{proposition}
\begin{proof}
	See Appendix \ref{appendix:iota_elements}.
\end{proof}

\begin{proposition}[\textit{Dual iota elements of a subset of $P$}]\label{iota_elements_dual}
	Similarly, for any subset $S \subseteq P$, we note $\overline{\iota}(S)$ the set containing the meet-irreducible elements of the smallest sublattice $L_S$ of $P$ containing $S$, i.e. $\overline{\iota}(S) = {^\wedge \mathcal{I}(L_S)}$. These so-called \textit{dual iota elements of $S$} can be obtained through the following equality:
	$$\overline{\iota}(S) = \left\lbrace \bigvee i^{\downarrow S}
	~/~ i\in {^\wedge\mathcal{I}(P)},~ i^{\downarrow S} \neq \emptyset
	\right\rbrace,$$
	where $i^{\downarrow S}$ is the lower closure of $i$ in $S$, i.e. $\{ s\in S ~/~ i \geq s\}$.
\end{proposition}
\begin{proof}
	Analog to the proof of Proposition \ref{iota_elements}, exploiting the dual definition of the distributive law in a lattice, i.e. Eq. (\ref{distrib_law_dual}).
\end{proof}

\begin{proposition}[\textit{Lattice support}]\label{supp_lattice}
	The smallest sublattice of $P$ containing both $^\wedge\supp{f}$ and $^\vee\supp{f}$, noted $^\mathcal{L}\supp{f}$, can be defined as:
	\begin{align*}
		^\mathcal{L}\supp{f} &= \left\lbrace \bigvee X ~/~ X \subseteq \iota(\supp{f}),~ X \neq \emptyset \right\rbrace \cup \left\lbrace \bigwedge \supp{f} \right\rbrace\\
		&= \left\lbrace \bigwedge X ~/~ X \subseteq \overline{\iota}(\supp{f}),~ X \neq \emptyset \right\rbrace \cup \left\lbrace \bigvee \supp{f} \right\rbrace.
	\end{align*}
	
	More specifically, $^\vee\supp{f}$ is contained in the upper closure $\supp{f}^{\uparrow {^\mathcal{L}\supp{f}}}$ of $\supp{f}$ in ${^\mathcal{L}\supp{f}}$: $$\supp{f}^{\uparrow {^\mathcal{L}\supp{f}}} = \{ x \in {^\mathcal{L}\supp{f}} ~/~ \exists s \in \supp{f},~ s \leq x \},$$ and $^\wedge\supp{f}$ is contained in the lower closure $\supp{f}^{\downarrow {^\mathcal{L}\supp{f}}}$ of $\supp{f}$ in ${^\mathcal{L}\supp{f}}$: $$\supp{f}^{\downarrow {^\mathcal{L}\supp{f}}} = \{ x \in {^\mathcal{L}\supp{f}} ~/~ 
	\exists s \in \supp{f},~ s \geq x \}.$$
	These sets can be computed in less than respectively $O(|\iota(\supp{f})|. |\supp{f}^{\uparrow {^\mathcal{L}\supp{f}}}|)$ and $O(|\iota(\supp{f})|. |\supp{f}^{\downarrow {^\mathcal{L}\supp{f}}}|)$, which is at most $O(|^\vee\mathcal{I}(P)|.|P|)$.
	
	\begin{proof}
		The proof is immediate here, considering Proposition \ref{iota_elements} and its proof, as well as Proposition \ref{iota_elements_dual}. In addition, since ${^\wedge\supp{f}}$ only contains the meet of elements of ${\supp{f}}$, all element of ${^\wedge\supp{f}}$ is less than at least one element of ${\supp{f}}$. Similarly, since ${^\vee\supp{f}}$ only contains the join of elements of ${\supp{f}}$, all element of ${^\vee\supp{f}}$ is greater than at least one element of ${\supp{f}}$. Hence $\supp{f}^{\uparrow {^\mathcal{L}\supp{f}}}$ and $\supp{f}^{\downarrow {^\mathcal{L}\supp{f}}}$.
	\end{proof}
\end{proposition}

As pointed out in \cite{kaski2016fast}, a special ordering of the join-irreducible elements of a lattice when using the Fast Zeta Transform \cite{bjorklund2016fast} leads to the optimal computation of its zeta and M\"obius transforms. Here, we use this ordering to build our EMT for finite distributive lattices in a way similar to \cite{kaski2016fast} but without the need to check the equality of the decompositions into the first $j$ join-irreducible elements at each step.

\begin{corollary}[\textit{Join-irreducible ordering}]\label{focal_atom_order}
	Let us consider a finite distributive lattice $(L, \leq)$ and let its join-irreducible elements ${^{\vee}\mathcal{I}(L)}$ be ordered such that $\forall i_k, i_{l} \in {^{\vee}\mathcal{I}(L)}$, $k < l \Rightarrow i_k \not\geq i_{l}$. If $L$ is a graded lattice (i.e. a lattice equipped with a rank function $\rho : L \rightarrow \mathbb{N}$), then $\rho(i_1) \leq \rho(i_2) \leq \dots \leq \rho(i_{|{^{\vee}\mathcal{I}(L)}|})$ implies this ordering. 
	We note 	${^{\vee}\mathcal{I}(L)}_{k} = \left\lbrace i_{1}, \dots, i_{k-1},~ i_{k}  \right\rbrace$.
	
	For all element $i_k \in {^{\vee}\mathcal{I}(L)}$, we have $i_k \not\leq \bigvee{^{\vee}\mathcal{I}(L)}_{k-1}$.
\end{corollary}
\begin{proof}
	Since the join-irreducible elements are ordered such that $\forall i_k, i_{l} \in {^{\vee}\mathcal{I}(L)}$, $k < l \Rightarrow i_k \not\geq i_{l}$, it is trivial to see that for any $i_l \in {^{\vee}\mathcal{I}(L)}$ and $i_k \in {^{\vee}\mathcal{I}(L)}_{l-1}$, we have $i_k \not\geq i_{l}$. Then, using Lemma \ref{safe_join} by recurrence, it is easy to get that $i_l \not\leq \bigvee{^{\vee}\mathcal{I}(L)}_{{l-1}}$.
\end{proof}
\begin{example}\label{ex:join_order}
	For example, in DST we work with $P=2^\Omega$, in which the rank function is the cardinality, i.e. for all $A \in P$, $\rho(A) = |A|$. So, with $(L, \subseteq)$ a subset lattice of $(2^\Omega, \subseteq)$, the ordering required in Corollary \ref{focal_atom_order} simply translates to sorting the join-irreducible elements of $L$ from the smallest set to the largest set. Thus, sorting ${^{\vee}\mathcal{I}(L)} = \{ i_1, i_2, \cdots, i_n \}$ such that $|i_1| \leq |i_2| \leq \cdots \leq |i_n|$, Corollary \ref{focal_atom_order} tells us that for any $k \in \llbracket 2, n \rrbracket$, we have $i_k \not\subseteq \bigcup {^{\vee}\mathcal{I}(L)}_{k-1}$, where
	${^{\vee}\mathcal{I}(L)}_{k} = \left\lbrace i_{1}, \dots, i_{k-1},~ i_{k}  \right\rbrace$.
\end{example}

\begin{corollary}[\textit{Meet-irreducible ordering}]\label{focal_atom_order_dual}
	Let us consider a finite distributive lattice $(L, \leq)$ and let its meet-irreducible elements ${^{\wedge}\mathcal{I}(L)}$ be ordered such that $\forall i_k, i_{l} \in {^{\vee}\mathcal{I}(L)}$, $k < l \Rightarrow i_k \not\leq i_{l}$. If $L$ is a graded lattice (i.e. a lattice equipped with a rank function $\rho : L \rightarrow \mathbb{N}$), then $\rho(i_1) \geq \rho(i_2) \geq \dots \geq \rho(i_{|{^{\wedge}\mathcal{I}(L)}|})$ implies this ordering. 
	We note ${^{\wedge}\mathcal{I}(L)}_{k} = \left\lbrace i_{1}, \dots, i_{k-1},~ i_{k}  \right\rbrace$.
	
	For all element $i_k \in {^{\wedge}\mathcal{I}(L)}$, we have $i_k \not\geq \bigwedge{^{\wedge}\mathcal{I}(L)}_{k-1}$.
\end{corollary}
\begin{proof}
	Since the meet-irreducible elements are ordered such that $\forall i_k, i_{l} \in {^{\wedge}\mathcal{I}(L)}$, $k < l \Rightarrow i_k \not\leq i_{l}$, it is trivial to see that for any $i_l \in {^{\wedge}\mathcal{I}(L)}$ and $i_k \in {^{\wedge}\mathcal{I}(L)}_{l-1}$, we have $i_k \not\leq i_{l}$. Then, using Lemma \ref{safe_meet} by recurrence, it is easy to get that $i_l \not\geq \bigwedge{^{\wedge}\mathcal{I}(L)}_{{l-1}}$.
\end{proof}
\begin{example}
	Taking back Example \ref{ex:join_order}, the ordering required in this Corollary \ref{focal_atom_order_dual} simply translates to sorting the meet-irreducible elements of $L$ from the largest set to the smallest set. Thus, sorting ${^{\wedge}\mathcal{I}(L)} = \{ i_1, i_2, \cdots, i_n \}$ such that $|i_1| \geq |i_2| \geq \cdots \geq |i_n|$, Corollary \ref{focal_atom_order_dual} tells us that for any $k \in \llbracket 2, n \rrbracket$, we have $i_k \not\supseteq \bigcap {^{\wedge}\mathcal{I}(L)}_{k-1}$, where
	${^{\wedge}\mathcal{I}(L)}_{k} = \left\lbrace i_{1}, \dots, i_{k-1},~ i_{k}  \right\rbrace$.
\end{example}

\subsection{Main results}

In this subsection, we present our two types of transformations computing the zeta and M\"obius transforms of a function $f:P\rightarrow \mathbb{R}$. The first one corresponds to Theorem \ref{mob_opti_L} and its corollaries and is based on a sublattice $L \subseteq P$, where $L$ contains the lattice support of $f$, i.e. $ {^\mathcal{L}\supp{f}} \subseteq L$.
The second one corresponds to Theorem \ref{mob_opti_F} and its corollary and is based on any lower subsemilattice of $P$ containing $^\wedge \supp{f}$. Its dual is presented in Corollary \ref{mob_opti_F_dual} and its corollary and is based on any upper subsemilattice of $P$ containing $^\vee \supp{f}$.

\begin{figure}[t]
	\centering
	\hspace{-1cm}
	\begin{tikzpicture}[scale=0.75, every node/.style={transform shape},
	node/.style={draw, dot,minimum size=0.2cm, inner sep=0pt},
	det/.style={draw, diamond,minimum size=1.1cm, inner sep=0pt},
	rect/.style={draw, rectangle,minimum size=1.1cm, inner sep=2pt}
	]
	
	\node (none) at (-5, 0) {$\emptyset$};
	\node (a) at (-3.5, 0) {\{a\}};
	\node (b) at (-2.1, 0) {\{d\}};
	\node (ab) at (-0.7, 0) {\{a,d\}};
	\node (c) at (0.7, 0) {\{c,d,f\}};
	\node (ac) at (2.1, 0) {\{a,c,d,f\}};
	\node (bc) at (3.5, 0) {$\Omega$};
	
	\node (none1) at (-5, -0.5) {$\bullet$};
	\node (a1) at (-3.5, -0.5) {$\bullet$};
	\node (b1) at (-2.1, -0.5) {$\bullet$};
	\node (ab1) at (-0.7, -0.5) {$\bullet$};
	\node (c1) at (0.7, -0.5) {$\bullet$};
	\node (ac1) at (2.1, -0.5) {$\bullet$};
	\node (bc1) at (3.5, -0.5) {$\bullet$};
	
	\draw [decorate,decoration={brace,amplitude=2pt},yshift=0pt]
	(-5.4,-1.5) -- (-5.4,-0.5) node [black,midway,xshift=-1.55cm] {\footnotesize
		$1: Y = X \cup \Omega$};
	
	\node (none2) at (-5, -1.5) {$\bullet$};
	\node (a2) at (-3.5, -1.5) {$\bullet$};
	\node (b2) at (-2.1, -1.5) {$\bullet$};
	\node (ab2) at (-0.7, -1.5) {$\bullet$};
	\node (c2) at (0.7, -1.5) {$\bullet$};
	\node (ac2) at (2.1, -1.5) {$\bullet$};
	\node (bc2) at (3.5, -1.5) {$\bullet$};
	
	\draw [decorate,decoration={brace,amplitude=2pt},yshift=0pt]
	(-5.4,-2.5) -- (-5.4,-1.5) node [black,midway,xshift=-1.55cm] {\footnotesize
		$2: Y = X \cup \{c,d,f\}$};
	
	\node (none3) at (-5, -2.5) {$\bullet$};
	\node (a3) at (-3.5, -2.5) {$\bullet$};
	\node (b3) at (-2.1, -2.5) {$\bullet$};
	\node (ab3) at (-0.7, -2.5) {$\bullet$};
	\node (c3) at (0.7, -2.5) {$\bullet$};
	\node (ac3) at (2.1, -2.5) {$\bullet$};
	\node (bc3) at (3.5, -2.5) {$\bullet$};
	
	\draw [decorate,decoration={brace,amplitude=2pt},yshift=0pt]
	(-5.4,-3.5) -- (-5.4,-2.5) node [black,midway,xshift=-1.55cm] {\footnotesize
		$3: Y = X \cup \{d\}$};
	
	\node (none4) at (-5, -3.5) {$\bullet$};
	\node (a4) at (-3.5, -3.5) {$\bullet$};
	\node (b4) at (-2.1, -3.5) {$\bullet$};
	\node (ab4) at (-0.7, -3.5) {$\bullet$};
	\node (c4) at (0.7, -3.5) {$\bullet$};
	\node (ac4) at (2.1, -3.5) {$\bullet$};
	\node (bc4) at (3.5, -3.5) {$\bullet$};
	
	\draw [decorate,decoration={brace,amplitude=2pt},yshift=0pt]
	(-5.4,-4.5) -- (-5.4,-3.5) node [black,midway,xshift=-1.55cm] {\footnotesize
		$4: Y = X \cup \{a\}$};
	
	\node (none5) at (-5, -4.5) {$\bullet$};
	\node (a5) at (-3.5, -4.5) {$\bullet$};
	\node (b5) at (-2.1, -4.5) {$\bullet$};
	\node (ab5) at (-0.7, -4.5) {$\bullet$};
	\node (c5) at (0.7, -4.5) {$\bullet$};
	\node (ac5) at (2.1, -4.5) {$\bullet$};
	\node (bc5) at (3.5, -4.5) {$\bullet$};
	
	\draw[->,>=latex] (none4) to (a5);
	\draw[->,>=latex] (b4) to (ab5);
	\draw[->,>=latex] (c4) to (ac5);
	
	\draw[->,>=latex] (none3) to (b4);
	\draw[->,>=latex] (a3) to (ab4);
	
	\draw[->,>=latex] (none2) to (c3);
	\draw[->,>=latex] (a2) to (ac3);
	\draw[->,>=latex] (b2) to (c3);
	\draw[->,>=latex] (ab2) to (ac3);
	
	\draw[->,>=latex] (none1) to (bc2);
	\draw[->,>=latex] (a1) to (bc2);
	\draw[->,>=latex] (b1) to (bc2);
	\draw[->,>=latex] (ab1) to (bc2);
	\draw[->,>=latex] (c1) to (bc2);
	\draw[->,>=latex] (ac1) to (bc2);
	
	\end{tikzpicture}
	\caption{\small{
			Illustration representing the paths generated by the arrows contained in the sequence $(H_k)_{k\in\llbracket 1, 4\rrbracket}$, where $H_k = (L, E_k)$ and $E_k = \{ (X,Y) \in L^2 ~/~ Y = X \cup i_{5-k} \}$ and $L=\{ \emptyset, \{ a \}, \{ d \}, \{a,d\}, \{ c,d,f \}, \{ a,c,d,f \}, \Omega \}$ and $\Omega = \{ a,b,c,d,e,f \}$ and $(i_k)_{k\in \llbracket 1, 4 \rrbracket} = (\{ a \}, \{ d \}, \{ c,d,f \}, \Omega)$. This sequence computes the same zeta transformations as $G_\subset = (L, E_\subset)$, where $E_\subset = \{ (X,Y) \in L^2 ~/~ X \subset Y \}$. 
	}}
	\label{fig:EMT_L_sub}
\end{figure}
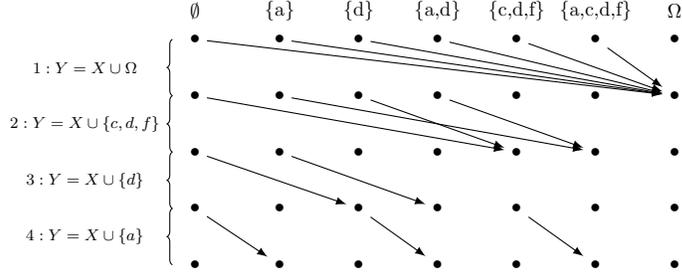

\newpage
\begin{theorem}[\textit{Efficient M\"obius Transformation in a distributive lattice}]\label{mob_opti_L}
	Let us consider a finite distributive lattice $L$ (such as $^\mathcal{L}\supp{f}$) and let its join-irreducible elements ${^{\vee}\mathcal{I}(L)} = \{ i_1, i_2, \cdots, i_n \}$ be ordered such that $\forall i_k, i_{l} \in {^{\vee}\mathcal{I}(L)}$, $k < l \Rightarrow i_k \not\geq i_{l}$.
	
	Consider the sequence $(H_k)_{k\in\llbracket 1, n\rrbracket}$, where $H_k = (L, E_k)$ and:
	$$E_k = \left\lbrace (x,y) \in L^2 ~/~ y = x \vee i_{n + 1 - k} \right\rbrace.$$
	This sequence computes the same zeta transformations as $G_< = (L, E_<)$, where $E_< = \{ (X,Y) \in L^2 ~/~ X < Y \}$.
	This sequence is illustrated in Fig. \ref{fig:EMT_L_sub}. The execution of any transformation based on this sequence is $O(n . |L|)$ in time and $O(|L|)$ in space.
\end{theorem}
	\begin{proof}
		See Appendix \ref{appendix:mob_opti_L}.
	\end{proof}

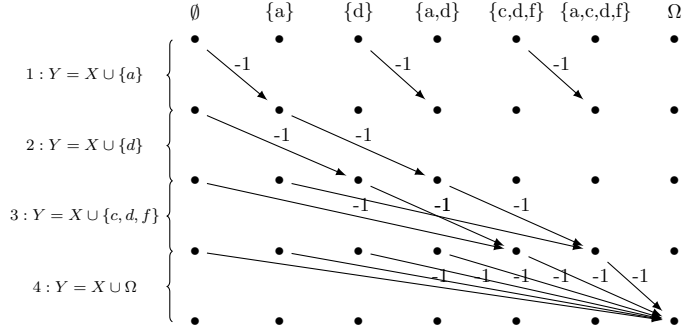
\begin{figure}[t]
	\centering
	\hspace{-1cm}
	\begin{tikzpicture}[scale=0.75, every node/.style={transform shape},
	node/.style={draw, dot,minimum size=0.2cm, inner sep=0pt},
	det/.style={draw, diamond,minimum size=1.1cm, inner sep=0pt},
	rect/.style={draw, rectangle,minimum size=1.1cm, inner sep=2pt}
	]
	
	\node (none) at (-5, 0.5) {$\emptyset$};
	\node (a) at (-3.5, 0.5) {\{a\}};
	\node (b) at (-2.1, 0.5) {\{d\}};
	\node (ab) at (-0.7, 0.5) {\{a,d\}};
	\node (c) at (0.7, 0.5) {\{c,d,f\}};
	\node (ac) at (2.1, 0.5) {\{a,c,d,f\}};
	\node (bc) at (3.5, 0.5) {$\Omega$};
	
	\node (none1) at (-5, 0) {$\bullet$};
	\node (a1) at (-3.5, 0) {$\bullet$};
	\node (b1) at (-2.1, 0) {$\bullet$};
	\node (ab1) at (-0.7, 0) {$\bullet$};
	\node (c1) at (0.7, 0) {$\bullet$};
	\node (ac1) at (2.1, 0) {$\bullet$};
	\node (bc1) at (3.5, 0) {$\bullet$};
	
	\draw [decorate,decoration={brace,amplitude=2pt},yshift=0pt]
	(-5.4,-\h) -- (-5.4,0) node [black,midway,xshift=-1.55cm] 
	{\footnotesize
		$1: Y = X \cup \{a\}$};
	
	\node (none2) at (-5, -\h) {$\bullet$};
	\node (a2) at (-3.5, -\h) {$\bullet$};
	\node (b2) at (-2.1, -\h) {$\bullet$};
	\node (ab2) at (-0.7, -\h) {$\bullet$};
	\node (c2) at (0.7, -\h) {$\bullet$};
	\node (ac2) at (2.1, -\h) {$\bullet$};
	\node (bc2) at (3.5, -\h) {$\bullet$};
	
	\draw [decorate,decoration={brace,amplitude=2pt},yshift=0pt]
	(-5.4,-2*\h) -- (-5.4,-\h) node [black,midway,xshift=-1.55cm] {\footnotesize
		$2: Y = X \cup \{d\}$};
	
	\node (none3) at (-5, -2*\h) {$\bullet$};
	\node (a3) at (-3.5, -2*\h) {$\bullet$};
	\node (b3) at (-2.1, -2*\h) {$\bullet$};
	\node (ab3) at (-0.7, -2*\h) {$\bullet$};
	\node (c3) at (0.7, -2*\h) {$\bullet$};
	\node (ac3) at (2.1, -2*\h) {$\bullet$};
	\node (bc3) at (3.5, -2*\h) {$\bullet$};
	
	\draw [decorate,decoration={brace,amplitude=2pt},yshift=0pt]
	(-5.4,-3*\h) -- (-5.4,-2*\h) node [black,midway,xshift=-1.55cm] {\footnotesize
		$3: Y= X \cup \{c,d,f\}$};
	
	\node (none4) at (-5, -3*\h) {$\bullet$};
	\node (a4) at (-3.5, -3*\h) {$\bullet$};
	\node (b4) at (-2.1, -3*\h) {$\bullet$};
	\node (ab4) at (-0.7, -3*\h) {$\bullet$};
	\node (c4) at (0.7, -3*\h) {$\bullet$};
	\node (ac4) at (2.1, -3*\h) {$\bullet$};
	\node (bc4) at (3.5, -3*\h) {$\bullet$};
	
	\draw [decorate,decoration={brace,amplitude=2pt},yshift=0pt]
	(-5.4,-4*\h) -- (-5.4,-3*\h) node [black,midway,xshift=-1.55cm] {\footnotesize
		$4: Y= X \cup \Omega$};
	
	\node (none5) at (-5, -4*\h) {$\bullet$};
	\node (a5) at (-3.5, -4*\h) {$\bullet$};
	\node (b5) at (-2.1, -4*\h) {$\bullet$};
	\node (ab5) at (-0.7, -4*\h) {$\bullet$};
	\node (c5) at (0.7, -4*\h) {$\bullet$};
	\node (ac5) at (2.1, -4*\h) {$\bullet$};
	\node (bc5) at (3.5, -4*\h) {$\bullet$};
	
	\draw[->,>=latex] (none1) to node[xshift=0.1cm, yshift=0.2cm] {-1}(a2);
	\draw[->,>=latex] (b1) to node[xshift=0.1cm, yshift=0.2cm] {-1}(ab2);
	\draw[->,>=latex] (c1) to node[xshift=0.1cm, yshift=0.2cm] {-1}(ac2);
	
	\draw[->,>=latex] (none2) to node[xshift=0.1cm, yshift=0.2cm] {-1}(b3);
	\draw[->,>=latex] (a2) to node[xshift=0.1cm, yshift=0.2cm] {-1}(ab3);
	
	\draw[->,>=latex] (none3) to node[xshift=0.1cm, yshift=0.2cm] {-1}(c4);
	\draw[->,>=latex] (a3) to node[xshift=0.1cm, yshift=0.2cm] {-1}(ac4);
	\draw[->,>=latex] (b3) to node[xshift=0.1cm, yshift=0.2cm] {-1}(c4);
	\draw[->,>=latex] (ab3) to node[xshift=0.1cm, yshift=0.2cm] {-1}(ac4);
	
	\draw[->,>=latex] (none4) to node[xshift=0.1cm, yshift=0.2cm] {-1}(bc5);
	\draw[->,>=latex] (a4) to node[xshift=0.1cm, yshift=0.2cm] {-1}(bc5);
	\draw[->,>=latex] (b4) to node[xshift=0.1cm, yshift=0.2cm] {-1}(bc5);
	\draw[->,>=latex] (ab4) to node[xshift=0.1cm, yshift=0.2cm] {-1}(bc5);
	\draw[->,>=latex] (c4) to node[xshift=0.1cm, yshift=0.2cm] {-1}(bc5);
	\draw[->,>=latex] (ac4) to node[xshift=0.1cm, yshift=0.2cm] {-1}(bc5);
	
	\end{tikzpicture}
	\caption{\small{Illustration representing the paths generated by the arrows contained in the sequence $(H_k)_{k\in\llbracket 1, 4\rrbracket}$, where $H_k = (L, E_k)$ and $E_k = \{ (X,Y) \in L^2 ~/~ Y = X \cup i_{k} \}$ and $L=\{ \emptyset, \{ a \}, \{ d \}, \{a,d\}, \{ c,d,f \}, \{ a,c,d,f \}, \Omega \}$ and $\Omega = \{ a,b,c,d,e,f \}$ and $(i_k)_{k\in \llbracket 1, 4 \rrbracket} = (\{ a \}, \{ d \}, \{ c,d,f \}, \Omega)$. This sequence computes the same M\"obius transformations as $G_\subset = (L, E_\subset)$, where $E_\subset = \{ (X,Y) \in L^2 ~/~ X \subset Y \}$. The ``-1'' labels emphasize the intended use of the operator $-$ with this sequence.}}
	\label{fig:EMT_L_sub_rev}
\end{figure}

\begin{corollary}\label{mob_opti_L_meet}
	It can be shown similarly that the sequence $(H_k)_{k\in\llbracket 1, n\rrbracket}$ computes the same zeta transforms as the sequence of Theorem \ref{mob_opti_L}, where $H_k = (L, E_k)$ and:
	\begin{align*}
	\hspace{-0.1cm}
	{E_k} = &\left\lbrace (x,y) \in L^2 ~/~ x = y \wedge \overline{i}_{{k}} \right\rbrace.
	\end{align*}
	with the meet-irreducible elements ${^{\wedge}\mathcal{I}(L)} = \{ \overline{i}_1, \overline{i}_2, \cdots, \overline{i}_n \}$ ordered such that $\forall \overline{i}_k, \overline{i}_{l} \in {^{\wedge}\mathcal{I}(L)}$, $k < l \Rightarrow \overline{i}_k \not\leq \overline{i}_{l}$, i.e. in reverse order compared to the join-irreducible elements of Theorem \ref{mob_opti_L}.
\end{corollary}

\begin{corollary}\label{mob_opti_L_rev}
	Consider the sequence $(H_k)_{k\in\llbracket 1, n\rrbracket}$, where $H_k = (L, E_k)$ and:
	$$E_k = \left\lbrace (x,y) \in L^2 ~/~ y = x \vee i_{k} \right\rbrace.$$
	This sequence computes the same M\"obius transformations as $G_< = (L, E_<)$, where $E_< = \{ (X,Y) \in L^2 ~/~ X < Y \}$. This sequence is illustrated in Fig. \ref{fig:EMT_L_sub_rev} and leads to the same complexities as the one presented in Theorem \ref{mob_opti_L}.
\end{corollary}
\begin{proof}
	For every arrow $(x, y) \in E_k$, if $i_k \not\leq x$, then there is an arrow such that $y = x \vee i_k \neq x$. However, there can be no arrow $(w, x) \in E_k$ since $x$ cannot be equal to $w \vee i_k$. Otherwise, if $i_k \leq x$, then $y = x \vee i_k = x$, i.e. $(x,y)$ is an identity arrow. Thus, for every arrow $(x, y) \in E_k\backslash I_P$, there is no arrow $(w, x)$ in $E_k\backslash I_P$, meaning that Theorem \ref{theorem:mobius} is satisfied. The sequence $(H_{n-k+1})_{k\in\llbracket 1, n\rrbracket}$ computes the same M\"obius transformations as $G_<$. 
\end{proof}

	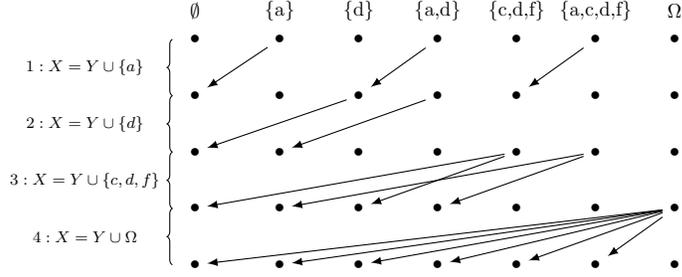
\begin{figure}[t]
	\centering
	\hspace{-1cm}
	\begin{tikzpicture}[scale=0.75, every node/.style={transform shape},
	node/.style={draw, dot,minimum size=0.2cm, inner sep=0pt},
	det/.style={draw, diamond,minimum size=1.1cm, inner sep=0pt},
	rect/.style={draw, rectangle,minimum size=1.1cm, inner sep=2pt}
	]
	
	\node (none) at (-5, 0) {$\emptyset$};
	\node (a) at (-3.5, 0) {\{a\}};
	\node (b) at (-2.1, 0) {\{d\}};
	\node (ab) at (-0.7, 0) {\{a,d\}};
	\node (c) at (0.7, 0) {\{c,d,f\}};
	\node (ac) at (2.1, 0) {\{a,c,d,f\}};
	\node (bc) at (3.5, 0) {$\Omega$};
	
	\node (none1) at (-5, -0.5) {$\bullet$};
	\node (a1) at (-3.5, -0.5) {$\bullet$};
	\node (b1) at (-2.1, -0.5) {$\bullet$};
	\node (ab1) at (-0.7, -0.5) {$\bullet$};
	\node (c1) at (0.7, -0.5) {$\bullet$};
	\node (ac1) at (2.1, -0.5) {$\bullet$};
	\node (bc1) at (3.5, -0.5) {$\bullet$};
	
	\draw [decorate,decoration={brace,amplitude=2pt},yshift=0pt]
	(-5.4,-1.5) -- (-5.4,-0.5) node [black,midway,xshift=-1.55cm] 
	{\footnotesize
		$1: X = Y \cup \{a\}$};
	
	\node (none2) at (-5, -1.5) {$\bullet$};
	\node (a2) at (-3.5, -1.5) {$\bullet$};
	\node (b2) at (-2.1, -1.5) {$\bullet$};
	\node (ab2) at (-0.7, -1.5) {$\bullet$};
	\node (c2) at (0.7, -1.5) {$\bullet$};
	\node (ac2) at (2.1, -1.5) {$\bullet$};
	\node (bc2) at (3.5, -1.5) {$\bullet$};
	
	\draw [decorate,decoration={brace,amplitude=2pt},yshift=0pt]
	(-5.4,-2.5) -- (-5.4,-1.5) node [black,midway,xshift=-1.55cm] {\footnotesize
		$2: X = Y \cup \{d\}$};
	
	\node (none3) at (-5, -2.5) {$\bullet$};
	\node (a3) at (-3.5, -2.5) {$\bullet$};
	\node (b3) at (-2.1, -2.5) {$\bullet$};
	\node (ab3) at (-0.7, -2.5) {$\bullet$};
	\node (c3) at (0.7, -2.5) {$\bullet$};
	\node (ac3) at (2.1, -2.5) {$\bullet$};
	\node (bc3) at (3.5, -2.5) {$\bullet$};
	
	\draw [decorate,decoration={brace,amplitude=2pt},yshift=0pt]
	(-5.4,-3.5) -- (-5.4,-2.5) node [black,midway,xshift=-1.55cm] {\footnotesize
		$3: X = Y \cup \{c,d,f\}$};
	
	\node (none4) at (-5, -3.5) {$\bullet$};
	\node (a4) at (-3.5, -3.5) {$\bullet$};
	\node (b4) at (-2.1, -3.5) {$\bullet$};
	\node (ab4) at (-0.7, -3.5) {$\bullet$};
	\node (c4) at (0.7, -3.5) {$\bullet$};
	\node (ac4) at (2.1, -3.5) {$\bullet$};
	\node (bc4) at (3.5, -3.5) {$\bullet$};
	
	\draw [decorate,decoration={brace,amplitude=2pt},yshift=0pt]
	(-5.4,-4.5) -- (-5.4,-3.5) node [black,midway,xshift=-1.55cm] {\footnotesize
		$4: X = Y \cup \Omega$};
	
	\node (none5) at (-5, -4.5) {$\bullet$};
	\node (a5) at (-3.5, -4.5) {$\bullet$};
	\node (b5) at (-2.1, -4.5) {$\bullet$};
	\node (ab5) at (-0.7, -4.5) {$\bullet$};
	\node (c5) at (0.7, -4.5) {$\bullet$};
	\node (ac5) at (2.1, -4.5) {$\bullet$};
	\node (bc5) at (3.5, -4.5) {$\bullet$};
	
	\draw[->,>=latex] (a1) to (none2);
	\draw[->,>=latex] (ab1) to (b2);
	\draw[->,>=latex] (ac1) to (c2);
	
	\draw[->,>=latex] (b2) to (none3);
	\draw[->,>=latex] (ab2) to (a3);
	
	\draw[->,>=latex] (c3) to (none4);
	\draw[->,>=latex] (ac3) to (a4);
	\draw[->,>=latex] (c3) to (b4);
	\draw[->,>=latex] (ac3) to (ab4);
	
	\draw[->,>=latex] (bc4) to (none5);
	\draw[->,>=latex] (bc4) to (a5);
	\draw[->,>=latex] (bc4) to (b5);
	\draw[->,>=latex] (bc4) to (ab5);
	\draw[->,>=latex] (bc4) to (c5);
	\draw[->,>=latex] (bc4) to (ac5);
	
	\end{tikzpicture}
	\caption{\small{Illustration representing the paths generated by the arrows contained in the sequence $(H_k)_{k\in\llbracket 1, 4\rrbracket}$, where $H_k = (L, E_k)$ and $E_k = \{ (X,Y) \in L^2 ~/~ X = Y \cup i_{k} \}$ and $L=\{ \emptyset, \{ a \}, \{ d \}, \{a,d\}, \{ c,d,f \}, \{ a,c,d,f \}, \Omega \}$ and $\Omega = \{ a,b,c,d,e,f \}$ and $(i_k)_{k\in \llbracket 1, 4 \rrbracket} = (\{ a \}, \{ d \}, \{ c,d,f \}, \Omega)$. This sequence computes the same zeta transformations as $G_\supset = (L, E_\supset)$, where $E_\supset = \{ (X,Y) \in L^2 ~/~ X \supset Y \}$. }}
	\label{fig:EMT_L_sup}
\end{figure}

\begin{corollary}\label{mob_opti_L_meet_rev}
	Again, it can be shown similarly that the sequence $(H_k)_{k\in\llbracket 1, n\rrbracket}$ computes the same M\"obius transforms as the sequence of Corollary \ref{mob_opti_L_rev}, where $H_k = (L, E_k)$ and:
	\begin{align*}
	\hspace{-0.1cm}
	{E_k} = &\left\lbrace (x,y) \in L^2 ~/~ x = y \wedge \overline{i}_{{n+1-k}} \right\rbrace.
	\end{align*}
	with the meet-irreducible elements ${^{\wedge}\mathcal{I}(L)} = \{ \overline{i}_1, \overline{i}_2, \cdots, \overline{i}_n \}$ ordered such that $\forall \overline{i}_k, \overline{i}_{l} \in {^{\wedge}\mathcal{I}(L)}$, $k < l \Rightarrow \overline{i}_k \not\leq \overline{i}_{l}$, i.e. in reverse order compared to the join-irreducible elements of Corollary \ref{mob_opti_L_rev}.
\end{corollary}

\begin{corollary}
	Dually, consider the sequence $(H_k)_{k\in\llbracket 1, n\rrbracket}$, where $H_k = (L, E_k)$ and:
	\begin{align*}
	\hspace{-0.1cm}
	{E_k} = &\left\lbrace (x,y) \in L^2 ~/~ x = y \vee i_{{k}} \right\rbrace.
	\end{align*}
	This sequence computes the same zeta transformations as $G_> = (L, E_>)$, where $E_> = \{ (X,Y) \in L^2 ~/~ X > Y \}$. This sequence is illustrated in Fig. \ref{fig:EMT_L_sup} and leads to the same complexities as its dual.
\end{corollary}

	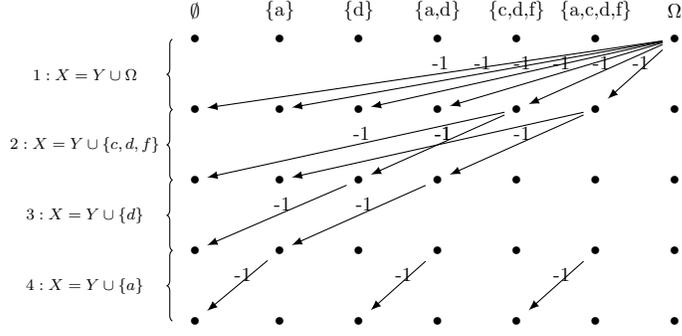
\begin{figure}[t]
	\centering
	\hspace{-1cm}
	\begin{tikzpicture}[scale=0.75, every node/.style={transform shape},
	node/.style={draw, dot,minimum size=0.2cm, inner sep=0pt},
	det/.style={draw, diamond,minimum size=1.1cm, inner sep=0pt},
	rect/.style={draw, rectangle,minimum size=1.1cm, inner sep=2pt}
	]
	
	\node (none) at (-5, 0.5) {$\emptyset$};
	\node (a) at (-3.5, 0.5) {\{a\}};
	\node (b) at (-2.1, 0.5) {\{d\}};
	\node (ab) at (-0.7, 0.5) {\{a,d\}};
	\node (c) at (0.7, 0.5) {\{c,d,f\}};
	\node (ac) at (2.1, 0.5) {\{a,c,d,f\}};
	\node (bc) at (3.5, 0.5) {$\Omega$};
	
	\node (none5) at (-5, 0) {$\bullet$};
	\node (a5) at (-3.5, 0) {$\bullet$};
	\node (b5) at (-2.1, 0) {$\bullet$};
	\node (ab5) at (-0.7, 0) {$\bullet$};
	\node (c5) at (0.7, 0) {$\bullet$};
	\node (ac5) at (2.1, 0) {$\bullet$};
	\node (bc5) at (3.5, 0) {$\bullet$};
	
	\draw [decorate,decoration={brace,amplitude=2pt},yshift=0pt]
	(-5.4,-\h) -- (-5.4,0) node [black,midway,xshift=-1.55cm] 
	{\footnotesize
		$1: X = Y \cup \Omega$};
	
	\node (none4) at (-5, -\h) {$\bullet$};
	\node (a4) at (-3.5, -\h) {$\bullet$};
	\node (b4) at (-2.1, -\h) {$\bullet$};
	\node (ab4) at (-0.7, -\h) {$\bullet$};
	\node (c4) at (0.7, -\h) {$\bullet$};
	\node (ac4) at (2.1, -\h) {$\bullet$};
	\node (bc4) at (3.5, -\h) {$\bullet$};
	
	\draw [decorate,decoration={brace,amplitude=2pt},yshift=0pt]
	(-5.4,-2*\h) -- (-5.4,-\h) node [black,midway,xshift=-1.55cm] {\footnotesize
		$2: X = Y \cup \{c,d,f\}$};
	
	\node (none3) at (-5, -2*\h) {$\bullet$};
	\node (a3) at (-3.5, -2*\h) {$\bullet$};
	\node (b3) at (-2.1, -2*\h) {$\bullet$};
	\node (ab3) at (-0.7, -2*\h) {$\bullet$};
	\node (c3) at (0.7, -2*\h) {$\bullet$};
	\node (ac3) at (2.1, -2*\h) {$\bullet$};
	\node (bc3) at (3.5, -2*\h) {$\bullet$};
	
	\draw [decorate,decoration={brace,amplitude=2pt},yshift=0pt]
	(-5.4,-3*\h) -- (-5.4,-2*\h) node [black,midway,xshift=-1.55cm] {\footnotesize
		$3: X = Y \cup \{d\}$};
	
	\node (none2) at (-5, -3*\h) {$\bullet$};
	\node (a2) at (-3.5, -3*\h) {$\bullet$};
	\node (b2) at (-2.1, -3*\h) {$\bullet$};
	\node (ab2) at (-0.7, -3*\h) {$\bullet$};
	\node (c2) at (0.7, -3*\h) {$\bullet$};
	\node (ac2) at (2.1, -3*\h) {$\bullet$};
	\node (bc2) at (3.5, -3*\h) {$\bullet$};
	
	\draw [decorate,decoration={brace,amplitude=2pt},yshift=0pt]
	(-5.4,-4*\h) -- (-5.4,-3*\h) node [black,midway,xshift=-1.55cm] {\footnotesize
		$4: X = Y \cup \{a\}$};
	
	\node (none1) at (-5, -4*\h) {$\bullet$};
	\node (a1) at (-3.5, -4*\h) {$\bullet$};
	\node (b1) at (-2.1, -4*\h) {$\bullet$};
	\node (ab1) at (-0.7, -4*\h) {$\bullet$};
	\node (c1) at (0.7, -4*\h) {$\bullet$};
	\node (ac1) at (2.1, -4*\h) {$\bullet$};
	\node (bc1) at (3.5, -4*\h) {$\bullet$};
	
	\draw[->,>=latex] (a2) to node[xshift=0.1cm, yshift=0.2cm] {-1}(none1);
	\draw[->,>=latex] (ab2) to node[xshift=0.1cm, yshift=0.2cm] {-1}(b1);
	\draw[->,>=latex] (ac2) to node[xshift=0.1cm, yshift=0.2cm] {-1}(c1);
	
	\draw[->,>=latex] (b3) to node[xshift=0.1cm, yshift=0.2cm] {-1}(none2);
	\draw[->,>=latex] (ab3) to node[xshift=0.1cm, yshift=0.2cm] {-1}(a2);
	
	\draw[->,>=latex] (c4) to node[xshift=0.1cm, yshift=0.2cm] {-1}(none3);
	\draw[->,>=latex] (ac4) to node[xshift=0.1cm, yshift=0.2cm] {-1}(a3);
	\draw[->,>=latex] (c4) to node[xshift=0.1cm, yshift=0.2cm] {-1}(b3);
	\draw[->,>=latex] (ac4) to node[xshift=0.1cm, yshift=0.2cm] {-1}(ab3);
	
	\draw[->,>=latex] (bc5) to node[xshift=0.1cm, yshift=0.2cm] {-1}(none4);
	\draw[->,>=latex] (bc5) to node[xshift=0.1cm, yshift=0.2cm] {-1}(a4);
	\draw[->,>=latex] (bc5) to node[xshift=0.1cm, yshift=0.2cm] {-1}(b4);
	\draw[->,>=latex] (bc5) to node[xshift=0.1cm, yshift=0.2cm] {-1}(ab4);
	\draw[->,>=latex] (bc5) to node[xshift=0.1cm, yshift=0.2cm] {-1}(c4);
	\draw[->,>=latex] (bc5) to node[xshift=0.1cm, yshift=0.2cm] {-1}(ac4);
	
	\end{tikzpicture}
	\caption{\small{Illustration representing the paths generated by the arrows contained in the sequence $(H_k)_{k\in\llbracket 1, 4\rrbracket}$, where $H_k = (L, E_k)$ and $E_k = \{ (X,Y) \in L^2 ~/~ X = Y \cup i_{5-k} \}$ and $L=\{ \emptyset, \{ a \}, \{ d \}, \{a,d\}, \{ c,d,f \}, \{ a,c,d,f \}, \Omega \}$ and $\Omega = \{ a,b,c,d,e,f \}$ and $(i_k)_{k\in \llbracket 1, 4 \rrbracket} = (\{ a \}, \{ d \}, \{ c,d,f \}, \Omega)$. This sequence computes the same M\"obius transformations as $G_\supset = (L, E_\supset)$, where $E_\supset = \{ (X,Y) \in L^2 ~/~ X \supset Y \}$. The ``-1'' labels emphasize the intended use of the operator $-$ with this sequence.}}
	\label{fig:EMT_L_sup_rev}
\end{figure}

\begin{corollary}
	Consider the sequence $(H_k)_{k\in\llbracket 1, n\rrbracket}$, where $H_k = (L, E_k)$ and:
	$$E_k = \left\lbrace (x,y) \in L^2 ~/~ x = y \vee i_{n+1-k} \right\rbrace.$$
	This sequence computes the same M\"obius transformations as $G_> = (L, E_>)$, where $E_> = \{ (X,Y) \in L^2 ~/~ X > Y \}$. This sequence 
	is illustrated in Fig. \ref{fig:EMT_L_sup_rev} and 
	leads to the same complexities as the one presented in Theorem \ref{mob_opti_L}.
\end{corollary}

The procedures exploiting the sequences of graphs described in Theorem \ref{mob_opti_L} and its corollaries to compute the zeta and M\"obius transforms of a function $f$ on $P$ is always less than $O(|{^\vee\mathcal{I}(P)}|.|P|)$. Their upper bound complexity for the distributive lattice $L={^\mathcal{L}\supp{f}}$ is $O(|{^\vee\mathcal{I}(L)}|.|{L}|)$, which is actually the optimal one for a lattice.

Yet, we can reduce this complexity even further if we have ${^\wedge\supp{f}}$ or ${^\vee\supp{f}}$. This is the motivation behind the sequence proposed in the following Theorem \ref{mob_opti_F}. As a matter of fact, \cite{bjorklund2016fast} proposed a meta-procedure producing an algorithm that computes the zeta and M\"obius transforms in an arbitrary intersection-closed family $F$ of sets of $2^\Omega$ with a circuit of size $O(|\Omega|.|F|)$. However, this meta-procedure is $O(|\Omega|.2^{|\Omega|})$. Here, Theorem \ref{mob_opti_F} provides a sequence of graphs that leads to procedures that directly compute the zeta and M\"obius transforms in $O(|\Omega|.|F|.\epsilon)$, where $\epsilon$ can be as low as 1. Besides, our method is far more general since it applies to any meet-closed or join-closed subset of a finite distributive lattice. The intuition behind it is that we can do the same computations as in Theorem \ref{mob_opti_L}, even in a subsemilattice, simply by bridging gaps, from the smallest gap to the biggest to make sure that all nodes are visited.

	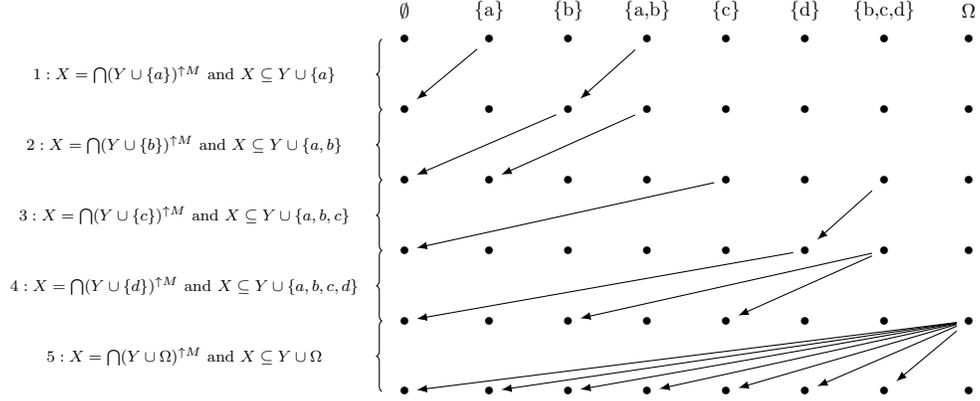
\begin{figure}[t]
	\centering
	\begin{tikzpicture}[scale=0.75, every node/.style={transform shape},
	node/.style={draw, dot,minimum size=0.2cm, inner sep=0pt},
	det/.style={draw, diamond,minimum size=1.1cm, inner sep=0pt},
	rect/.style={draw, rectangle,minimum size=1.1cm, inner sep=2pt}
	]
	
	\node (none) at (-5, 0.5) {$\emptyset$};
	\node (a) at (-3.5, 0.5) {\{a\}};
	\node (b) at (-2.1, 0.5) {\{b\}};
	\node (ab) at (-0.7, 0.5) {\{a,b\}};
	\node (c) at (0.7, 0.5) {\{c\}};
	\node (ac) at (2.1, 0.5) {\{d\}};
	\node (bc) at (3.5, 0.5) {\{b,c,d\}};
	\node (abc) at (5, 0.5) {$\Omega$};
	
	\node (none1) at (-5, 0) {$\bullet$};
	\node (a1) at (-3.5, 0) {$\bullet$};
	\node (b1) at (-2.1, 0) {$\bullet$};
	\node (ab1) at (-0.7, 0) {$\bullet$};
	\node (c1) at (0.7, 0) {$\bullet$};
	\node (ac1) at (2.1, 0) {$\bullet$};
	\node (bc1) at (3.5, 0) {$\bullet$};
	\node (abc1) at (5, 0) {$\bullet$};
	
	\draw [decorate,decoration={brace,amplitude=2pt},yshift=0pt]
	(-5.4,-\h) -- (-5.4,0) node [black,midway,xshift=-3.5cm] {\footnotesize
		$1: X = \bigcap (Y \cup \{a\})^{\uparrow M} \text{ and } X \subseteq Y \cup \{a\}$};
	
	\node (none2) at (-5, -\h) {$\bullet$};
	\node (a2) at (-3.5, -\h) {$\bullet$};
	\node (b2) at (-2.1, -\h) {$\bullet$};
	\node (ab2) at (-0.7, -\h) {$\bullet$};
	\node (c2) at (0.7, -\h) {$\bullet$};
	\node (ac2) at (2.1, -\h) {$\bullet$};
	\node (bc2) at (3.5, -\h) {$\bullet$};
	\node (abc2) at (5, -\h) {$\bullet$};
	
	\draw [decorate,decoration={brace,amplitude=2pt},yshift=0pt]
	(-5.4,-2*\h) -- (-5.4,-\h) node [black,midway,xshift=-3.5cm] {\footnotesize
		$2: X = \bigcap (Y \cup \{b\})^{\uparrow M} \text{ and } X \subseteq Y \cup \{a, b\}$};
	
	\node (none3) at (-5, -2*\h) {$\bullet$};
	\node (a3) at (-3.5, -2*\h) {$\bullet$};
	\node (b3) at (-2.1, -2*\h) {$\bullet$};
	\node (ab3) at (-0.7, -2*\h) {$\bullet$};
	\node (c3) at (0.7, -2*\h) {$\bullet$};
	\node (ac3) at (2.1, -2*\h) {$\bullet$};
	\node (bc3) at (3.5, -2*\h) {$\bullet$};
	\node (abc3) at (5, -2*\h) {$\bullet$};
	
	\draw [decorate,decoration={brace,amplitude=2pt},yshift=0pt]
	(-5.4,-3*\h) -- (-5.4,-2*\h) node [black,midway,xshift=-3.5cm] {\footnotesize
		$3: X = \bigcap (Y \cup \{c\})^{\uparrow M} \text{ and } X \subseteq Y \cup \{a,b,c\}$};
	
	\node (none4) at (-5, -3*\h) {$\bullet$};
	\node (a4) at (-3.5, -3*\h) {$\bullet$};
	\node (b4) at (-2.1, -3*\h) {$\bullet$};
	\node (ab4) at (-0.7, -3*\h) {$\bullet$};
	\node (c4) at (0.7, -3*\h) {$\bullet$};
	\node (ac4) at (2.1, -3*\h) {$\bullet$};
	\node (bc4) at (3.5, -3*\h) {$\bullet$};
	\node (abc4) at (5, -3*\h) {$\bullet$};
	
	\draw [decorate,decoration={brace,amplitude=2pt},yshift=0pt]
	(-5.4,-4*\h) -- (-5.4,-3*\h) node [black,midway,xshift=-3.5cm] {\footnotesize
		$4: X = \bigcap (Y \cup \{d\})^{\uparrow M} \text{ and } X \subseteq Y \cup \{a,b,c,d\}$};
	
	\node (none5) at (-5, -4*\h) {$\bullet$};
	\node (a5) at (-3.5, -4*\h) {$\bullet$};
	\node (b5) at (-2.1, -4*\h) {$\bullet$};
	\node (ab5) at (-0.7, -4*\h) {$\bullet$};
	\node (c5) at (0.7, -4*\h) {$\bullet$};
	\node (ac5) at (2.1, -4*\h) {$\bullet$};
	\node (bc5) at (3.5, -4*\h) {$\bullet$};
	\node (abc5) at (5, -4*\h) {$\bullet$};
	
	\draw [decorate,decoration={brace,amplitude=2pt},yshift=0pt]
	(-5.4,-5*\h) -- (-5.4,-4*\h) node [black,midway,xshift=-3.5cm] {\footnotesize
		$5: X = \bigcap (Y \cup \Omega)^{\uparrow M} \text{ and } X \subseteq Y \cup \Omega$};
	
	\node (none6) at (-5, -5*\h) {$\bullet$};
	\node (a6) at (-3.5, -5*\h) {$\bullet$};
	\node (b6) at (-2.1, -5*\h) {$\bullet$};
	\node (ab6) at (-0.7, -5*\h) {$\bullet$};
	\node (c6) at (0.7, -5*\h) {$\bullet$};
	\node (ac6) at (2.1, -5*\h) {$\bullet$};
	\node (bc6) at (3.5, -5*\h) {$\bullet$};
	\node (abc6) at (5, -5*\h) {$\bullet$};
	
	\draw[->,>=latex] (a1) to (none2);
	\draw[->,>=latex] (ab1) to (b2);
	
	\draw[->,>=latex] (b2) to (none3);
	\draw[->,>=latex] (ab2) to (a3);
	
	\draw[->,>=latex] (c3) to (none4);
	\draw[->,>=latex] (bc3) to (ac4);
	
	\draw[->,>=latex] (ac4) to (none5);
	\draw[->,>=latex] (bc4) to (b5);
	\draw[->,>=latex] (bc4) to (c5);
	
	\draw[->,>=latex] (abc5) to (none6);
	\draw[->,>=latex] (abc5) to (a6);
	\draw[->,>=latex] (abc5) to (b6);
	\draw[->,>=latex] (abc5) to (c6);
	\draw[->,>=latex] (abc5) to (ab6);
	\draw[->,>=latex] (abc5) to (ac6);
	\draw[->,>=latex] (abc5) to (bc6);
	
	\end{tikzpicture}
	\caption{\small{
			Illustration representing the paths generated by the arrows contained in the sequence $(H_k)_{k\in\llbracket 1, 5\rrbracket}$, where $H_k = (M, E_k)$ and $E_k = \left\lbrace (X, Y) \in M^2 ~/~ X = \bigcap (Y \cup i_k)^{\uparrow M} \text{ and } X \subseteq Y \cup \bigcup\iota(M)_k \right\rbrace$
			and $\iota(M)_k = \{ {i}_1, {i}_2, \dots, {i}_{k} \}$ and $M=\{ \emptyset, \{ a \}, \{ b \}, \{ a,b \}, \{ c \}, \{ d \}, \{b,c,d\}, \Omega \}$ and $\Omega = \{ a,b,c,d,e,f \}$ and $(i_k)_{k\in \llbracket 1, 5 \rrbracket} = (\{ a \}, \{ b \}, \{ c \}, \{ d \}, \Omega)$. This sequence computes the same zeta transformations as $G_\supset = (M, E_\supset)$, where $E_\supset = \{ (X,Y) \in M^2 ~/~ X \supset Y \}$. Actually, since there is no order between any two iota elements of $\iota(M)\backslash\{\Omega\}$ in this example, the order chosen here is arbitrary. Any order would compute the same zeta transformations as $G_\supset$, as long as $\Omega$ is the last iota element to consider.}
	}
	\label{fig:EMT_F_sup}
\end{figure}

\begin{theorem}[\textit{Efficient M\"obius Transformation in a join-closed or meet-closed subset of $P$}]\label{mob_opti_F}
	Let us consider a meet-closed subset $M$ of $P$ (such as $^\wedge\supp{f}$). Also, let the iota elements $\iota(M) = \{ i_1, i_2, \cdots, i_n \}$ be ordered such that $\forall i_k, i_{l} \in \iota(M)$, $k < l \Rightarrow i_k \not\geq i_{l}$.
	
	Consider the sequence $(H_k)_{k\in\llbracket 1, n\rrbracket}$, where $H_k = (M, E_k)$ and:
	$$E_k = \left\lbrace (x,y) \in M^2 ~/~ x = \bigwedge (y \vee i_k)^{\uparrow M} \text{ and } x \leq y \vee \bigvee\iota(M)_k \right\rbrace,$$
	where $\iota(M)_k = \{ {i}_1, {i}_2, \dots, {i}_{k} \}$.
	This sequence computes the same zeta transformations as $G_> = (M, E_>)$, where $E_> = \{ (X,Y) \in M^2 ~/~ X > Y \}$. This sequence is illustrated in Fig. \ref{fig:EMT_F_sup}. The execution of any transformation based on this sequence is at most $O(|\iota(M)| . |M|)$ in space and $O(|\iota(M)| . |M|.\epsilon)$ in time, where $\epsilon$ represents the average number of operations required to ``bridge a gap'', i.e. to find the minimum of $(y \vee i_k)^{\uparrow M}$. 
\end{theorem}
\begin{proof}
	See Appendix \ref{appendix:mob_opti_F}
\end{proof}

\begin{remark}
	This number $\epsilon$ is hard to evaluate beforehand since it depends on both the number of ``gaps to bridge'' and the average number of elements that can be greater than some element. If there is no gap, or if there is only one greater element, this $\epsilon$ can be as low as 1. Moreover, the choice of data structures may greatly reduce this $\epsilon$. For instance, if $P=2^\Omega$, then dynamic binary trees may be employed to avoid the consideration of elements that cannot be less than some already found element in $(y \cup i_k)^{\uparrow M}$, by cutting some branches.
\end{remark}


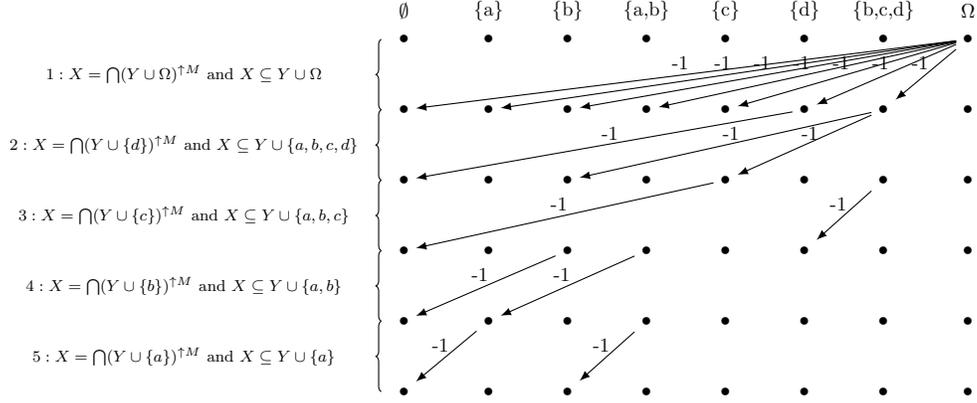
\begin{figure}[t]
	\centering
	\hspace{-0.5cm}
	\begin{tikzpicture}[scale=0.75, every node/.style={transform shape},
	node/.style={draw, dot,minimum size=0.2cm, inner sep=0pt},
	det/.style={draw, diamond,minimum size=1.1cm, inner sep=0pt},
	rect/.style={draw, rectangle,minimum size=1.1cm, inner sep=2pt}
	]
	
	\node (none) at (-5, 0.5+\h) {$\emptyset$};
	\node (a) at (-3.5, 0.5+\h) {\{a\}};
	\node (b) at (-2.1, 0.5+\h) {\{b\}};
	\node (ab) at (-0.7, 0.5+\h) {\{a,b\}};
	\node (c) at (0.7, 0.5+\h) {\{c\}};
	\node (ac) at (2.1, 0.5+\h) {\{d\}};
	\node (bc) at (3.5, 0.5+\h) {\{b,c,d\}};
	\node (abc) at (5, 0.5+\h) {$\Omega$};

	\node (none6) at (-5, \h) {$\bullet$};
	\node (a6) at (-3.5, \h) {$\bullet$};
	\node (b6) at (-2.1, \h) {$\bullet$};
	\node (ab6) at (-0.7, \h) {$\bullet$};
	\node (c6) at (0.7, \h) {$\bullet$};
	\node (ac6) at (2.1, \h) {$\bullet$};
	\node (bc6) at (3.5, \h) {$\bullet$};
	\node (abc6) at (5, \h) {$\bullet$};
	
	\draw [decorate,decoration={brace,amplitude=2pt},yshift=0pt]
	(-5.4,0) -- (-5.4,\h) node [black,midway,xshift=-3.5cm] {\footnotesize
		$1: X = \bigcap (Y \cup \Omega)^{\uparrow M} \text{ and } X \subseteq Y \cup \Omega$};
	
	\node (none1) at (-5, 0) {$\bullet$};
	\node (a1) at (-3.5, 0) {$\bullet$};
	\node (b1) at (-2.1, 0) {$\bullet$};
	\node (ab1) at (-0.7, 0) {$\bullet$};
	\node (c1) at (0.7, 0) {$\bullet$};
	\node (ac1) at (2.1, 0) {$\bullet$};
	\node (bc1) at (3.5, 0) {$\bullet$};
	\node (abc1) at (5, 0) {$\bullet$};
	
	\draw [decorate,decoration={brace,amplitude=2pt},yshift=0pt]
	(-5.4,-\h) -- (-5.4,0) node [black,midway,xshift=-3.5cm] {\footnotesize
		$2: X = \bigcap (Y \cup \{d\})^{\uparrow M} \text{ and } X \subseteq Y \cup \{a,b,c,d\}$};
	
	\node (none2) at (-5, -\h) {$\bullet$};
	\node (a2) at (-3.5, -\h) {$\bullet$};
	\node (b2) at (-2.1, -\h) {$\bullet$};
	\node (ab2) at (-0.7, -\h) {$\bullet$};
	\node (c2) at (0.7, -\h) {$\bullet$};
	\node (ac2) at (2.1, -\h) {$\bullet$};
	\node (bc2) at (3.5, -\h) {$\bullet$};
	\node (abc2) at (5, -\h) {$\bullet$};
	
	\draw [decorate,decoration={brace,amplitude=2pt},yshift=0pt]
	(-5.4,-2*\h) -- (-5.4,-\h) node [black,midway,xshift=-3.5cm] {\footnotesize
		$3: X = \bigcap (Y \cup \{c\})^{\uparrow M} \text{ and } X \subseteq Y \cup \{a,b,c\}$};
	
	\node (none3) at (-5, -2*\h) {$\bullet$};
	\node (a3) at (-3.5, -2*\h) {$\bullet$};
	\node (b3) at (-2.1, -2*\h) {$\bullet$};
	\node (ab3) at (-0.7, -2*\h) {$\bullet$};
	\node (c3) at (0.7, -2*\h) {$\bullet$};
	\node (ac3) at (2.1, -2*\h) {$\bullet$};
	\node (bc3) at (3.5, -2*\h) {$\bullet$};
	\node (abc3) at (5, -2*\h) {$\bullet$};
	
	\draw [decorate,decoration={brace,amplitude=2pt},yshift=0pt]
	(-5.4,-3*\h) -- (-5.4,-2*\h) node [black,midway,xshift=-3.5cm] {\footnotesize
		$4: X = \bigcap (Y \cup \{b\})^{\uparrow M} \text{ and } X \subseteq Y \cup \{a, b\}$};
	
	\node (none4) at (-5, -3*\h) {$\bullet$};
	\node (a4) at (-3.5, -3*\h) {$\bullet$};
	\node (b4) at (-2.1, -3*\h) {$\bullet$};
	\node (ab4) at (-0.7, -3*\h) {$\bullet$};
	\node (c4) at (0.7, -3*\h) {$\bullet$};
	\node (ac4) at (2.1, -3*\h) {$\bullet$};
	\node (bc4) at (3.5, -3*\h) {$\bullet$};
	\node (abc4) at (5, -3*\h) {$\bullet$};
	
	\draw [decorate,decoration={brace,amplitude=2pt},yshift=0pt]
	(-5.4,-4*\h) -- (-5.4,-3*\h) node [black,midway,xshift=-3.5cm] {\footnotesize
		$5: X = \bigcap (Y \cup \{a\})^{\uparrow M} \text{ and } X \subseteq Y \cup \{a\}$};
	
	\node (none5) at (-5, -4*\h) {$\bullet$};
	\node (a5) at (-3.5, -4*\h) {$\bullet$};
	\node (b5) at (-2.1, -4*\h) {$\bullet$};
	\node (ab5) at (-0.7, -4*\h) {$\bullet$};
	\node (c5) at (0.7, -4*\h) {$\bullet$};
	\node (ac5) at (2.1, -4*\h) {$\bullet$};
	\node (bc5) at (3.5, -4*\h) {$\bullet$};
	\node (abc5) at (5, -4*\h) {$\bullet$};
	
	\draw[->,>=latex] (a4) to node[xshift=-0.1cm, yshift=0.2cm] {-1}(none5);
	\draw[->,>=latex] (ab4) to node[xshift=-0.1cm, yshift=0.2cm] {-1}(b5);
	
	\draw[->,>=latex] (b3) to node[xshift=-0.1cm, yshift=0.2cm] {-1}(none4);
	\draw[->,>=latex] (ab3) to node[xshift=-0.1cm, yshift=0.2cm] {-1}(a4);
	
	\draw[->,>=latex] (c2) to node[xshift=-0.1cm, yshift=0.2cm] {-1}(none3);
	\draw[->,>=latex] (bc2) to node[xshift=-0.1cm, yshift=0.2cm] {-1}(ac3);
	
	\draw[->,>=latex] (ac1) to node[xshift=0.1cm, yshift=0.2cm] {-1}(none2);
	\draw[->,>=latex] (bc1) to node[xshift=0.1cm, yshift=0.2cm] {-1}(b2);
	\draw[->,>=latex] (bc1) to node[xshift=0.1cm, yshift=0.2cm] {-1}(c2);
	
	\draw[->,>=latex] (abc6) to node[xshift=-0.1cm, yshift=0.2cm] {-1}(none1);
	\draw[->,>=latex] (abc6) to node[xshift=-0.1cm, yshift=0.2cm] {-1}(a1);
	\draw[->,>=latex] (abc6) to node[xshift=-0.1cm, yshift=0.2cm] {-1}(b1);
	\draw[->,>=latex] (abc6) to node[xshift=-0.1cm, yshift=0.2cm] {-1}(c1);
	\draw[->,>=latex] (abc6) to node[xshift=-0.1cm, yshift=0.2cm] {-1}(ab1);
	\draw[->,>=latex] (abc6) to node[xshift=-0.1cm, yshift=0.2cm] {-1}(ac1);
	\draw[->,>=latex] (abc6) to node[xshift=-0.1cm, yshift=0.2cm] {-1}(bc1);
	
	\end{tikzpicture}
	\caption{\small{
			Illustration representing the paths generated by the arrows contained in the sequence $(H_k)_{k\in\llbracket 1, 5\rrbracket}$, where $H_k = (M, E_k)$ and $E_k = \left\lbrace (X, Y) \in M^2 ~/~ X = \bigcap (Y \cup i_{6-k})^{\uparrow M} \text{ and } X \subseteq Y \cup \bigcup\iota(M)_{6-k} \right\rbrace$
			and $\iota(M)_k = \{ {i}_1, {i}_2, \dots, {i}_{k} \}$ and $M=\{ \emptyset, \{ a \}, \{ b \}, \{ a,b \}, \{ c \}, \{ d \}, \{b,c,d\}, \Omega \}$ and $\Omega = \{ a,b,c,d,e,f \}$ and $(i_k)_{k\in \llbracket 1, 5 \rrbracket} = (\{ a \}, \{ b \}, \{ c \}, \{ d \}, \Omega)$. This sequence computes the same M\"obius transformations as $G_\supset = (M, E_\supset)$, where $E_\supset = \{ (X,Y) \in M^2 ~/~ X \supset Y \}$. The ``-1'' labels emphasize the intended use of the operator $-$ with this sequence. Actually, since there is no order between any two iota elements of $\iota(M)\backslash\{\Omega\}$ in this example, the order chosen in Fig. \ref{fig:EMT_F_sup} is arbitrary. Thus, any order here would compute the same M\"obius transformations as $G_\supset$, as long as $\Omega$ is the first iota element to consider.
	}}
	\label{fig:EMT_F_sup_rev}
\end{figure}

\begin{corollary}
	Consider the sequence $(H_k)_{k\in\llbracket 1, n\rrbracket}$, where $H_k = (M, E_k)$ and:
	$$E_k = \left\lbrace (x,y) \in M^2 ~/~ x = \bigwedge (y \vee i_{n+1-k})^{\uparrow M} \text{ and } x \leq y \vee \bigvee\iota(M)_{n+1-k} \right\rbrace.$$
	This sequence computes the same M\"obius transformations as $G_> = (M, E_>)$, where $E_> = \{ (X,Y) \in M^2 ~/~ X > Y \}$. This sequence is illustrated in Fig. \ref{fig:EMT_F_sup_rev} and leads to the same complexities as the one presented in Theorem \ref{mob_opti_F}.
\end{corollary}
\begin{proof}
	Analog to the proof of Corollary \ref{mob_opti_L_rev}.
\end{proof}

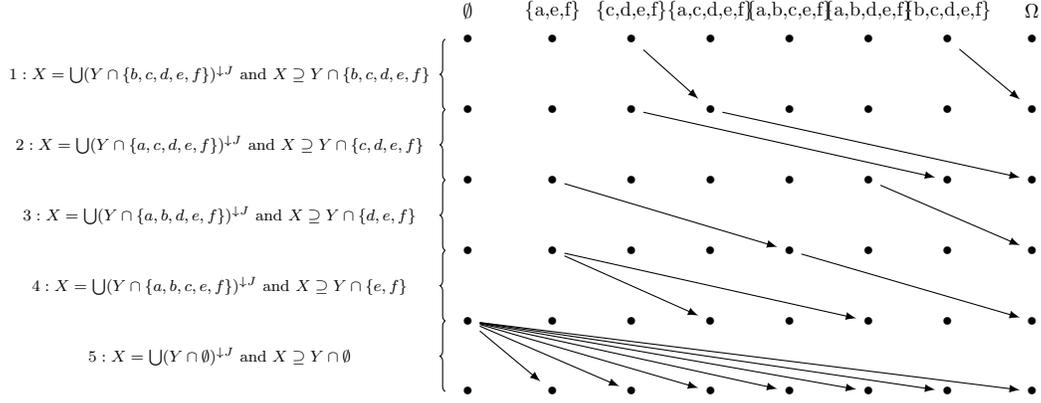
\begin{figure}[t]
	\centering
	\hspace{-0.4cm}
	\begin{tikzpicture}[scale=0.75, every node/.style={transform shape},
	node/.style={draw, dot,minimum size=0.2cm, inner sep=0pt},
	det/.style={draw, diamond,minimum size=1.1cm, inner sep=0pt},
	rect/.style={draw, rectangle,minimum size=1.1cm, inner sep=2pt}
	]
	
	\node (none) at (-5, 0.5) {$\emptyset$};
	\node (a) at (-3.5, 0.5) {\{a,e,f\}};
	\node (b) at (-2.1, 0.5) {\{c,d,e,f\}};
	\node (ab) at (-0.7, 0.5) {\{a,c,d,e,f\}};
	\node (c) at (0.7, 0.5) {\{a,b,c,e,f\}};
	\node (ac) at (2.1, 0.5) {\{a,b,d,e,f\}};
	\node (bc) at (3.5, 0.5) {\{b,c,d,e,f\}};
	\node (abc) at (5, 0.5) {$\Omega$};
	
	\node (none1) at (-5, 0) {$\bullet$};
	\node (a1) at (-3.5, 0) {$\bullet$};
	\node (b1) at (-2.1, 0) {$\bullet$};
	\node (ab1) at (-0.7, 0) {$\bullet$};
	\node (c1) at (0.7, 0) {$\bullet$};
	\node (ac1) at (2.1, 0) {$\bullet$};
	\node (bc1) at (3.5, 0) {$\bullet$};
	\node (abc1) at (5, 0) {$\bullet$};
	
	\draw [decorate,decoration={brace,amplitude=2pt},yshift=0pt]
	(-5.4,-\h) -- (-5.4,0) node [black,midway,xshift=-4cm] {\footnotesize
		$1: X = \bigcup (Y \cap \{b,c,d,e,f\})^{\downarrow J} \text{ and } X \supseteq Y \cap \{b,c,d,e,f\}$};
	
	\node (none2) at (-5, -\h) {$\bullet$};
	\node (a2) at (-3.5, -\h) {$\bullet$};
	\node (b2) at (-2.1, -\h) {$\bullet$};
	\node (ab2) at (-0.7, -\h) {$\bullet$};
	\node (c2) at (0.7, -\h) {$\bullet$};
	\node (ac2) at (2.1, -\h) {$\bullet$};
	\node (bc2) at (3.5, -\h) {$\bullet$};
	\node (abc2) at (5, -\h) {$\bullet$};
	
	\draw [decorate,decoration={brace,amplitude=2pt},yshift=0pt]
	(-5.4,-2*\h) -- (-5.4,-\h) node [black,midway,xshift=-4cm] {\footnotesize
		$2: X = \bigcup (Y \cap \{a,c,d,e,f\})^{\downarrow J} \text{ and } X \supseteq Y \cap \{c,d,e,f\}$};
	
	\node (none3) at (-5, -2*\h) {$\bullet$};
	\node (a3) at (-3.5, -2*\h) {$\bullet$};
	\node (b3) at (-2.1, -2*\h) {$\bullet$};
	\node (ab3) at (-0.7, -2*\h) {$\bullet$};
	\node (c3) at (0.7, -2*\h) {$\bullet$};
	\node (ac3) at (2.1, -2*\h) {$\bullet$};
	\node (bc3) at (3.5, -2*\h) {$\bullet$};
	\node (abc3) at (5, -2*\h) {$\bullet$};
	
	\draw [decorate,decoration={brace,amplitude=2pt},yshift=0pt]
	(-5.4,-3*\h) -- (-5.4,-2*\h) node [black,midway,xshift=-4cm] {\footnotesize
		$3: X = \bigcup (Y \cap \{a,b,d,e,f\})^{\downarrow J} \text{ and } X \supseteq Y \cap \{d,e,f\}$};
	
	\node (none4) at (-5, -3*\h) {$\bullet$};
	\node (a4) at (-3.5, -3*\h) {$\bullet$};
	\node (b4) at (-2.1, -3*\h) {$\bullet$};
	\node (ab4) at (-0.7, -3*\h) {$\bullet$};
	\node (c4) at (0.7, -3*\h) {$\bullet$};
	\node (ac4) at (2.1, -3*\h) {$\bullet$};
	\node (bc4) at (3.5, -3*\h) {$\bullet$};
	\node (abc4) at (5, -3*\h) {$\bullet$};
	
	\draw [decorate,decoration={brace,amplitude=2pt},yshift=0pt]
	(-5.4,-4*\h) -- (-5.4,-3*\h) node [black,midway,xshift=-4cm] {\footnotesize
		$4: X = \bigcup (Y \cap \{a,b,c,e,f\})^{\downarrow J} \text{ and } X \supseteq Y \cap \{e,f\}$};
	
	\node (none5) at (-5, -4*\h) {$\bullet$};
	\node (a5) at (-3.5, -4*\h) {$\bullet$};
	\node (b5) at (-2.1, -4*\h) {$\bullet$};
	\node (ab5) at (-0.7, -4*\h) {$\bullet$};
	\node (c5) at (0.7, -4*\h) {$\bullet$};
	\node (ac5) at (2.1, -4*\h) {$\bullet$};
	\node (bc5) at (3.5, -4*\h) {$\bullet$};
	\node (abc5) at (5, -4*\h) {$\bullet$};
	
	\draw [decorate,decoration={brace,amplitude=2pt},yshift=0pt]
	(-5.4,-5*\h) -- (-5.4,-4*\h) node [black,midway,xshift=-4cm] {\footnotesize
		$5: X = \bigcup (Y \cap \emptyset)^{\downarrow J} \text{ and } X \supseteq Y \cap \emptyset$};
	
	\node (none6) at (-5, -5*\h) {$\bullet$};
	\node (a6) at (-3.5, -5*\h) {$\bullet$};
	\node (b6) at (-2.1, -5*\h) {$\bullet$};
	\node (ab6) at (-0.7, -5*\h) {$\bullet$};
	\node (c6) at (0.7, -5*\h) {$\bullet$};
	\node (ac6) at (2.1, -5*\h) {$\bullet$};
	\node (bc6) at (3.5, -5*\h) {$\bullet$};
	\node (abc6) at (5, -5*\h) {$\bullet$};
	
	\draw[->,>=latex] (b1) to (ab2);
	\draw[->,>=latex] (bc1) to (abc2);
	
	\draw[->,>=latex] (ab2) to (abc3);
	\draw[->,>=latex] (b2) to (bc3);
	
	\draw[->,>=latex] (ac3) to (abc4);
	\draw[->,>=latex] (a3) to (c4);
	
	\draw[->,>=latex] (c4) to (abc5);
	\draw[->,>=latex] (a4) to (ab5);
	\draw[->,>=latex] (a4) to (ac5);
	
	\draw[->,>=latex] (none5) to (abc6);
	\draw[->,>=latex] (none5) to (a6);
	\draw[->,>=latex] (none5) to (b6);
	\draw[->,>=latex] (none5) to (c6);
	\draw[->,>=latex] (none5) to (ab6);
	\draw[->,>=latex] (none5) to (ac6);
	\draw[->,>=latex] (none5) to (bc6);
	
	\end{tikzpicture}
	\caption{\small{
			Illustration representing the paths generated by the arrows contained in the sequence $(H_k)_{k\in\llbracket 1, 5\rrbracket}$, where $H_k = (J, E_k)$ and $E_k = \left\lbrace (X, Y) \in J^2 ~/~ X = \bigcup (Y \cap \overline{i}_k)^{\downarrow J} \text{ and } X \supseteq Y \cap \bigcap\overline{\iota}(J)_k \right\rbrace$
			and $\overline{\iota}(J)_k = \{ \overline{i}_1, \overline{i}_2, \dots, \overline{i}_{k} \}$ and $J=\{ \emptyset, \{a,e,f\}, \{c,d,e,f\}, \{a,c,d,e,f\}, \{a,b,c,e,f\}, \{a,b,d,e,f\}, \{b,c,d,e,f\}, \Omega \}$ and $\Omega = \{ a,b,c,d,e,f \}$ and $(\overline{i}_k)_{k\in \llbracket 1, 5 \rrbracket} = (\{ b,c,d,e,f \}, \{ a,c,d,e,f \}, \{ a,b,d,e,f \}, \{ a,b,c,e,f \}, \emptyset)$. This sequence computes the same zeta transformations as $G_\subset = (J, E_\subset)$, where $E_\subset = \{ (X,Y) \in J^2 ~/~ X \subset Y \}$. Actually, since there is no order between any two dual iota elements of $\overline{\iota}(J)\backslash\{\emptyset\}$ in this example, the order chosen here is arbitrary. Any order would compute the same zeta transformations as $G_\subset$, as long as $\emptyset$ is the last dual iota element to consider.	
	}}
	\label{fig:EMT_F_sub}
\end{figure}

\begin{corollary}\label{mob_opti_F_dual}
	Dually, let us consider a join-closed subset $J$ of $P$ (such as $^\vee\supp{f}$). Also, let the dual iota elements $\overline{\iota}(J) = \{ \overline{i}_1, \overline{i}_2, \cdots, \overline{i}_n \}$ be ordered such that $\forall \overline{i}_k, \overline{i}_{l} \in \overline{\iota}(J)$, $k < l \Rightarrow \overline{i}_k \not\leq \overline{i}_{l}$, i.e. in reverse order compared to the iota elements of Theorem \ref{mob_opti_F}.
	
	In the direct line of Corollary \ref{mob_opti_L_meet}, consider the sequence $(H_k)_{k\in\llbracket 1, n\rrbracket}$, where $H_k = (J, E_k)$ and:
	$$E_k = \bigg\lbrace (x,y) \in J^2 ~/~ x = \hspace{-0.cm}\bigvee(y \wedge \overline{i}_k)^{\downarrow J} \text{ and } x \geq y \wedge \bigwedge\overline{\iota}(J)_k \bigg\rbrace,$$
	where $\overline{\iota}(J)_k = \{ \overline{i}_1, \overline{i}_2, \dots, \overline{i}_{k} \}$.
	This sequence computes the same zeta transformations as $G_< = (J, E_<)$, where $E_< = \{ (X,Y) \in J^2 ~/~ X < Y \}$. This sequence is illustrated in Fig. \ref{fig:EMT_F_sub}. The execution of any transformation based on this sequence is at most $O(|\overline{\iota}(J)| . |J|)$ in space and $O(|\overline{\iota}(J)| . |J|.\epsilon)$ in time, where $\epsilon$ represents the average number of operations required to ``bridge a gap'', i.e. to find the maximum of $(y \wedge \overline{i}_k)^{\downarrow M}$. 
\end{corollary}


\begin{figure}[t]
	\centering
	\hspace{-0.4cm}
	\begin{tikzpicture}[scale=0.75, every node/.style={transform shape},
	node/.style={draw, dot,minimum size=0.2cm, inner sep=0pt},
	det/.style={draw, diamond,minimum size=1.1cm, inner sep=0pt},
	rect/.style={draw, rectangle,minimum size=1.1cm, inner sep=2pt}
	]
	
	\node (none) at (-5, 0.5) {$\emptyset$};
	\node (a) at (-3.5, 0.5) {\{a,e,f\}};
	\node (b) at (-2.1, 0.5) {\{c,d,e,f\}};
	\node (ab) at (-0.7, 0.5) {\{a,c,d,e,f\}};
	\node (c) at (0.7, 0.5) {\{a,b,c,e,f\}};
	\node (ac) at (2.1, 0.5) {\{a,b,d,e,f\}};
	\node (bc) at (3.5, 0.5) {\{b,c,d,e,f\}};
	\node (abc) at (5, 0.5) {$\Omega$};
	
	\node (none1) at (-5, 0) {$\bullet$};
	\node (a1) at (-3.5, 0) {$\bullet$};
	\node (b1) at (-2.1, 0) {$\bullet$};
	\node (ab1) at (-0.7, 0) {$\bullet$};
	\node (c1) at (0.7, 0) {$\bullet$};
	\node (ac1) at (2.1, 0) {$\bullet$};
	\node (bc1) at (3.5, 0) {$\bullet$};
	\node (abc1) at (5, 0) {$\bullet$};
	
	\draw [decorate,decoration={brace,amplitude=2pt},yshift=0pt]
	(-5.4,-\h) -- (-5.4,0) node [black,midway,xshift=-4cm] {\footnotesize
		$1: X = \bigcup (Y \cap \emptyset)^{\downarrow J} \text{ and } X \supseteq Y \cap \emptyset$};
	
	\node (none2) at (-5, -\h) {$\bullet$};
	\node (a2) at (-3.5, -\h) {$\bullet$};
	\node (b2) at (-2.1, -\h) {$\bullet$};
	\node (ab2) at (-0.7, -\h) {$\bullet$};
	\node (c2) at (0.7, -\h) {$\bullet$};
	\node (ac2) at (2.1, -\h) {$\bullet$};
	\node (bc2) at (3.5, -\h) {$\bullet$};
	\node (abc2) at (5, -\h) {$\bullet$};
	
	\draw [decorate,decoration={brace,amplitude=2pt},yshift=0pt]
	(-5.4,-2*\h) -- (-5.4,-\h) node [black,midway,xshift=-4cm] {\footnotesize
		$2: X = \bigcup (Y \cap \{a,b,c,e,f\})^{\downarrow J} \text{ and } X \supseteq Y \cap \{e,f\}$};
	
	\node (none3) at (-5, -2*\h) {$\bullet$};
	\node (a3) at (-3.5, -2*\h) {$\bullet$};
	\node (b3) at (-2.1, -2*\h) {$\bullet$};
	\node (ab3) at (-0.7, -2*\h) {$\bullet$};
	\node (c3) at (0.7, -2*\h) {$\bullet$};
	\node (ac3) at (2.1, -2*\h) {$\bullet$};
	\node (bc3) at (3.5, -2*\h) {$\bullet$};
	\node (abc3) at (5, -2*\h) {$\bullet$};
	
	\draw [decorate,decoration={brace,amplitude=2pt},yshift=0pt]
	(-5.4,-3*\h) -- (-5.4,-2*\h) node [black,midway,xshift=-4cm] {\footnotesize
		$3: X = \bigcup (Y \cap \{a,b,d,e,f\})^{\downarrow J} \text{ and } X \supseteq Y \cap \{d,e,f\}$};
	
	\node (none4) at (-5, -3*\h) {$\bullet$};
	\node (a4) at (-3.5, -3*\h) {$\bullet$};
	\node (b4) at (-2.1, -3*\h) {$\bullet$};
	\node (ab4) at (-0.7, -3*\h) {$\bullet$};
	\node (c4) at (0.7, -3*\h) {$\bullet$};
	\node (ac4) at (2.1, -3*\h) {$\bullet$};
	\node (bc4) at (3.5, -3*\h) {$\bullet$};
	\node (abc4) at (5, -3*\h) {$\bullet$};
	
	\draw [decorate,decoration={brace,amplitude=2pt},yshift=0pt]
	(-5.4,-4*\h) -- (-5.4,-3*\h) node [black,midway,xshift=-4cm] {\footnotesize
		$4: X = \bigcup (Y \cap \{a,c,d,e,f\})^{\downarrow J} \text{ and } X \supseteq Y \cap \{c,d,e,f\}$};
	
	\node (none5) at (-5, -4*\h) {$\bullet$};
	\node (a5) at (-3.5, -4*\h) {$\bullet$};
	\node (b5) at (-2.1, -4*\h) {$\bullet$};
	\node (ab5) at (-0.7, -4*\h) {$\bullet$};
	\node (c5) at (0.7, -4*\h) {$\bullet$};
	\node (ac5) at (2.1, -4*\h) {$\bullet$};
	\node (bc5) at (3.5, -4*\h) {$\bullet$};
	\node (abc5) at (5, -4*\h) {$\bullet$};
	
	\draw [decorate,decoration={brace,amplitude=2pt},yshift=0pt]
	(-5.4,-5*\h) -- (-5.4,-4*\h) node [black,midway,xshift=-4cm] {\footnotesize
		$5: X = \bigcup (Y \cap \{b,c,d,e,f\})^{\downarrow J} \text{ and } X \supseteq Y \cap \{b,c,d,e,f\}$};
	
	\node (none6) at (-5, -5*\h) {$\bullet$};
	\node (a6) at (-3.5, -5*\h) {$\bullet$};
	\node (b6) at (-2.1, -5*\h) {$\bullet$};
	\node (ab6) at (-0.7, -5*\h) {$\bullet$};
	\node (c6) at (0.7, -5*\h) {$\bullet$};
	\node (ac6) at (2.1, -5*\h) {$\bullet$};
	\node (bc6) at (3.5, -5*\h) {$\bullet$};
	\node (abc6) at (5, -5*\h) {$\bullet$};
	
	\draw[->,>=latex] (b5) to node[xshift=-0.1cm, yshift=0.2cm] {-1}(ab6);
	\draw[->,>=latex] (bc5) to node[xshift=-0.1cm, yshift=0.2cm] {-1}(abc6);
	
	\draw[->,>=latex] (ab4) to node[xshift=-0.1cm, yshift=0.2cm] {-1}(abc5);
	\draw[->,>=latex] (b4) to node[xshift=-0.1cm, yshift=0.2cm] {-1}(bc5);
	
	\draw[->,>=latex] (ac3) to node[xshift=-0.1cm, yshift=0.2cm] {-1}(abc4);
	\draw[->,>=latex] (a3) to node[xshift=-0.1cm, yshift=0.2cm] {-1}(c4);
	
	\draw[->,>=latex] (c2) to node[xshift=-0.1cm, yshift=0.2cm] {-1}(abc3);
	\draw[->,>=latex] (a2) to node[xshift=-0.1cm, yshift=0.2cm] {-1}(ab3);
	\draw[->,>=latex] (a2) to node[xshift=-0.1cm, yshift=0.2cm] {-1}(ac3);
	
	\draw[->,>=latex] (none1) to node[xshift=-0.1cm, yshift=0.2cm] {-1}(abc2);
	\draw[->,>=latex] (none1) to node[xshift=-0.1cm, yshift=0.2cm] {-1}(a2);
	\draw[->,>=latex] (none1) to node[xshift=-0.1cm, yshift=0.2cm] {-1}(b2);
	\draw[->,>=latex] (none1) to node[xshift=-0.1cm, yshift=0.2cm] {-1}(c2);
	\draw[->,>=latex] (none1) to node[xshift=-0.1cm, yshift=0.2cm] {-1}(ab2);
	\draw[->,>=latex] (none1) to node[xshift=-0.1cm, yshift=0.2cm] {-1}(ac2);
	\draw[->,>=latex] (none1) to node[xshift=-0.1cm, yshift=0.2cm] {-1}(bc2);
	\end{tikzpicture}
	\caption{\small{
			Illustration representing the paths generated by the arrows contained in the sequence $(H_k)_{k\in\llbracket 1, 5\rrbracket}$, where $H_k = (J, E_k)$ and $E_k = \left\lbrace (X, Y) \in J^2 ~/~ X = \bigcup (Y \cap \overline{i}_{6-k})^{\downarrow J} \text{ and } X \supseteq Y \cap \bigcap\overline{\iota}(J)_{6-k} \right\rbrace$
			and $\overline{\iota}(J)_k = \{ \overline{i}_1, \overline{i}_2, \dots, \overline{i}_{k} \}$ and $J=\{ \emptyset, \{a,e,f\}, \{c,d,e,f\}, \{a,c,d,e,f\}, \{a,b,c,e,f\}, \{a,b,d,e,f\}, \{b,c,d,e,f\}, \Omega \}$ and $\Omega = \{ a,b,c,d,e,f \}$ and $(\overline{i}_k)_{k\in \llbracket 1, 5 \rrbracket} = (\{ b,c,d,e,f \}, \{ a,c,d,e,f \}, \{ a,b,d,e,f \}, \{ a,b,c,e,f \}, \emptyset)$. This sequence computes the same M\"obius transformations as $G_\subset = (J, E_\subset)$, where $E_\subset = \{ (X,Y) \in J^2 ~/~ X \subset Y \}$. The ``-1'' labels emphasize the intended use of the operator $-$ with this sequence. Actually, since there is no order between any two dual iota elements of $\overline{\iota}(J)\backslash\{\emptyset\}$ in this example, the order chosen in Fig. \ref{fig:EMT_F_sub} is arbitrary. Thus, any order would compute the same M\"obius transformations as $G_\subset$, as long as $\emptyset$ is the first dual iota element to consider.
}}\label{fig:EMT_F_sub_rev}
\end{figure}
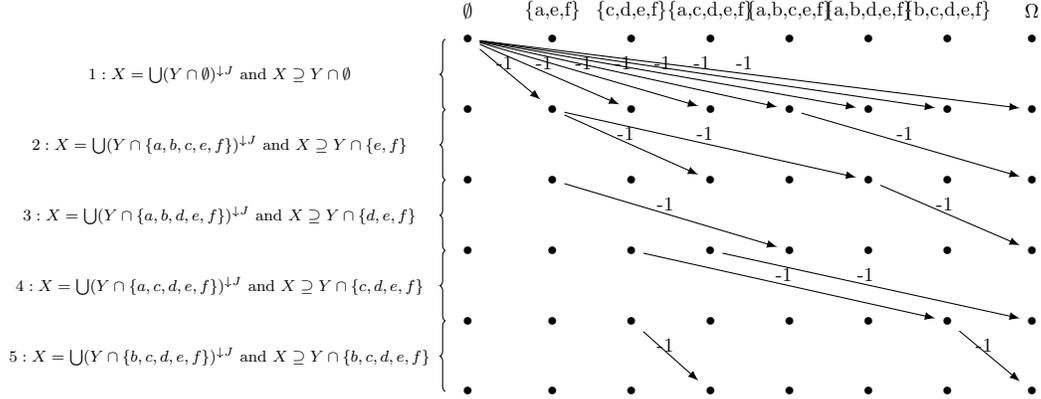

\begin{corollary}
	Finally, in the direct line of Corollary \ref{mob_opti_L_meet_rev}, consider the sequence $(H_k)_{k\in\llbracket 1, n\rrbracket}$, where $H_k = (J, E_k)$ and:
	$$E_k = \bigg\lbrace (x,y) \in J^2 ~/~ x = \hspace{-0.cm}\bigvee(y \wedge \overline{i}_{n+1-k})^{\downarrow J} \text{ and } x \geq y \wedge \bigwedge\overline{\iota}(J)_{n+1-k} \bigg\rbrace.$$
	This sequence computes the same M\"obius transformations as $G_< = (J, E_<)$, where $E_< = \{ (X,Y) \in J^2 ~/~ X < Y \}$. This sequence is illustrated in Fig. \ref{fig:EMT_F_sub_rev} and leads to the same complexities as the one presented in Corollary \ref{mob_opti_F_dual}.
\end{corollary}

\section{Discussions}\label{discussion}

\subsection{General usage}

If $|\supp{f}|$ is of same order of magnitude as $|^\vee\mathcal{I}(P)|$ or lower, then we can directly compute the focal points $^\wedge\supp{f}$ or $^\vee\supp{f}$. Next, with $^\wedge\supp{f}$, we can compute Efficient M\"obius Transformations based on Theorem \ref{mob_opti_F} to get the zeta or M\"obius transform of any function $f$ in $(P, \geq)$.

Let us take the sequence $(H_k)_{k\in\llbracket 1, n\rrbracket}$ from Theorem \ref{mob_opti_F}, where $H_k = (^\wedge\supp{f}, E_k)$ and:
$$E_k = \left\lbrace (x,y) \in {^\wedge\supp{f}}^2 ~/~ x = \bigwedge (y \vee i_k)^{\uparrow {^\wedge\supp{f}}} \text{ and } x \leq y \vee \bigvee\iota(^\wedge\supp{f})_k \right\rbrace,$$
where $\iota(^\wedge\supp{f})_k = \{ {i}_1, {i}_2, \dots, {i}_{k} \}$ such that $\forall i_k, i_{l} \in \iota(^\wedge\supp{f})$, $k < l \Rightarrow i_k \not\geq i_{l}$.
\begin{example}\label{SUM:discussion_1}
	Consider the mass function $m$ and the commonality function $q$ from Example \ref{SUM:q_from_m}. The transformation $((H_k)_{k\in\llbracket 1, n\rrbracket}, m, +)$, where $^\wedge\supp{f}={^\wedge\supp{m}}$, computes the commonality function $q$.
\end{example}
\begin{example}
	Consider the mass function $m$ and the commonality function $q$ from Example \ref{SUM:m_from_q}. The transformation $((H_{n+1-k})_{k\in\llbracket 1, n\rrbracket}, q, -)$, where $^\wedge\supp{f}={^\wedge\supp{m}}$, computes the mass function $m$.
\end{example}
\begin{example}
	Consider the conjunctive weight function $w$ and the commonality function $q$ from Example \ref{SUM:q_from_w}. The transformation $((H_k)_{k\in\llbracket 1, n\rrbracket}, w^{-1}, \times)$, where $^\wedge\supp{f} = {^\wedge\supp{w-1}}$, computes the commonality function $q$.
\end{example}
\begin{example}
	Consider the conjunctive weight function $w$ and the commonality function $q$ from Example \ref{SUM:w_from_q}. The transformation $((H_{n+1-k})_{k\in\llbracket 1, n\rrbracket}, w^{-1}, /)$, where $^\wedge\supp{f} = {^\wedge\supp{w-1}}$, computes the conjunctive weight function $w$.
\end{example}

Let us now take the sequence $(H_k)_{k\in\llbracket 1, n\rrbracket}$ from Corollary \ref{mob_opti_F_dual}, where $H_k = (^\vee\supp{f}, E_k)$ and:
$$E_k = \left\lbrace (x,y) \in {^\vee\supp{f}}^2 ~/~ x = \bigvee (y \wedge \overline{i}_k)^{\downarrow {^\vee\supp{f}}} \text{ and } x \geq y \wedge \bigwedge\overline{\iota}(^\vee\supp{f})_k \right\rbrace,$$
where $\overline{\iota}(^\vee\supp{f})_k = \{ \overline{i}_1, \overline{i}_2, \dots, \overline{i}_{k} \}$ and such that $\forall \overline{i}_k, \overline{i}_{l} \in \overline{\iota}(^\vee\supp{f})$, $k < l \Rightarrow \overline{i}_k \not\leq \overline{i}_{l}$.
\begin{example}
	Consider the mass function $m$ and the implicability function $b$ from Example \ref{SUM:b_from_m}. The transformation $((H_k)_{k\in\llbracket 1, n\rrbracket}, m, +)$, where $^\vee\supp{f}={^\vee\supp{m}}$, computes the implicability function $b$.
\end{example}
\begin{example}
	Consider the mass function $m$ and the implicability function $b$ from Example \ref{SUM:m_from_b}. The transformation $((H_{n+1-k})_{k\in\llbracket 1, n\rrbracket}, b, -)$, where $^\vee\supp{f}={^\vee\supp{m}}$, computes the mass function $m$.
\end{example}
\begin{example}
	Consider the disjunctive weight function $v$ and the implicability function $b$ from Example \ref{SUM:b_from_v}. The transformation $((H_k)_{k\in\llbracket 1, n\rrbracket}, v^{-1}, \times)$, where $^\vee\supp{f} = {^\vee\supp{v-1}}$, computes the implicability function $b$.
\end{example}
\begin{example}\label{SUM:discussion_last}
	Consider the disjunctive weight function $v$ and the implicability function $b$ from Example \ref{SUM:v_from_b}. The transformation $((H_{n+1-k})_{k\in\llbracket 1, n\rrbracket}, v^{-1}, /)$, where $^\vee\supp{f} = {^\vee\supp{v-1}}$, computes the disjunctive weight function $v$.
\end{example}

These transformations can be computed in at most $O(|{^\vee \mathcal{I}(P)}|.|\supp{f}| + |I(\supp{f})|.|^o\supp{f}|)$ operations, where $I \in \{\iota, \overline{\iota} \}$ and $o \in \{ \wedge, \vee \}$, which is at most $O(|{^\vee \mathcal{I}(P)}|.|P|)$.

Otherwise, if $|\supp{f}| \gg |^\vee\mathcal{I}(P)|$, then we can compute $\supp{f}^{\uparrow {^\mathcal{L}\supp{f}}}$ or $\supp{f}^{\downarrow {^\mathcal{L}\supp{f}}}$ from the lattice support of Proposition \ref{supp_lattice}, and then compute Efficient M\"obius Transformations based on Theorem \ref{mob_opti_L}. Doing so, computing the same transforms can be done in at most $O(|^\vee\mathcal{I}(P)|.|\supp{f}| + |I(\supp{f})|.|\supp{f}^{A {^\mathcal{L}\supp{f}}}|)$ operations, where $I \in \{\iota, \overline{\iota} \}$ and $A \in \{ \uparrow, \downarrow \}$, which is at most $O(|^\vee\mathcal{I}(P)|.|P|)$.

Either way, it is always possible to compute zeta and M\"obius transforms in a distributive lattice in less than $O(|^\vee\mathcal{I}(P)|.|P|)$ in time and space.

\subsection{Dempster-Shafer Theory}

So, we can always compute most DST transformations (See Examples \ref{SUM:discussion_1} to \ref{SUM:discussion_last}), wherever the FMT applies, in less than $O(|\Omega|.2^{|\Omega|})$ operations in the general case. The EMT are always more efficient than the FMT.

Moreover, $\supp{f}^{\downarrow {^\mathcal{L}\supp{f}}}$ can be optimized if $\Omega \in \supp{f}$. Indeed, in this case, we have $\supp{f}^{\downarrow {^\mathcal{L}\supp{f}}} = {^\mathcal{L}\supp{f}}$, while there may be a lot less elements in $(\supp{f}\backslash\{\Omega \})^{\downarrow {^\mathcal{L}\supp{f}}}$. If so, one can equivalently compute the down set $(\supp{f}\backslash\{\Omega \})^{\downarrow {^\mathcal{L}\supp{f}}}$, execute an EMT with Theorem \ref{mob_opti_L}, and then add the value on $\Omega$ to the value on all sets of $(\supp{f}\backslash\{\Omega \})^{\downarrow {^\mathcal{L}\supp{f}}}$. 

The same can be done with $(\supp{f}\backslash\{\emptyset \})^{\uparrow {^\mathcal{L}\supp{f}}}$. This trick can be particularly useful in the case of the conjunctive and disjunctive weight function, which require that $\supp{f}$ contains respectively $\Omega$ and $\emptyset$.

Also, optimizations built for the FMT, such as the reduction of $\Omega$ to the core $\mathcal{C}$ or its optimal coarsened version $\Omega'$, are already encoded in the use of the function $\iota$ (see Example \ref{consonant_case}). On the other hand, optimizations built for the evidence-based approach, such as approximations by reduction of the number of focal sets, i.e. reducing the size of $\supp{f}$, can still greatly enhance the EMT. 

Finally, it was proposed in \cite{FMT} to fuse two mass functions $m_1$ and $m_2$ using Dempster's rule by computing the corresponding commonality functions $q_1$ and $q_2$ in $O(|\Omega|.2^{|\Omega|})$, then computing $q_{12}=q_1.q_2$ in $O(2^{|\Omega|})$ and finally computing back the fused mass function $m_{12}$ from $q_{12}$ in $O(|\Omega|.2^{|\Omega|})$. 
Here, we propose to compute the same detour but only on the elements of $\supp{m_{12}} \subseteq (\supp{m_1} \cup \supp{m_2})^{\downarrow {^\mathcal{L}\supp{f}}}$. Indeed, notice that ${\supp{m_{12}}} \subseteq {^\wedge(\supp{m_1} \cup \supp{m_2})}$, which implies that ${\supp{m_{12}}}^{\downarrow {^\mathcal{L}\supp{f}}} \subseteq {(\supp{m_1} \cup \supp{m_2})}^{\downarrow {^\mathcal{L}\supp{f}}}$. Thus, noting $L = {(\supp{m_1} \cup \supp{m_2})}^{\downarrow {^\mathcal{L}\supp{f}}}$, we compute the corresponding commonality functions $q_1$ and $q_2$ in $O(|\iota(L)|.|L|)$, then compute $q_{12}=q_1.q_2$ in $O(|L|)$ and finally compute back the fused mass function $m_{12}$ from $q_{12}$ in $O(|\iota(L)|.|L|)$, where $\iota(L) = \iota(\supp{m_1} \cup \supp{m_2})$. 


\begin{example}[\textit{Coarsening in the consonant case}]\label{consonant_case}
	Let $\supp{f} = \{ F_1,~ F_2,~ \dots,~ F_K \}$ such that $F_1 \subset F_2 \subset \dots \subset F_K$. A coarsening $\Omega'$ of $\Omega$ is a mapping from disjoint groups of elements of $\Omega$ to elements of $\Omega'$. The set $\Omega'$ can be seen as a partition of $\Omega$. The goal of this coarsening of $\Omega$ is to provide a reduced powerset $2^{\Omega'}$. The best coarsening in this example would create as much elements in $\Omega'$ as there are elements in $\supp{f}$. Thus, the best coarsening would give us a powerset of size $2^{|\supp{f}|}$.
	
	On the other hand, our iota elements $\iota(\supp{f})$ are the join-irreducible elements of the smallest sublattice of $2^{\Omega}$ containing $\supp{f}$. This lattice is what we called the \textit{lattice support of $f$} and noted $^{\mathcal{L}}\supp{f}$. By definition, we necessarily have $|{^{\mathcal{L}}\supp{f}}| \leq 2^{|\supp{f}|}$.
	More precisely here, all elements of $\supp{f}$ are both focal points and join-irreducible elements of $^{\mathcal{L}}\supp{f}$, i.e. $\iota(\supp{f}) = \supp{f} = {^\vee \supp{f}} = {^\wedge \supp{f}}$, if $\emptyset \not\in \supp{f}$ (Otherwise, we have $\iota(\supp{f}) = \supp{f}\backslash\{\emptyset\}$). In fact, since our iota elements are not mapped elements of a reduced set $\Omega'$ but instead raw sets from $2^{\Omega}$, combinations of joins lead to a vastly different lattice. In this example, we have $^{\mathcal{L}}\supp{f} = \supp{f}$, instead of the $2^{\supp{f}}$ given by coarsening.
\end{example}

\section{Conclusion}\label{SUM:conclusion}

In this paper, we proposed the \textit{Efficient M\"obius Transformations} (EMT), which are general procedures to compute the zeta and M\"obius transforms of any function defined on any finite distributive lattice with optimal complexity. They are based on our reformulation of the M\"obius inversion theorem with focal points only, featured in our previous work \cite{me_journal}. The EMT optimally exploit the information contained in both the support of this function and the structure of distributive lattices. Doing so, the EMT always perform better than the optimal complexity for an algorithm considering the whole lattice, such as the FMT. Following these findings, it remains to propose explicit algorithms and implementation guidelines. We provide this for the powerset lattice, for DST, in Appendix \ref{implementation}. We plan to release both a C++ (roughly presented in Appendix) and a Python open-source DST implementation in the near future.

\begin{appendices}
	

\section{Proofs about the Efficient M\"obius Transformations}

\subsection{\autoref{iota_elements}}\label{appendix:iota_elements}

\begin{proof}	
	Let us define $L_S$ as the set containing the join of all combinations of iota elements of $S$, in addition to $\bigwedge$. Formally, we have:
	$$L_S = \left\lbrace \bigvee X ~/~ X \subseteq \iota(S),~ X \neq \emptyset \right\rbrace \cup \left\lbrace \bigwedge S \right\rbrace.$$
	By construction, we already know that the join of any number (except 0) of elements from $L_S$ is also in $L_S$. So, $L_S$ is an upper-subsemilattice of $P$. In the following, we will prove that it is also a lower-subsemilattice of $P$.
	
	But, before anything, notice that $\bigwedge \cdot^{\uparrow S}: P \rightarrow P$ is a closure operator, i.e. for any elements $x, y \in P$, we have:
	\begin{align}
	x &\leq \bigwedge x^{\uparrow S}\label{extensive}\\
	x \leq y ~&\Rightarrow~ \bigwedge x^{\uparrow S} \leq \bigwedge y^{\uparrow S}\label{increasing}\\
	\bigwedge\left(\bigwedge x^{\uparrow S}\right)^{\uparrow S} &= \bigwedge x^{\uparrow S}\label{idempotent}
	\end{align}
	
	Now, consider the meet of two iota elements $\iota_1 \wedge \iota_2$, where $\iota_1, \iota_2 \in \iota(S)$. By definition, there are two join-irreducible elements $x,y \in {^\vee \mathcal{I}(P)}$ such that $\iota_1 \wedge \iota_2 = \bigwedge x^{\uparrow S} \wedge \bigwedge y^{\uparrow S} = \bigwedge (x^{\uparrow S} \cup y^{\uparrow S})$.
	Let us note $\delta = \bigwedge (x^{\uparrow S} \cup y^{\uparrow S})$.
	Since $S \supseteq x^{\uparrow S} \cup y^{\uparrow S}$, we know that $\bigwedge S \leq \delta$. If $\delta = \bigwedge S$, then $\delta \in L_S$.
	Otherwise, given that $P$ is a lattice, we know that $\delta$ is equal to the join of all the join-irreducible elements of $P$ that are less than $\delta$, i.e. $\bigvee \delta^{\downarrow{^\vee \mathcal{I}(P)}} = \delta$.
	
	For all join-irreducible elements $i \in \delta^{\downarrow{^\vee \mathcal{I}(P)}}$, we have $i \leq \delta$.  By Eq. (\ref{increasing}), we also have $\bigwedge i^{\uparrow S} \leq \bigwedge \delta^{\uparrow S}$.
	Moreover, we know that $\delta^{\uparrow S} \supseteq x^{\uparrow S} \cup y^{\uparrow S}$ because $\delta = \bigwedge (x^{\uparrow S} \cup y^{\uparrow S})$. This implies that we have 
	$\bigwedge\delta^{\uparrow S} \leq \delta$,
	and so $\bigwedge i^{\uparrow S} \leq \delta$.

	In addition, by Eq. (\ref{extensive}), we get that $i \leq \bigwedge i^{\uparrow S}$, which means that, for all join-irreducible elements $i \in \delta^{\downarrow{^\vee \mathcal{I}(P)}}$, we have $i \leq \bigwedge i^{\uparrow S} \leq \delta$. Therefore, combined with the fact that $\bigvee \delta^{\downarrow{^\vee \mathcal{I}(P)}} = \delta$, we finally obtain that $\bigvee \delta^{\downarrow{\iota(S)}} = \delta$. In plain English, this means that $\delta$ is equal to the join of all the iota elements of $S$ that are less than $\delta$. So, by definition of $L_S$, we get that $\delta \in L_S$. Thus, the meet of two iota elements $\iota_1 \wedge \iota_2$, where $\iota_1, \iota_2 \in \iota(S)$, is in $L_S$.
	
	It only remains to consider the meet of two arbitrary elements of $L_S$, i.e. $x \wedge y$, where $x, y \in L_S$. Notice that if $x = \bigwedge S$ or $y = \bigwedge S$, then $x \wedge y = \bigwedge S \in L_S$. Otherwise, it can be decomposed as follows:
	$$
	x \wedge y = \left(\bigvee x^{\downarrow{\iota(S)}}\right) \wedge \left(\bigvee y^{\downarrow{\iota(S)}} \right)
	$$
	Since $P$ follows the distributive law Eq. (\ref{distrib_law}), we can rewrite this equation as:
	\begin{align*}
	x \wedge y &= \bigvee_{\iota_1 \in x^{\downarrow{\iota(S)}}} \bigvee_{\iota_2 \in y^{\downarrow{\iota(S)}}} \left( \iota_1 \wedge \iota_2 \right)
	\end{align*}
	The meet of two iota elements of $S$ being in $L_S$, we get that $x \wedge y$ is equal to the join of elements that are all in $L_S$. As we already established that $L_S$ is an upper-subsemilattice of $P$, we get that the meet of two arbitrary elements of $L_S$ is also in $L_S$. Thus, $L_S$ is a lower-subsemilattice of $P$ as well and therefore a sublattice of $P$.
	
	In addition, notice that for all element $s \in S$ and for all $i\in s^{\downarrow{^\vee\mathcal{I}(P)}}$, we have by construction $i \leq \bigwedge i^{\uparrow S} \leq s$. Therefore, $P$ being a lattice, we have $s = \bigvee s^{\downarrow{^\vee\mathcal{I}(P)}} = \bigvee s^{\downarrow \iota(S)}$, i.e. $s \in L_S$. Besides, if $\bigwedge P \in S$, then it is equal to $\bigwedge S$, which is also in $L_S$ by construction. So, $S \subseteq L_S$. It follows that the meet or join of every nonempty subset of $S$ is in $L_S$, i.e. $M_S \subseteq L_S$ and $J_S \subseteq L_S$, where $M_S$ is the smallest meet-closed subset of $P$ containing $S$ and $J_S$ is the smallest join-closed subset of $P$ containing $S$. Furthermore, iota elements are defined as the meet of a set of elements of $S$, which implies that they are necessarily all contained in $M_S$, i.e. $\iota(S) \subseteq M_S$. This means that we cannot build a smaller sublattice of $P$ containing $S$. Therefore, $L_S$ is the smallest sublattice of $P$ containing $S$. 
	
	Finally, let us verify that all iota elements are join-irreducible elements of $L_S$. For any join-irreducible element $i \in {^\vee\mathcal{I}(P)}$, assume there are two distinct elements $x,y \in L_S$ such that $x < \bigwedge i^{\uparrow S}$ and $y < \bigwedge i^{\uparrow S}$. This implies by Eq. (\ref{increasing}) and (\ref{idempotent}) that $\bigwedge x^{\uparrow S} \leq \bigwedge i^{\uparrow S}$ and $\bigwedge y^{\uparrow S} \leq \bigwedge i^{\uparrow S}$. Moreover, if $i \leq x$ and $i \leq y$, then by Eq. (\ref{increasing}), we get that $\bigwedge i^{\uparrow S} \leq \bigwedge x^{\uparrow S}$ and $\bigwedge i^{\uparrow S} \leq \bigwedge y^{\uparrow S}$, which means that $x = y = \bigwedge i^{\uparrow S}$. This contradicts the fact that $x \neq y$. Thus, we get that $i \not\leq x$ and $i \not\leq y$. By Lemma \ref{safe_join}, this implies that $i \not\leq x \vee y$. Since $i \leq \bigwedge i^{\uparrow S}$, we have necessarily $x \vee y \neq \bigwedge i^{\uparrow S}$ and so $x \vee y < \bigwedge i^{\uparrow S}$. Therefore, $\bigwedge i^{\uparrow S}$ is a join-irreducible element of $L_S$, i.e. $\iota(S)$ is the set containing only the join-irreducible elements of $L_S$. 
\end{proof}
	
	\subsection{\autoref{mob_opti_L}}\label{appendix:mob_opti_L}
	
	\begin{proof}
		Consider the sequence $(H_k)_{k\in\llbracket 1, n\rrbracket}$, where $H_k = (L, E_k)$ and:
		$$E_k = \left\lbrace (x,y) \in L^2 ~/~ y = x \vee i_{n + 1 - k} \right\rbrace.$$
		By definition, for all $k\in \llbracket 1, n \rrbracket$ and $\forall (x,y) \in {{E}}_k$, we have $x,y \in L$ and $x \leq y$, i.e. $(x,y) \in E_\leq$.
		Reciprocally, every arrow $e \in E_\leq$ can be decomposed as a unique path $(e_1, e_2, \dots, e_{n}) \in A_1 \times A_2 \times \dots \times A_{n}$, where $A_k = E_k \cup I_P$:
		
		Similarly to the FMT, this sequence builds unique paths simply by generating the whole lattice $L$ step by step with each join-irreducible element of $L$. However, unlike the FMT, the join-irreducible elements of $L$ are not necessarily atoms. Doing so, pairs of join-irreducible elements may be ordered, causing the sequence to skip or double some elements. And even if all the join-irreducible elements of $L$ are atoms, since $L$ is not necessarily a Boolean lattice, the join of two atoms may be greater than a third atom (e.g. if $L$ is the diamond lattice), leading to the same issue. Indeed, it is easy to build a path between two elements $x,y$ of $L$ such that $x \leq y$: At step 1, we take the arrow $(x, x\vee i_n)$ if $i_n \leq y$ (we take the identity arrow $(x,x)$ otherwise). At step 2, we take the arrow $(p, p \vee i_{n-1})$ if $i_{n-1} \leq y$, where $p = x\vee i_n$ or $p=x$ depending on whether or not $i_{n} \leq y$, and so on until we get to $y$. Obviously, we cannot get to $y$ if we take an arrow $(p, p\vee i)$ where $i \not\leq y$. So, any path from $x$ to $y$ only consists of arrows obtained with join-irreducible elements that are less than $y$. But, are they all necessary to reach $y$ from $x$?
		Let $i_k$ be a join-irreducible element such that $i_k \leq y$. This join-irreducible element is only considered in $E_{n+1-k}$. Suppose we do not take the arrow at step ${n+1-k}$, i.e. we replace it by an identity arrow $(p, p)$, where $p$ was reached through a path from $x$ with arrows at steps 1 to ${n-k}$. 
		Since $L$ is a distributive lattice, and since its join-irreducible elements are ordered such that $\forall i_j, i_{l} \in {^{\vee}\mathcal{I}(L)}$, $j < l \Rightarrow i_j \not\geq i_{l}$, we have by Corollary \ref{focal_atom_order} that for any $k\in \llbracket 1, n \rrbracket$, $i_k \not\leq \bigvee{^{\vee}\mathcal{I}(L)}_{{k-1}}$. So, if $i_k \not\leq p$, then by Lemma \ref{safe_join}, we also have $i_k \not\leq p \vee\bigvee{^{\vee}\mathcal{I}(L)}_{{k-1}}$. Since ${^{\vee}\mathcal{I}(L)}_{{k-1}}$ contains all join-irreducible elements considered after step $n+1-k$, this implies that there is no path from $p$ to an element greater than $i_k$. Yet, $i_k \leq y$. Thus, if $i_k \not\leq p$, then $y$ can only be reached from $p$ through a path containing the arrow $(p, p\vee i_k)$. Otherwise, if $i_k \leq p$, then $p = p \vee i_k$, which means that only an identity arrow can be taken at step $n+1-k$ anyway. Either way, there is only one arrow that can be taken at step $n+1-k$ to build a path between $p$ and $y$. All join-irreducible elements less than $y$ must be used to build a path from $x$ to $y$. Thereby, this path is unique,
		%
		%
		meaning that Theorem \ref{theorem:FMT} is satisfied. The sequence $(H_k)_{k\in\llbracket 1, n\rrbracket}$ computes the same zeta transformations as $G_<$. 
	\end{proof}
	
	\subsection{\autoref{mob_opti_F}}\label{appendix:mob_opti_F}
	
	\begin{proof}
		Consider the sequence $(H_k)_{k\in\llbracket 1, n\rrbracket}$, where $H_k = (M, E_k)$ and:
		$$E_k = \left\lbrace (x,y) \in M^2 ~/~ x = \bigwedge (y \vee i_k)^{\uparrow M} \text{ and } x \leq y \vee \bigvee\iota(M)_k \right\rbrace,$$
		where $\iota(M)_k = \{ {i}_1, {i}_2, \dots, {i}_{k} \}$.
		
		By definition, for all $k\in \llbracket 1, n \rrbracket$ and $\forall (x,y) \in {{E}}_k$, we have $x,y \in M$ and $x \geq y$, i.e. $(x,y) \in E_\geq$. 
		Reciprocally, every arrow $e \in E_\geq$ can be decomposed as a unique path $(e_1, e_2, \dots, e_{n}) \in A_1 \times A_2 \times \dots \times A_{n}$, where $A_k = E_k \cup I_P$:
		
		Recall the procedure, described in Theorem \ref{mob_opti_L}, that builds unique paths simply by generating all elements of a finite distributive lattice $L \supseteq M$, based on the join of its join-irreducible elements, step by step. Here, the idea is to do the same, except that we remove all elements that are not in $M$. Doing so, the only difference is that the join $y \vee i_k$ of an element $y\in M$ with a join-irreducible element $i_k\in \iota(M)$ of this hypothetical lattice $L$ may not be in $M$. However, thanks to the meet-closure of $M$, we can ``jump the gap" between two elements $y$ and $p$ of $M$, should they be separated by elements of $L\backslash M$. Indeed, for all join-irreducible element $i_k \in \iota(M)$, if $x \geq y \vee i_k$, then since $M$ is meet-closed and $x\in M$, there is a unique element $p\in M$ that we call \textit{proxy} such that $p = \bigwedge (y \vee i_k)^{\uparrow M}$. In complement, the synchronizing condition $p \leq y \vee \bigvee\iota(M)_k$ ensures the unicity of this jump, and so
		the unicity of the path between any two elements $x$ and $y$ of $M$.

		Finding a path from $x$ to $y$ is easy: take all iota elements less than $x$, and simply compute successive joins with them, starting from $y$. At step $n$, we take the arrow $(\bigwedge (y \vee i_n)^{\uparrow M}, y)$ if $i_n \leq x$, and $(y, y)$ otherwise. At step $n-1$, we take the arrow $(\bigwedge (p \vee i_{n-1})^{\uparrow M}, p)$ if $i_{n-1} \leq x$, where $p = \bigwedge (y \vee i_n)^{\uparrow M}$ or $p = y$, depending on whether or not $i_n \leq x$. Proceeding as such until $x$ is reached guarantees the existence of a path, since the synchronizing condition is always satisfied. Then, the question is: Are there other paths?
		
		Obviously, for some $p \in M$ such that $x \geq p$, no path from $x$ to $y$ can contain arrows $(\bigwedge (p \vee i)^{\uparrow M}, p)$ if $i \not\leq x$. Thus, it contains only arrows corresponding to joins with iota elements that are less than $x$. Next, let us consider 
		some element $p \in M$.
		If $i_{k-1} \leq p$, then $p = \bigwedge (p \vee i_{k-1})^{\uparrow M}$, which means that only an identity arrow $(p, p)$ can be taken at step $k-1$. Otherwise, if $i_{k-1} \not\leq p$, then the arrow $(\bigwedge (p \vee i_{k-1})^{\uparrow M}, p)$ can only exist in the sequence if $\bigwedge (p \vee i_{k-1})^{\uparrow M} \leq p \vee \bigvee \iota(M)_{k-1}$. We know by Corollary \ref{focal_atom_order} that $i_k \not\leq \bigvee \iota(M)_{k-1}$. By Lemma \ref{safe_join}, this implies that if $i_k \not\leq p$, then $i_k \not\leq p \vee \bigvee \iota(M)_{k-1}$, which means that $i_k \not\leq \bigwedge (p \vee i_{k-1})^{\uparrow M}$. Through the same reasoning, this means that $i_k \not\leq \bigwedge\left(\bigwedge (p \vee i_{k-1})^{\uparrow M} \vee i_{k-2}\right)^{\uparrow M}$. In fact, by recurrence, we have $i_k \not\leq \bigwedge\left(\cdots\bigwedge\left(\bigwedge (p \vee i_{k-1})^{\uparrow M} \vee i_{k-2}\right)^{\uparrow M} \cdots \vee i_1\right)^{\uparrow M}$. Thus, if $i_k \not\leq p$, then no element greater than $i_k$ can be at the tail of a path leading to $p$ through arrows of the sequence. 
		This means that on a path from $x$ to $y$, if $i_k \leq x$, then at step $k$, either $i_k \leq p$ (i.e. we can only take the identity arrow $(p, p)$) or $i_k \not\leq p$, which implies that we must take the arrow $(\bigwedge (p \vee i_{k})^{\uparrow M}, p)$.
		This implies that all iota elements less than $x$ must be used in the joins corresponding to the arrows of the path from $x$ to $y$. Therefore, the path described above is unique, which satisfies Theorem \ref{theorem:FMT}.
		The sequence $(H_k)_{k\in\llbracket 1, n\rrbracket}$ computes the same zeta transformations as $G_>$.
	\end{proof}
	
	\section{Implementation of the Efficient M\"obius Transformations (EMT)}\label{implementation}
	
	We present in this section evidence-based algorithms for the computation of DST transformations such as the commonality function $q$ and the implicability function $b$, as well as their inversions, namely the mass function $m$ and the conjunctive and disjunctive weight functions $w$ and $v$. All these algorithms have better complexities than the FMT, i.e. less than $O(|\Omega|.2^{|\Omega|})$.
	
	
	
	We implemented these algorithms as a general-purpose C++ \textit{evidence-based} framework along with combination rules from DST. We plan to transpose this implementation as a Python package in the near future to ease its usage.
	Code and implementation details can be found at \cite{my_impl}.

	\subsection{Data structure}
	
	\subsubsection{Overview}
	
	The core of our implementation uses the class \textit{bitset} (a contiguous sequence of bits) from the standard C++ library, along with a special dynamic binary tree of our design presented in section \ref{data_struct}. This tree allows us to search for supersets or subsets without having to consider all sets. For simple look-ups, i.e. just to get the value associated with a set, we use Hashmaps, since they feature constant time complexities for look-up and insertion.
	
	Each representation of evidence (mass function, commonality function, implicability function, conjunctive weight function, disjunctive weight functions, etc) has its own class. 
	A mass function is an object containing a tree of values different from 0, i.e. corresponding to focal sets. It inherits the abstract class \textit{mobius\_transform} which simply defines the behavior of storing values such as focal sets. An object inheriting this abstract class can be created either directly, by providing a key-value object such as an Hashmap or a tree, or indirectly, through inversion of a given zeta transform and a given order relation ($\subseteq$ or $\supseteq$). It also features methods to remove negligeable values and renormalize. The conjunctive and disjunctive weight functions also inherit this abstract class. For these last classes, all set that is not present in their tree is associated with 1.
	
	Other representations inherit the abstract class \textit{zeta\_transform}. This abstract class defines the behavior of computing focal points from focal sets, given some order relation, and storing their values. It also keeps its mobius\_transform object in memory for eventual additional projections, i.e. to get the value associated with other sets than focal points. An object inheriting this abstract class can be created either directly, by providing another zeta\_transform object, or indirectly, by providing an object inheriting the mobius\_transform abstract class, an order relation ($\subseteq$ or $\supseteq$) and an operation ($+$ or $\times$). For the latter, you may also provide a specific computation scheme (without structure, as a semilattice or as a lattice). This class also features methods to find the value associated with a non-focal point.
	The commonality, implicability and plausibility functions inherit this abstract class.

	\subsubsection{Frame of discernment}
	
	A frame of discernment (FOD), noted $\Omega$, is represented as an object containing an array of labels and a Hashmap that enables one to find the index associated with a particular label.
	
	\subsubsection{Powerset function}\label{data_struct}
	
	\begin{figure}[t]
		\centering
		\hspace{-1.cm}
		\begin{tikzpicture}[scale=0.75, every node/.style={transform shape},
		node/.style={draw, circle,minimum size=1.3cm, inner sep=0pt},
		det/.style={draw, diamond,minimum size=1.1cm, inner sep=0pt},
		rect/.style={draw, rectangle,minimum size=1.1cm, inner sep=2pt}
		]
		
		\node[node] (none) at (0, 2) {$\emptyset$};
		
		\node[node] (a) at (0, 0) {\{a\}};
		
		\node[node] (b) at (-2, -1.5) {\{b\}};
		\node[node] (ab) at (2, -1.5) {\{a, b\}};
		
		\node[node] (c) at (-4, -3) {\{c\}};
		\node[node] (ac) at (1, -3) {\{a, c\}};
		\node[node] (bc) at (-1, -3) {\{b, c\}};
		\node[node] (abc) at (4, -3) {\{a, b, c\}};
		
		\draw[->,>=latex] (a) to node[xshift=-0.1cm, yshift=0.2cm] {0} (b);
		\draw[->,>=latex] (a) to node[xshift=0.1cm, yshift=0.2cm] {1} (ab);
		
		\draw[->,>=latex] (b) to node[xshift=-0.1cm, yshift=0.2cm] {0} (c);
		\draw[->,>=latex] (b) to node[xshift=0.1cm, yshift=0.2cm] {1} (bc);
		
		\draw[->,>=latex] (ab) to node[xshift=-0.1cm, yshift=0.2cm] {0} (ac);
		\draw[->,>=latex] (ab) to node[xshift=0.1cm, yshift=0.2cm] {1} (abc);
		
		\draw[red, dashed, very thick] (-5, -0.75) node[xshift=0.2cm, yshift=0.2cm] {a} to (5, -0.75);
		\draw[red, dashed, very thick] (-5, -2.25) node[xshift=0.2cm, yshift=0.2cm] {b} to (5, -2.25);
		\draw[red, dashed, very thick] (-5, -3.75) node[xshift=0.2cm, yshift=0.2cm] {c} to (5, -3.75);
		
		\draw [decorate,decoration={brace,amplitude=10pt},yshift=0pt]
		(-5.25,-0.75) -- (-5.25,2.75) node [black,midway,xshift=-0.8cm] {\footnotesize
			$2^{\{a\}}$};
		
		\draw [decorate,decoration={brace,amplitude=10pt, mirror},yshift=0pt]
		(5.25,-2.25) -- (5.25,2.75) node [black,midway,xshift=0.9cm] {\footnotesize
			$2^{\{a, b\}}$};
		
		\draw [decorate,decoration={brace,amplitude=10pt},yshift=0pt]
		(-6.25,-3.75) -- (-6.25,2.75) node [black,midway,xshift=-1cm] {\footnotesize
			$2^{\{a, b, c\}}$};
		
		\draw[rounded corners, red, rotate around={-37:(0,0)}] (-0.75,-0.75) rectangle (5.75, 0.75);
		
		\node[red] (singletons) at (-4.25, -4.5) {singletons};
		
		\draw[rounded corners, red, rotate around={37:(0,0)}] (0.75,-0.75) rectangle (-5.75, 0.75);
		
		\node[red] (sub-FODs) at (4.25, -4.5) {sub-FODs};
		\end{tikzpicture}
		\caption{Illustration depicting our data structure for powerset functions on the frame of discernment $\Omega = \{a, b, c\}$.
			It is a binary tree with values on nodes and leaves. For the sake of clarity, the structure is shown as if it was static, where all elements from $2^\Omega$ are considered special elements. We see clearly that it reproduces the natural generation of the powerset lattice, encapsulating the powerset of every $\{\omega_1, \dots, \omega_{n}\}$ in the powerset of $\{\omega_1, \dots, \omega_{n+1}\}$. As a consequence, the search for all singletons and all these sub-FODs can be restrained respectiveley to the left and right chain of nodes.
		}
		\label{fig:data_structure}
	\end{figure}
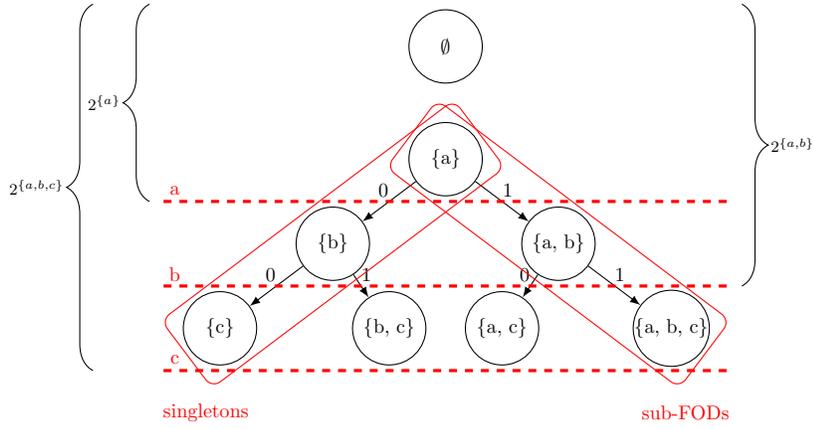
	
	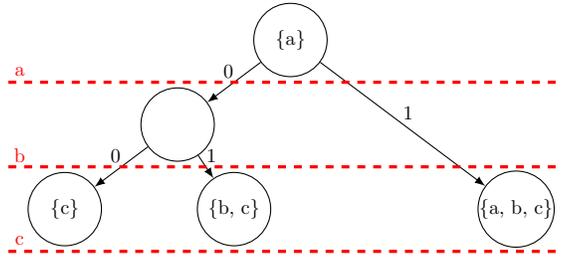
\begin{figure}[t]
		\centering
		\hspace{-0cm}
		\begin{tikzpicture}[scale=0.75, every node/.style={transform shape},
		node/.style={draw, circle,minimum size=1.3cm, inner sep=0pt},
		det/.style={draw, diamond,minimum size=1.1cm, inner sep=0pt},
		rect/.style={draw, rectangle,minimum size=1.1cm, inner sep=2pt}
		]
		
		\node[node] (a) at (0, 0) {\{a\}};
		
		\node[node] (b) at (-2, -1.5) {};
		
		\node[node] (c) at (-4, -3) {\{c\}};
		\node[node] (bc) at (-1, -3) {\{b, c\}};
		\node[node] (abc) at (4, -3) {\{a, b, c\}};
		
		\draw[->,>=latex] (a) to node[xshift=-0.1cm, yshift=0.2cm] {0} (b);
		\draw[->,>=latex] (a) to node[xshift=0.1cm, yshift=0.2cm] {1} (abc);
		
		\draw[->,>=latex] (b) to node[xshift=-0.1cm, yshift=0.2cm] {0} (c);
		\draw[->,>=latex] (b) to node[xshift=0.1cm, yshift=0.2cm] {1} (bc);
		
		\draw[red, dashed, very thick] (-5, -0.75) node[xshift=0.2cm, yshift=0.2cm] {a} to (5, -0.75);
		\draw[red, dashed, very thick] (-5, -2.25) node[xshift=0.2cm, yshift=0.2cm] {b} to (5, -2.25);
		\draw[red, dashed, very thick] (-5, -3.75) node[xshift=0.2cm, yshift=0.2cm] {c} to (5, -3.75);
		\end{tikzpicture}
		\caption{
			Illustration depicting our data structure for powerset functions on the frame of discernment $\Omega = \{a, b, c\}$. Here, special sets are $\{a\}$, $\{c\}$, $\{b,c\}$ and $\{a,b,c\}$. The blank node is a disjunction node.
		}
		\label{fig:data_structure_dyn}
	\end{figure}
	
	\begin{figure}[t]
		\centering
		\hspace{0cm}
		\begin{tikzpicture}[scale=0.75, every node/.style={transform shape},
		node/.style={draw, circle,minimum size=1.3cm, inner sep=0pt},
		det/.style={draw, diamond,minimum size=1.1cm, inner sep=0pt},
		rect/.style={draw, rectangle,minimum size=1.1cm, inner sep=2pt}
		]
		
		\node[node] (a) at (0, 0) {};
		
		\node[node] (b) at (-2, -1.5) {};
		\node[node] (ab) at (2, -1.5) {};
		
		\node[node] (c) at (-4, -3) {};
		\node[node] (ac) at (1, -3) {};
		\node[node] (bc) at (-1, -3) {};
		\node[node] (abc) at (4, -3) {};
		
		\node[node] (none) at (-5, -5) {$\emptyset$};
		
		\node[node] (a') at (0.7, -5) {\{a\}};
		
		\node[node] (b') at (-2.1, -5) {\{b\}};
		\node[node] (ab') at (3.5, -5) {\{a, b\}};
		
		\node[node] (c') at (-3.5, -5) {\{c\}};
		\node[node] (ac') at (2.1, -5) {\{a, c\}};
		\node[node] (bc') at (-0.7, -5) {\{b, c\}};
		\node[node] (abc') at (5, -5) {\{a, b, c\}};
		
		\draw[->,>=latex] (a) to node[xshift=-0.1cm, yshift=0.2cm] {0} (b);
		\draw[->,>=latex] (a) to node[xshift=0.1cm, yshift=0.2cm] {1} (ab);
		
		\draw[->,>=latex] (b) to node[xshift=-0.1cm, yshift=0.2cm] {0} (c);
		\draw[->,>=latex] (b) to node[xshift=0.1cm, yshift=0.2cm] {1} (bc);
		
		\draw[->,>=latex] (ab) to node[xshift=-0.1cm, yshift=0.2cm] {0} (ac);
		\draw[->,>=latex] (ab) to node[xshift=0.1cm, yshift=0.2cm] {1} (abc);
		
		\draw[->,>=latex] (c) to node[xshift=-0.1cm, yshift=0.2cm] {0} (none);
		\draw[->,>=latex] (c) to node[xshift=0.1cm, yshift=0.2cm] {1} (c');
		
		\draw[->,>=latex] (bc) to node[xshift=-0.1cm, yshift=0.2cm] {0} (b');
		\draw[->,>=latex] (bc) to node[xshift=0.1cm, yshift=0.2cm] {1} (bc');
		
		\draw[->,>=latex] (ac) to node[xshift=-0.1cm, yshift=0.2cm] {0} (a');
		\draw[->,>=latex] (ac) to node[xshift=0.1cm, yshift=0.2cm] {1} (ac');
		
		\draw[->,>=latex] (abc) to node[xshift=-0.1cm, yshift=0.2cm] {0} (ab');
		\draw[->,>=latex] (abc) to node[xshift=0.1cm, yshift=0.2cm] {1} (abc');
		
		\draw[red, dashed, very thick] (-5.5, -0.75) node[xshift=0.2cm, yshift=0.2cm] {a} to (5.5, -0.75);
		\draw[red, dashed, very thick] (-5.5, -2.25) node[xshift=0.2cm, yshift=0.2cm] {b} to (5.5, -2.25);
		\draw[red, dashed, very thick] (-5.5, -4) node[xshift=0.2cm, yshift=0.2cm] {c} to (5.5, -4);
		
		\draw [decorate,decoration={brace,amplitude=10pt},yshift=0pt, rotate around={90:(0,0)}]
		(-5.75,2.75) -- (-5.75,5.75) node [black,midway,xshift=-0.7cm,rotate around={-90:(0,0)}] {\footnotesize
			$2^{\{c\}}$};
		
		\draw [decorate,decoration={brace,amplitude=10pt},yshift=0pt, rotate around={90:(0,0)}]
		(-6.5,0) -- (-6.5,5.75) node [black,midway,xshift=-0.7cm,rotate around={-90:(0,0)}] {\footnotesize
			$2^{\{b, c\}}$};
		
		\draw [decorate,decoration={brace,amplitude=10pt},yshift=0pt, rotate around={90:(0,0)}]
		(-7.25,-5.75) -- (-7.25,5.75) node [black,midway,xshift=-0.7cm,rotate around={-90:(0,0)}] {\footnotesize
			$2^{\{a, b, c\}}$};
		
		\end{tikzpicture}
		\caption{Same example as in Fig. \ref{fig:data_structure} with an analogous data structure proposed by Wilson \cite{wilson2000}. It is a binary tree in which values are only assigned to terminal leaves, not intermediate nodes. As in Fig. \ref{fig:data_structure}, for the sake of clarity, the structure is shown as if it was static, where all elements from $2^\Omega$ are considered special elements.
		}
		\label{fig:data_wilson}
	\end{figure}
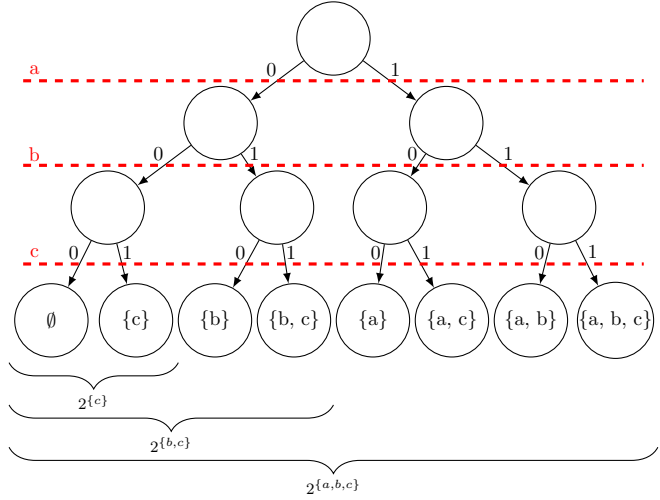

\begin{algorithm}[h]
	\KwIn{$S$, $\leq$}
	\KwOut{$F$, $F_{\text{Map}}$}
	$F \leftarrow S$\;
	$F_{\text{Map}} \leftarrow \text{Hashmap$<$bitset, float$>$}$\;
	\For{$i = 1$ \KwTo $|S|$}{
		$F_{\text{Map}}[S[i]] \leftarrow 0$\;
	}
	\If{$\leq~ = ~\subseteq$}{
		\tcp{$|$ is the bitwise OR operator}
		$\cdot \leftarrow |$\;
	}\Else{
		\tcp{\& is the bitwise AND operator} 
		$\cdot \leftarrow$ \&\;
	}

	\For{$i = 1$ \KwTo $|S|$}{
		\For{$ii = i + 1$ \KwTo $|F|$}{
			$A \leftarrow F[ii] \cdot S[i]$\;
			\If{$A \not\in F_{\text{Map}}.\text{keys()}$}{
				append $A$ to $F$\;
				$F_{\text{Map}}[A] \leftarrow 0$\;
			}
		}
	}
	\caption{Computation of the focal points associated with $(S, \leq)$, where $\leq \;\in \{\subseteq, \supseteq \}$.}\label{focal_points}
\end{algorithm}
	
	Our data structure for \textit{powerset functions} (i.e. functions that assign values to elements of $2^\Omega$) is based on the representation of sets as binary strings, as in \cite{wilson2000} and \cite{haenni03}, and on the binary tree depicted in Fig. \ref{fig:data_structure}. It is a dynamic powerset binary tree, only storing nodes corresponding to \textit{special sets}. All other set not present in the tree is assumed to be associated with a fixed value (i.e. 0 for a mass function, 1 for a weight function). Thus, special sets include focal sets but may not all be focal sets, as is the case with the lattice support. Each node in the tree contains a boolean value to indicate whether it has been set to a value or not, an eventual \textit{value}, pointers to parent and children, its depth index and a bitset representing an element from $2^\Omega$.
	
	The creation of this tree is incremental, starting with the singleton $\{\omega_1\}$, where $\Omega = [\omega_1, \dots, \omega_{n}]$, as root, whether it is a special set or not. The next special set is inserted to its right or left given that it contains $\omega_1$ or not. In fact, for a pair ($A$, \textit{value}), where $A$ is a special set, this procedure will assign \textit{value} to a node of depth equal to the greatest index in $\llbracket 1, n \rrbracket$ corresponding to an element of $A$. This node is found following the binary code that represents $A$ to navigate the tree until the last element index is encountered in $A$, i.e. until we reach the last bit set to $1$ in $A$.
	So, for some other special set $B$, if $B$ contains all elements of $A$, in addition to other ones including one of greater maximum index, then $B$ is inserted at the right of $A$. If $B$ has all elements of $A$, in addition to other ones not including one greater than the maximum index of $A$, then $B$ is not on the same branch as $A$ and will be assigned to a node of same depth as $A$. All supersets of $A$ have a depth equal to or greater than its.
	Of course, this also means that all subsets of $A$ have a depth equal to or less than its. Moreover, sets that are at the left of $A$ are not contained in $A$ and do not contain $A$, since they have all elements of $A$ but the one of greatest index, in addition to others of greater index than the maximum index of $A$. 
	Furthermore, if $B$ diverges with $A$ before its depth, it may be necessary to create a \textit{disjunction}. A disjunction node (i.e. a node that does not hold any value) is inserted to split the branch in two at the depth equal to the first element index not common to both $A$ and $B$. The one that does not contain it will be placed at the left of this disjunction node, and the one that does will be placed at its right. This behavior is illustrated in Fig. \ref{fig:data_structure_dyn}.
	
	A similar binary tree for powerset functions has been proposed in \cite{wilson2000}. It is illustrated in Fig. \ref{fig:data_wilson}. 
	While both our binary tree and theirs are dynamic, theirs does not exploit depths and requires to store $2F -1$ elements, where $F$ is the number of special elements to store. 
	Ours needs to store at most $F + \frac{F}{2}$ elements, and only $F$ if every disjunction node between two special sets is also a special set (or if there is simply no disjunction between special sets, e.g. in the consonant case). Furthermore, it features interesting properties like the fact that the search for all singletons and some sub-FODs can be restrained respectiveley to the left and right chains of nodes. As mentioned above, having depths allows us to exploit the fact that subsets can only be of lower depth, while supersets can only be of greater depth. This form of powerset binary tree also speeds up the search for any value since we have to check at most $n$ booleans to find a value, where $n \in [1, |\Omega|]$, while the version of \cite{wilson2000} always have to check $n$ booleans. 
	
	Recently, another similar
	structure to ours has been proposed in \cite{polpitiya16,polpitiya17}. However, their data structure is not dynamic, i.e. it stores all subsets instead of only special sets. Doing so, they do not have to store a binary string in each node, but they always have an overall exponential spatial complexity (and of course, a time complexity at least exponential accordingly).
	
	In the case of an infinite FOD, the idea is to represent it as an ever changing FOD containing a special element $\omega_\infty$ as first element that symbolizes the \textit{rest} of the FOD. As we know that $\omega_\infty$ will always be an element of this FOD, having $\{\omega_\infty\}$ as the root node of its powerset binary tree reduces the number of operations of reorganization after addition or removal of any FOD element.

\subsection{Procedures computing focal points}\label{algos_F}

In this section, we present algorithms computing all focal points, given focal sets. 

\subsubsection{General procedure}\label{focal_pt_imp}


Here, we apply Property 2 of section 3.4 of \cite{me_journal} to the case where $P=2^\Omega$, leading to Algorithm \ref{focal_points}.
For a set $S \subseteq 2^\Omega$, e.g. $\supp{f}$, this algorithm computes its focal points $^oS$, where $o\in \{ \wedge, \vee \}$ with a complexity less than $O(|S|~.~|{^oS}|)$ in time and $O(|{^oS}|)$ in space.

\subsubsection{Linear analysis}\label{linal}

There are some cases in which a linear pre-analysis of complexity $O(|S|)$ both in time and space is sufficient to find all focal points.
This analysis focuses on the progressive union of all focal sets in a linear run.
Formally, let $S\backslash\{\Omega\} = \{A_1, A_2, \dots, A_{K} \}$ and $S\backslash\{\Omega\}_k = \{A_1, A_2, \dots, A_{k} \}$. This analysis focuses on $I_i = U_{i-1} \cap A_i$ such that $i \in \llbracket 2, K \rrbracket,$ and $U_{i-1} = \bigcup S\backslash\{\Omega\}_{i-1}$ and $I_1 = A_1$.
More precisely, we have
\begin{align*}
I_i &= U_{i-1} \cap A_i \\
&= \left(\textstyle\bigcup S\backslash\{\Omega\}_{i-1}\right) ~\cap A_i\\
&= \textstyle\bigcup_{F \in \{A_1, \dots, A_{i-1} \}} \left(F \cap A_i \right).
\end{align*}
In other words, these intersections $I_i$ contain all focal points based on pairs of focal sets.

In most cases, without testing each pair of focal sets, we cannot know which focal points are generated based solely on $I_i$ since several combinations of sets in $2^{I_i}$ can lead to $I_i$ as union.
For example, if $\Omega = \{a,b,c,d\}$ and $S\backslash\{\Omega\} = \{\{a,b\}, \{a,c\}, \{b,c,d\} \}$, then $U_2 = \{a,b,c\}$ and $I_3 = \{a,b,c\} \cap \{b,c,d\} = \{b,c\}$. Yet, $\{b,c\}$ is not a focal point. It is the result of the union of the focal points $\{a,b\} \cap \{b,c,d\}= \{b\}$ and $\{a,c\} \cap \{b,c,d\}=\{c\}$.

However, there are two special cases in which we know that $I_i$ is a focal point:
$(a)$ if $|I_i| = 0,$ then $2^{I_i} = \{I_i \}$,
and $(b)$ if $|I_i| = 1,$ then $2^{I_i} = \{I_i, \emptyset \}$.

So, if $\forall i \in \llbracket 2, K \rrbracket,$ $|I_i| = 0$, then $\forall F_1,F_2 \in S\backslash\{\Omega\},$ $F_1 \cap F_2 = \emptyset$. This is the \textit{quasi-Bayesian} case that has already been treated in Proposition 1 of \cite{disj_dec_cautious_bold}.
By definition of this case, we have $\forall F \in S\backslash\{\emptyset, \Omega\},$ $F^{\uparrow {^\wedge S}} = \{\Omega \}$ and the only possible focal point other than the focal sets is $\emptyset$, i.e. ${^\wedge S}\backslash \{\emptyset \} = S\backslash \{\emptyset \}$. This means that for any powerset function $f$ such that $\supp{f} \subseteq S$, the computation of its zeta and M\"obius transforms in $(2^\Omega, \supseteq)$ is always $O(|\supp{f}|)$, where $|\supp{f}| \leq |\Omega|+2$.

In addition, this analysis points to a slightly more general case that contains the quasi-Bayesian one 
in which $\forall i \in \llbracket 2, K \rrbracket,$ $|I_i| \leq 1$.
Indeed, when $|I_i| = 1$, we have $2^{I_i} = \{I_i, \emptyset \}$, which means that all focal points composing $I_i$ are in $\{I_i, \emptyset\}$ and at least one of them is the singleton $I_i$, since $\emptyset$ is the neutral element for the union of sets.
Also, the only new focal point that could be generated based on the intersection of the singleton $I_i$ with another focal point is $\emptyset$. 


Algorithm \ref{focal_points_linear} sums up this procedure for $^\wedge S$.
This linear analysis can be performed for the dual order $\subseteq$ as well, focusing on $S\backslash\{\emptyset\} = \{A_1, A_2, \dots, A_{K} \}$, with $\overline{I_i} = \overline{U_{i-1}} \cup A_i$ such that $i \in \llbracket 2, K \rrbracket,$ and $\overline{U_{i-1}} = \bigcap S\backslash\{\emptyset\}_{i-1}$ and $\overline{I_1} = A_1$. 
This dual procedure is provided by Algorithm \ref{focal_points_linear_sub}.

\begin{algorithm}
	\KwIn{$S\backslash\{\Omega\} = \{A_1, A_2, \dots, A_{K} \}$}
	\KwOut{$^\wedge S$, \textit{is\_almost\_bayesian}}
	$^\wedge S \leftarrow S$\;
	$U \leftarrow A_1$\;
	$\textit{is\_almost\_bayesian} \leftarrow \text{True}$\;
	\For{$i= 2 ~\KwTo~ K$}{
		$I \leftarrow U \cap A_i$\;
		\If{$|I| > 1$}{
			$\textit{is\_almost\_bayesian} \leftarrow \text{False}$\;
			\textbf{break}\;
		}\ElseIf{$|I| = 1$}{
			\If{$I \not\in {^\wedge S}$}{
				append $I$ to $^\wedge S$\;
			}
		}
		\If{$\emptyset \not\in {^\wedge S}$}{
			append $\emptyset$ to $^\wedge S$\;
		}
		$U \leftarrow U \cup A_i$\;
	}
	\caption{Linear computation of $^\wedge S$ based on $S$.}\label{focal_points_linear}
\end{algorithm}

\begin{algorithm}
	\KwIn{$S\backslash\{\emptyset\} = \{A_1, A_2, \dots, A_{K} \}$}
	\KwOut{${^\vee S}$, \textit{is\_almost\_bayesian}}
	${^\vee S} \leftarrow S$\;
	$U \leftarrow A_1$\;
	$\textit{is\_almost\_bayesian} \leftarrow \text{True}$\;
	\For{$i= 2 ~\KwTo~ K$}{
		$I \leftarrow U \cup A_i$\;
		\If{$|I| < |\Omega| - 1$}{
			$\textit{is\_almost\_bayesian} \leftarrow \text{False}$\;
			\textbf{break}\;
		}\ElseIf{$|I| = |\Omega| - 1$}{
			\If{$I \not\in {^\vee S}$}{
				append $I$ to ${^\vee S}$\;
			}
		}
		\If{$\Omega \not\in {^\vee S}$}{
			append $\Omega$ to ${^\vee S}$\;
		}
		$U \leftarrow U \cap A_i$\;
	}
	\caption{Linear computation of ${^\vee S}$ based on $S$.}\label{focal_points_linear_sub}
\end{algorithm}

\subsection{Computation of iota elements}

Computation of the iota elements (See \autoref{iota_elements}) and dual iota elements (See \autoref{iota_elements_dual}) of $S$ are presented respectively in Algorithms \ref{algo:dotOmega} and \ref{algo:dotOmega_dual}. 

\begin{algorithm}[h!]
	\KwIn{$S$}
	\KwOut{$I$}
	
	\ForEach{$\omega \in \Omega$}{
		$i \leftarrow \Omega$\;
		\ForEach{$F \in S$, such that $F \supseteq \{\omega\}$}{
			$\textit{included} \leftarrow \text{True}$\;
			$i \leftarrow i \cap F$\;
			\If{$i = \{\omega\}$}{
				\textbf{break}\;
			}
		}
		\If{\textit{included}}{
			append $i$ to $I$\;
		}
	}
	\caption{Computation of $\iota(S)$.}\label{algo:dotOmega}
\end{algorithm}

\begin{algorithm}[h!]
	\KwIn{$S$}
	\KwOut{$I$}
	
	\ForEach{$\omega \in \Omega$}{
		$i \leftarrow \emptyset$\;
		\ForEach{$F \in S$, such that $F \subseteq \Omega\backslash\{\omega\}$}{
			$\textit{included} \leftarrow \text{True}$\;
			$i \leftarrow i \cup F$\;
			\If{$i = \Omega\backslash\{\omega\}$}{
				\textbf{break}\;
			}
		}
		\If{\textit{included}}{
			append $i$ to $I$\;
		}
	}
	\caption{Computation of $\overline{\iota}(S)$.}\label{algo:dotOmega_dual}
\end{algorithm}

\subsection{Computation of the lattice support}

Computation of the upper and lower closure in the lattice support of $S$ (See \autoref{supp_lattice}) are presented respectively in Algorithms \ref{algo:dotLsub} and \ref{algo:dotLsup}. 

\begin{algorithm}[h!]
	\KwIn{$\iota(S)$, $S$}
	\KwOut{$L$}
	$L \leftarrow S$\;
	\ForEach{$i \in \iota(S)$}{
		\ForEach{$A \in L$}{
			$B \leftarrow A \cup i$\;
			\If{$B \not\in L$}{
				append $B$ to $L$\;
			}
		}
	}
	\caption{Computation of $S^{\uparrow{^{\mathcal{L}} S}}$ based on $\iota(S)$ and $S$.}\label{algo:dotLsub}
\end{algorithm}

\begin{algorithm}[h!]
	\KwIn{$\overline{\iota}(S)$, $S$}
	\KwOut{$L$}
	$L \leftarrow S$\;
	\ForEach{$i \in \overline{\iota}(S)$}{
		\ForEach{$A \in L$}{
			$B \leftarrow A \cap i$\;
			\If{$B \not\in L$}{
				append $B$ to $L$\;
			}
		}
	}
	\caption{Computation of $S^{\downarrow{^{\mathcal{L}} S}}$ based on $\overline{\iota}(S)$ and $S$.}\label{algo:dotLsup}
\end{algorithm}

\subsection{Computation of DST transformations in the consonant case}\label{algos_consonance}

The consonant case has already been treated in Proposition 2 of \cite{disj_dec_cautious_bold}.
A consonant structure of evidence is a nested structure where each focal set is contained in every focal set of greater or equal cardinality. Formally, $\forall F_i,F_j \in \mathcal{F},$ if $|F_i| \leq |F_j|$ then $F_i \subseteq F_j$. By definition, there can be only one focal set of each cardinality in $\llbracket 0, |\Omega| \rrbracket$, each of them being a subset of every focal set of greater cardinality. This means that there can be no focal point other than focal sets and that each focal point has a proxy focal point.
Let us note $|F_1| < |F_2| < \dots < |F_{K}|$ the focal elements in $S$.
To check if this structure is consonant, we simply have to check that $\forall i \in \llbracket 1, K-1 \rrbracket,$ $F_i \subset F_{i+1}$. This analysis is performed by Algorithm \ref{consonance_check}.

Algorithms \ref{algo:consonance_b} to \ref{algo:consonance_w_rev} present procedures in the consonant case to compute the following transformations: 
\begin{multicols}{4}
	\begin{itemize}
		\item $m$ to $b$, 
		\item $b$ to $m$, 
		\item $m$ to $q$, 
		\item $q$ to $m$, 
		\item $b$ to $v$, 
		\item $v$ to $b$, 
		\item $q$ to $w$, 
		\item $w$ to $q$.
	\end{itemize}
\end{multicols}
Their complexity in time is $\min\left[O(|\Omega|), O\left(|S|.\log \left(|S|\right)\right) \right]$ (which corresponds to the complexity of the sorting algorithm used on $S$). Their space complexity is $O(|S|)$.

\begin{algorithm}[h]
	\KwIn{$S$}
	\KwOut{\textit{is\_consonant}}
	
	sort $S$ such that $S = \{F_1, F_2, \dots, F_{K}\},$ where $|F_1| \leq |F_2| \leq \dots \leq |F_{K}|$\;
	$\textit{is\_consonant} \leftarrow$ True\;
	\For{$i \in \llbracket 1, K-1 \rrbracket$}{
		\If{$F_i \not\subseteq F_{i+1}$}{
			$\textit{is\_consonant} \leftarrow$ False\;
			\textbf{break}\;
		}
	}
	\caption{
		Consonance check.}\label{consonance_check}
\end{algorithm}

\subsection{Computation of DST transformations in a semilattice}\label{algos_dotF}

Algorithms \ref{algo:b_dotF} to \ref{algo:w_dotF_rev} present procedures using $^\wedge S$ or $^\vee S$ to compute the following transformations: 
\begin{multicols}{4}
	\begin{itemize}
		\item $m$ to $b$, 
		\item $b$ to $m$, 
		\item $m$ to $q$, 
		\item $q$ to $m$, 
		\item $b$ to $v$, 
		\item $v$ to $b$, 
		\item $q$ to $w$, 
		\item $w$ to $q$.
	\end{itemize}
\end{multicols}
They all use the sequences of graphs of Theorem \ref{mob_opti_F} and Corollary \ref{mob_opti_F_dual}. Their complexity in time is $O(I(S).|{^o S}|.\epsilon)$, where $I \in \{ \iota, \overline{\iota} \}$ and $o \in \{ \wedge, \vee \}$ and where $\epsilon$ represents the average number of operations required to ``bridge a gap'' (See Theorem \ref{mob_opti_F}). This is always less than $O(|\Omega|.2^{\Omega})$. Their space complexity is $O(|{^o S}|)$.

\subsection{Computation of DST transformations in a lattice}\label{dotOmega}

The general procedure to compute ${^o S}$ is less than $O(|S|.|{^o S}|)$. But, if one wishes to be certain that our algorithms are at least as efficient as the FMT, i.e. at most $O(|\Omega|.2^{|\Omega|})$, then one has to look at $|S|$. Indeed, if $O(|S|) \leq O(|\Omega|)$ (e.g. $|S| < 10.|\Omega|$), then $O(|S|.|{^o S}|) \leq O(|\Omega|.2^{|\Omega|})$. Otherwise, one can rely on the lattice support ${^{\mathcal{L}} S}$, since $\iota(S)$ can be computed in less than $O(|\Omega|.|S|)$ with Algorithm \ref{algo:dotOmega} and is then used to compute ${^o S}$ in $O(\iota(S).|{^{\mathcal{L}} S}|)$ in Algorithms \ref{algo:dotLsub} and \ref{algo:dotLsup}, where $|\iota(S)| \leq |{\Omega}|$.

Algorithms \ref{algo:b_dotL} to \ref{algo:w_dotL_rev} present procedures using ${^{\mathcal{L}} S}$ (more precisely, the upper and lower closures $S^{\uparrow{^{\mathcal{L}} S}}$ and $S^{\downarrow{^{\mathcal{L}} S}}$) to compute the following transformations: 
\begin{multicols}{4}
	\begin{itemize}
		\item $m$ to $b$, 
		\item $b$ to $m$, 
		\item $m$ to $q$, 
		\item $q$ to $m$, 
		\item $b$ to $v$, 
		\item $v$ to $b$, 
		\item $q$ to $w$, 
		\item $w$ to $q$.
	\end{itemize}
\end{multicols}
They all use the sequences of graphs of Theorem \ref{mob_opti_L} and its corollaries. Their complexity in time is less than $O(I(S).|{^{\mathcal{L}} S}|)$, where $I \in \{ \iota, \overline{\iota} \}$ and $o \in \{ \wedge, \vee \}$, which is always less than $O(|\Omega|.2^{|\Omega|})$. Their space complexity is less than $O(|{^{\mathcal{L}} S}|)$.

\subsection{Computation of DST transformations independently from $\Omega$}\label{algos_indep_N}

If $\supp{f}$ is almost Bayesian or if $|\Omega|$ happens to be considerable to the point that one would like to compute the previous DST transformations independently from $|\Omega|$, Algorithms \ref{algo:b_dotF_noN} to \ref{algo:w_dotF_noN_rev} present procedures of time complexities in $[O(|{^oS}|), O(|{^oS}|^2)]$. Their complexity in space is $O(|{^oS}|)$.
They 
present procedures for the following transformations:
\begin{multicols}{4}
	\begin{itemize}
		\item $m$ to $b$, 
		\item $b$ to $m$, 
		\item $m$ to $q$, 
		\item $q$ to $m$, 
		\item $b$ to $v$, 
		\item $v$ to $b$, 
		\item $q$ to $w$, 
		\item $w$ to $q$.
	\end{itemize}
\end{multicols}

\begin{figure}
	\centering
	\hspace{-1.2cm}
	\newdimen\nodeDist
	\nodeDist=2.5cm
	\begin{tikzpicture}[
	action/.style={%
		draw,
		rectangle,
	},
	question/.style={%
		draw,
		diamond,
		inner sep=0pt
	},
	]
	
%
	\node [action] (A) {$\substack{\text{Linear analysis}\\\text{(section \ref{linal})}}$};
%
	
	\path (A) ++(-90:2cm) node [question] (D) {$\substack{\textit{\small is\_almost\_bayesian}}$};
	
	
	\path (D) ++(-130:2.5cm) node [question] (E) {$\substack{\textit{\small is\_consonant}}$};
	
	\path (E) ++(-45:2.5cm) node [action] (J) {$\substack{\text{Use Alg. from}\\\text{section \ref{algos_consonance}}}$};
	
	\path (E) ++(-130:2.3cm) node [question] (C) {$\substack{O(S) ~\leq~ O(|\Omega|)\\\text{\tiny(e.g. $S < 10.|\Omega|$)}}$};
	
	\path (C) ++(-55:\nodeDist) node [action] (C') {$\substack{\text{Compute $^o S$}\\\text{(section \ref{focal_pt_imp})}}$};
	
	\path (C') ++(-90:3cm) node [question] (H) {$\substack{O(|{^o S}|) ~\leq~ O(|\Omega|)\\\text{\tiny(e.g. $|{^o S}| < 10. |\Omega|$)}}$};
	
	\path (H) ++(-45:2.5cm) node [action] (G) {$\substack{\text{Use Alg. from}\\\text{section \ref{algos_indep_N}}}$};
	
	\path (H) ++(-135:2.5cm) node [action] (I) {$\substack{\text{Use $^o S$ and $I(S)$}\\\text{(section \ref{algos_dotF})}}$};
	
	\path (C) ++(-130:2.3cm) node [action] (B) {$\substack{\text{Use ${^\mathcal{L} S}$ and $I(S)$}\\\text{(section \ref{dotOmega})}}$};

	\path (B) ++(-90:1.3cm) node (X) {$O(|I(S).|{^\mathcal{L} S}|)$};
	
	\path (J) ++(-90:1.2cm) node (Y) {$\min\left[O(|\Omega|), O\left(|S|.\log \left(|S|\right)\right) \right]$};
	
	\path (G) ++(-90:1.2cm) node (Z) {$\in \left[O(S), O(|^o S|^2)\right]$};
	
	\path (I) ++(-90:1.2cm) node (ZZ) {$O(|I(S)|.|^o S|.\epsilon)$};
	
	
	\draw[->,>=latex] (A) to (D);
	\draw[->,>=latex] (C) to node [left,pos=0.25] {no} (B);
	\draw[->,>=latex] (C) to node [right,pos=0.1] {yes} (C');
	\draw[->,>=latex] (C') to (H);
	\draw[->,>=latex] (D) to node [left,pos=0.25] {no} (E);
	\draw[->,>=latex] (E) to node [left,pos=0.25] {no} (C);
	\draw[->,>=latex] (H) to node [right,pos=0.25] {yes} (G);
	\draw[->,>=latex] (D) to [bend right=-50] node [right,pos=0.02]{yes} (G);
	\draw[->,>=latex] (E) to node [right,pos=0.25] {yes} (J);
	\draw[->,>=latex] (H) to node [left,pos=0.25] {no} (I);
	\draw[->,>=latex] (B) to (X);
	\draw[->,>=latex] (J) to (Y);
	\draw[->,>=latex] (G) to (Z);
	\draw[->,>=latex] (I) to (ZZ);
	\end{tikzpicture}
	\caption{Decision tree for the choice of which algorithms of section \ref{implementation} to use. Of course, this decision tree can be implemented as an algorithm in order to create a unique general procedure automatically choosing the type of algorithm to use, no matter the DST transformation. Diamond nodes represent Boolean tests. Rectangle nodes represent the chosen action. Finally, terminal nodes (leaves) indicate the final time complexity of the whole procedure, including the computation of any DST transformation, where $o\in \{ \wedge, \vee \}$ and $I \in \{ \iota, \overline{\iota} \}$.
	In particular, when \textit{is\_almost\_Bayesian} is true, the complexity of the whole procedure is $O(S)$.
	Notice that all these complexities are less than $O(|\Omega|.2^{|\Omega|})$.}\label{fig:decision_tree}
\end{figure}
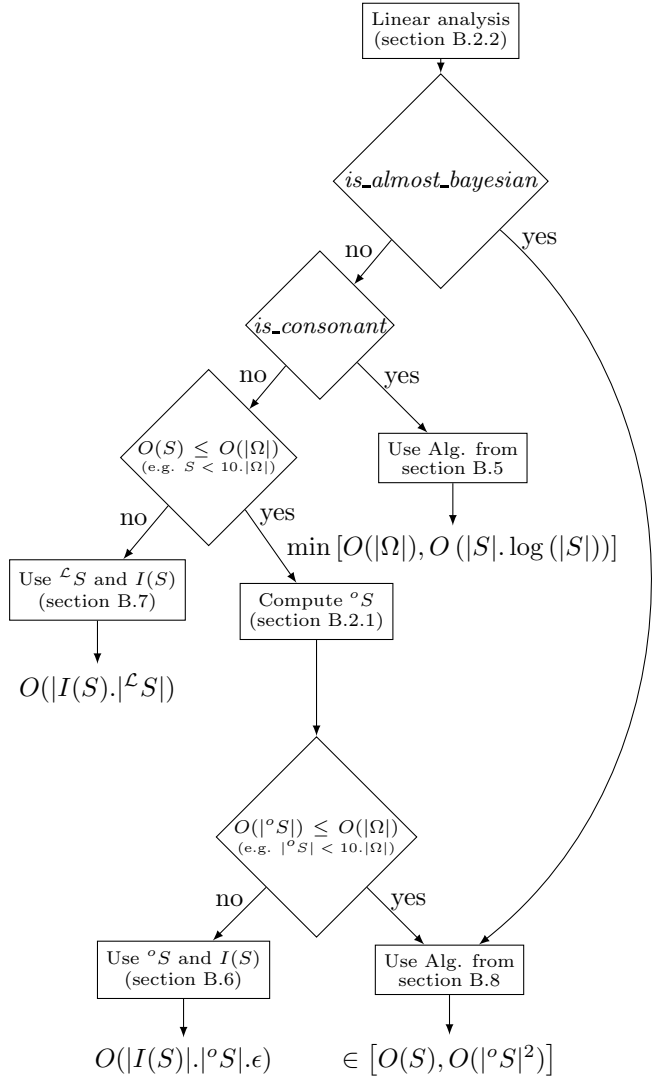

Fig. \ref{fig:decision_tree} offers a decision tree to help the reader in choosing the right algorithm for their case. Of course, this decision tree can be implemented as an algorithm in order to create a unique general procedure automatically choosing the type of algorithm to use, no matter the DST transformation.

\begin{algorithm}
	\KwIn{$\{m(B) ~/~ B \in S \}$, \textit{is\_consonant}, $S = \{F_1, F_2, \dots, F_{K}\},$ where $|F_1| \leq |F_2| \leq \dots \leq |F_{K}|$}
	\KwOut{$\{b(B) ~/~ B \in S \}$}
	\If{$\textit{is\_consonant}$}{
		$b(F_1) \leftarrow m(F_1)$\;
		\For{$i = 2 ~\KwTo~ K$}{
			$b(F_i) \leftarrow m(F_i) + b(F_{i-1})$\;
		}
	}
	\caption{Computation of $\{b(B) ~/~ B \in S \}$ based on $\{m(B) ~/~ B \in S \}$ in the consonant case.}\label{algo:consonance_b}
\end{algorithm}

\begin{algorithm}
	\KwIn{$\{b(B) ~/~ B \in S \}$, \textit{is\_consonant}, $S = \{F_1, F_2, \dots, F_{K}\},$ where $|F_1| \leq |F_2| \leq \dots \leq |F_{K}|$}
	\KwOut{$\{m(B) ~/~ B \in S \}$}
	\If{$\textit{is\_consonant}$}{
		$m(F_1) \leftarrow b(F_1)$\;
		\For{$i = 2 ~\KwTo~ K$}{
			$m(F_i) \leftarrow b(F_i) - b(F_{i-1})$\;
		}
	}
	\caption{Computation of $\{m(B) ~/~ B \in S \}$ based on $\{b(B) ~/~ B \in S \}$ in the consonant case.}\label{algo:consonance_b_rev}
\end{algorithm}

\begin{algorithm}
	\KwIn{$\{m(B) ~/~ B \in S \}$, \textit{is\_consonant}, $S = \{F_1, F_2, \dots, F_{K}\},$ where $|F_1| \leq |F_2| \leq \dots \leq |F_{K}|$}
	\KwOut{$\{q(B) ~/~ B \in S \}$}
	\If{$\textit{is\_consonant}$}{
		$q(F_K) \leftarrow m(F_K)$\;
		\For{$i = K-1 ~\KwTo~ 1$}{
			$q(F_i) \leftarrow m(F_i) + q(F_{i+1})$\;
		}
	}
	\caption{Computation of $\{q(B) ~/~ B \in S \}$ based on $\{m(B) ~/~ B \in S \}$ in the consonant case.}\label{algo:consonance_q}
\end{algorithm}

\begin{algorithm}
	\KwIn{$\{q(B) ~/~ B \in S \}$, \textit{is\_consonant}, $S = \{F_1, F_2, \dots, F_{K}\},$ where $|F_1| \leq |F_2| \leq \dots \leq |F_{K}|$}
	\KwOut{$\{m(B) ~/~ B \in S \}$}
	\If{$\textit{is\_consonant}$}{
		$m(F_K) \leftarrow q(F_K)$\;
		\For{$i = K-1 ~\KwTo~ 1$}{
			$m(F_i) \leftarrow q(F_i) - q(F_{i+1})$\;
		}
	}
	\caption{Computation of $\{m(B) ~/~ B \in S \}$ based on $\{q(B) ~/~ B \in S \}$ in the consonant case.}\label{algo:consonance_q_rev}
\end{algorithm}

\begin{algorithm}
	\KwIn{$\{b(B) ~/~ B \in S \}$, \textit{is\_consonant}, $S = \{F_1, F_2, \dots, F_{K}\},$ where $|F_1| \leq |F_2| \leq \dots \leq |F_{K}|$}
	\KwOut{$\{v(B) ~/~ B \in S \}$}
	\If{$\textit{is\_consonant}$}{
		$v(F_1) \leftarrow b(F_1)^{-1}$\;
		\For{$i = 2 ~\KwTo~ K$}{
			$v(F_i) \leftarrow b(F_i)^{-1}.~b(F_{i-1})$\;
		}
	}
	\caption{Computation of $\{v(B) ~/~ B \in S \}$ based on $\{b(B) ~/~ B \in S \}$ in the consonant case.}\label{algo:consonance_v}
\end{algorithm}

\begin{algorithm}
	\KwIn{$\{v(B) ~/~ B \in S \}$, \textit{is\_consonant}, $S = \{F_1, F_2, \dots, F_{K}\},$ where $|F_1| \leq |F_2| \leq \dots \leq |F_{K}|$}
	\KwOut{$\{b(B) ~/~ B \in S \}$}
	\If{$\textit{is\_consonant}$}{
		$b(F_1) \leftarrow v(F_1)^{-1}$\;
		\For{$i = 2 ~\KwTo~ K$}{
			$b(F_i) \leftarrow v(F_i)^{-1}.~b(F_{i-1})$\;
		}
	}
	\caption{Computation of $\{b(B) ~/~ B \in S \}$ based on $\{v(B) ~/~ B \in S \}$ in the consonant case.}\label{algo:consonance_v_rev}
\end{algorithm}

\begin{algorithm}
	\KwIn{$\{q(B) ~/~ B \in S \}$, \textit{is\_consonant}, $S = \{F_1, F_2, \dots, F_{K}\},$ where $|F_1| \leq |F_2| \leq \dots \leq |F_{K}|$}
	\KwOut{$\{w(B) ~/~ B \in S \}$}
	\If{$\textit{is\_consonant}$}{
		$w(F_K) \leftarrow q(F_K)^{-1}$\;
		\For{$i = K-1 ~\KwTo~ 1$}{
			$w(F_i) \leftarrow q(F_i)^{-1}.~q(F_{i+1})$\;
		}
	}
	\caption{Computation of $\{w(B) ~/~ B \in S \}$ based on $\{q(B) ~/~ B \in S \}$ in the consonant case.}\label{algo:consonance_w}
\end{algorithm}

\begin{algorithm}
	\KwIn{$\{w(B) ~/~ B \in S \}$, \textit{is\_consonant}, $S = \{F_1, F_2, \dots, F_{K}\},$ where $|F_1| \leq |F_2| \leq \dots \leq |F_{K}|$}
	\KwOut{$\{q(B) ~/~ B \in S \}$}
	\If{$\textit{is\_consonant}$}{
		$q(F_K) \leftarrow w(F_K)^{-1}$\;
		\For{$i = K-1 ~\KwTo~ 1$}{
			$q(F_i) \leftarrow w(F_i)^{-1}.~q(F_{i+1})$\;
		}
	}
	\caption{Computation of $\{q(B) ~/~ B \in S \}$ based on $\{w(B) ~/~ B \in S \}$ in the consonant case.}\label{algo:consonance_w_rev}
\end{algorithm}

\begin{algorithm}
	\KwIn{$\{m(B) ~/~ B \in {^\vee S} \}$, ${^\vee S}$, $\overline{\iota}(S)$}
	\KwOut{$\{b(B) ~/~ B \in {^\vee S} \}$}
	sort $\overline{\iota}(S)$ such that $\overline{\iota}(S) = \{i_1, i_2, \dots, i_{K}\},$ where $|i_1| \leq |i_2| \leq \dots \leq |i_{K}|$\;
	\ForEach{$A \in {^\vee S}$}{
		$b(A) \leftarrow m(A)$\;
	}
	$\Omega_{cum} \leftarrow \Omega$\;
	\For{$k = K ~\KwTo~ 1$}{
		$\Omega_{cum} \leftarrow \Omega_{cum} \cap i_k$\;
		\ForEach{$A \in {^\vee S}$}{
			$B \leftarrow A \cap i_k$\;
			\If{$B \neq A$}{
				$X \leftarrow \displaystyle\argmax_{C \in B^{\downarrow{^\vee S}}} \hspace{-0.cm}\left(|C| \right)$\;
				\If{$X \neq$ \text{NULL} \textbf{and} $X \supseteq A \cap \Omega_{cum}$}{
					$b(A) \leftarrow b(A) + b(X)$\;
				}
			}
		}
	}
	\caption{Computation of $\{b(B) ~/~ B \in {^\vee S} \}$ based on $\{m(B) ~/~ B \in {^\vee S} \}$.}\label{algo:b_dotF}
\end{algorithm}

\begin{algorithm}
	\KwIn{$\{b(B) ~/~ B \in {^\vee S} \}$, ${^\vee S}$, $\overline{\iota}(S)$}
	\KwOut{$\{m(B) ~/~ B \in {^\vee S} \}$}
	sort $\overline{\iota}(S)$ such that $\overline{\iota}(S) = \{i_1, i_2, \dots, i_{K}\},$ where $|i_1| \leq |i_2| \leq \dots \leq |i_{K}|$\;
	\ForEach{$A \in {^\vee S}$}{
		$m(A) \leftarrow b(A)$\;
	}
	$\Omega_{cum} \leftarrow \Omega$\;
	\For{$k = 1$ \KwTo $K$}{
		$\Omega_{cum} \leftarrow \Omega_{cum} \cap i_k$\;
		\ForEach{$A \in {^\vee S}$}{
			$B \leftarrow A \cap i_k$\;
			\If{$B \neq A$}{
				$X \leftarrow \displaystyle\argmax_{C \in B^{\downarrow{^\vee S}}} \hspace{-0.cm}\left(|C| \right)$\;
				\If{$X \neq$ \text{NULL} \textbf{and} $X \supseteq A \cap \Omega_{cum}$}{
					$m(A) \leftarrow m(A) - m(X)$\;
				}
			}
		}
	}
	\caption{Computation of $\{m(B) ~/~ B \in {^\vee S} \}$ based on $\{b(B) ~/~ B \in {^\vee S} \}$.}\label{algo:b_dotF_rev}
\end{algorithm}

\begin{algorithm}
	\KwIn{$\{m(B) ~/~ B \in {^\wedge S} \}$, ${^\wedge S}$, $\iota(S)$}
	\KwOut{$\{q(B) ~/~ B \in {^\wedge S} \}$}
	sort ${\iota}(S)$ such that ${\iota}(S) = \{i_1, i_2, \dots, i_{K}\},$ where $|i_1| \leq |i_2| \leq \dots \leq |i_{K}|$\;
	\ForEach{$A \in {^\wedge S}$}{
		$q(A) \leftarrow m(A)$\;
	}
	$\Omega_{cum} \leftarrow \emptyset$\;
	\For{$k = 1$ \KwTo $K$}{
		$\Omega_{cum} \leftarrow \Omega_{cum} \cup i_k$\;
		\ForEach{$A \in {^\wedge S}$}{
			$B \leftarrow A \cup i_k$\;
			\If{$B \neq A$}{
				$X \leftarrow \displaystyle\argmin_{C \in B^{\uparrow{^\wedge S}}} \hspace{-0.cm}\left(|C| \right)$\;
				\If{$X \neq$ \text{NULL} \textbf{and} $X \subseteq A \cup \Omega_{cum}$}{
					$q(A) \leftarrow q(A) + q(X)$\;
				}
			}
		}
	}
	\caption{Computation of $\{q(B) ~/~ B \in {^\wedge S} \}$ based on $\{m(B) ~/~ B \in {^\wedge S} \}$.}\label{algo:q_dotF}
\end{algorithm}

\begin{algorithm}
	\KwIn{$\{q(B) ~/~ B \in {^\wedge S} \}$, ${^\wedge S}$, $\iota(S)$}
	\KwOut{$\{m(B) ~/~ B \in {^\wedge S} \}$}
	sort ${\iota}(S)$ such that ${\iota}(S) = \{i_1, i_2, \dots, i_{K}\},$ where $|i_1| \leq |i_2| \leq \dots \leq |i_{K}|$\;
	\ForEach{$A \in {^\wedge S}$}{
		$m(A) \leftarrow q(A)$\;
	}
	$\Omega_{cum} \leftarrow \emptyset$\;
	\For{$k = K$ \KwTo $1$}{
		$\Omega_{cum} \leftarrow \Omega_{cum} \cup i_k$\;
		\ForEach{$A \in {^\wedge S}$}{
			$B \leftarrow A \cup i_k$\;
			\If{$B \neq A$}{
				$X \leftarrow \displaystyle\argmin_{C \in B^{\uparrow{^\wedge S}}} \hspace{-0.cm}\left(|C| \right)$\;
				\If{$X \neq$ \text{NULL} \textbf{and} $X \subseteq A \cup \Omega_{cum}$}{
					$m(A) \leftarrow m(A) - m(X)$\;
				}
			}
		}
	}
	\caption{Computation of $\{m(B) ~/~ B \in {^\wedge S} \}$ based on $\{q(B) ~/~ B \in {^\wedge S} \}$.}\label{algo:q_dotF_rev}
\end{algorithm}

\begin{algorithm}
	\KwIn{$\{b(B) ~/~ B \in {^\vee S} \}$, ${^\vee S}$, $\overline{\iota}(S)$}
	\KwOut{$\{v(B) ~/~ B \in {^\vee S} \}$}
	sort $\overline{\iota}(S)$ such that $\overline{\iota}(S) = \{i_1, i_2, \dots, i_{K}\},$ where $|i_1| \leq |i_2| \leq \dots \leq |i_{K}|$\;
	\ForEach{$A \in {^\vee S}$}{
		$v(A) \leftarrow b(A)^{-1}$\;
	}
	$\Omega_{cum} \leftarrow \Omega$\;
	\For{$k = 1$ \KwTo $K$}{
		$\Omega_{cum} \leftarrow \Omega_{cum} \cap i_k$\;
		\ForEach{$A \in {^\vee S}$}{
			$B \leftarrow A \cap i_k$\;
			\If{$B \neq A$}{
				$X \leftarrow \displaystyle\argmax_{C \in B^{\downarrow{^\vee S}}} \hspace{-0.cm}\left(|C| \right)$\;
				\If{$X \neq$ \text{NULL} \textbf{and} $X \supseteq A \cap \Omega_{cum}$}{
					$v(A) \leftarrow v(A) / v(X)$\;
				}
			}
		}
	}
	\caption{Computation of $\{v(B) ~/~ B \in {^\vee S} \}$ based on $\{b(B) ~/~ B \in {^\vee S} \}$.}\label{algo:v_dotF}
\end{algorithm}

\begin{algorithm}
	\KwIn{$\{v(B) ~/~ B \in {^\vee S} \}$, ${^\vee S}$, $\overline{\iota}(S)$}
	\KwOut{$\{b(B) ~/~ B \in {^\vee S} \}$}
	sort $\overline{\iota}(S)$ such that $\overline{\iota}(S) = \{i_1, i_2, \dots, i_{K}\},$ where $|i_1| \leq |i_2| \leq \dots \leq |i_{K}|$\;
	\ForEach{$A \in {^\vee S}$}{
		$b(A) \leftarrow v(A)^{-1}$\;
	}
	$\Omega_{cum} \leftarrow \Omega$\;
	\For{$k = K$ \KwTo $1$}{
		$\Omega_{cum} \leftarrow \Omega_{cum} \cap i_k$\;
		\ForEach{$A \in {^\vee S}$}{
			$B \leftarrow A \cap i_k$\;
			\If{$B \neq A$}{
				$X \leftarrow \displaystyle\argmax_{C \in B^{\downarrow{^\vee S}}} \hspace{-0.cm}\left(|C| \right)$\;
				\If{$X \neq$ \text{NULL} \textbf{and} $X \supseteq A \cap \Omega_{cum}$}{
					$b(A) \leftarrow b(A) . b(X)$\;
				}
			}
		}
	}
	\caption{Computation of $\{b(B) ~/~ B \in {^\vee S} \}$ based on $\{v(B) ~/~ B \in {^\vee S} \}$.}\label{algo:v_dotF_rev}
\end{algorithm}

\begin{algorithm}
	\KwIn{$\{q(B) ~/~ B \in {^\wedge S} \}$, ${^\wedge S}$, $\iota(S)$}
	\KwOut{$\{w(B) ~/~ B \in {^\wedge S} \}$}
	sort ${\iota}(S)$ such that ${\iota}(S) = \{i_1, i_2, \dots, i_{K}\},$ where $|i_1| \leq |i_2| \leq \dots \leq |i_{K}|$\;
	\ForEach{$A \in {^\wedge S}$}{
		$w(A) \leftarrow q(A)^{-1}$\;
	}
	$\Omega_{cum} \leftarrow \emptyset$\;
	\For{$k = K$ \KwTo $1$}{
		$\Omega_{cum} \leftarrow \Omega_{cum} \cup i_k$\;
		\ForEach{$A \in {^\wedge S}$}{
			$B \leftarrow A \cup i_k$\;
			\If{$B \neq A$}{
				$X \leftarrow \displaystyle\argmin_{C \in B^{\uparrow{^\wedge S}}} \hspace{-0.cm}\left(|C| \right)$\;
				\If{$X \neq$ \text{NULL} \textbf{and} $X \subseteq A \cup \Omega_{cum}$}{
					$w(A) \leftarrow w(A) / w(X)$\;
				}
			}
		}
	}
	\caption{Computation of $\{w(B) ~/~ B \in {^\wedge S} \}$ based on $\{q(B) ~/~ B \in {^\wedge S} \}$.}\label{algo:w_dotF}
\end{algorithm}

\begin{algorithm}
	\KwIn{$\{w(B) ~/~ B \in {^\wedge S} \}$, ${^\wedge S}$, $\iota(S)$}
	\KwOut{$\{q(B) ~/~ B \in {^\wedge S} \}$}
	sort ${\iota}(S)$ such that ${\iota}(S) = \{i_1, i_2, \dots, i_{K}\},$ where $|i_1| \leq |i_2| \leq \dots \leq |i_{K}|$\;
	\ForEach{$A \in {^\wedge S}$}{
		$q(A) \leftarrow w(A)^{-1}$\;
	}
	$\Omega_{cum} \leftarrow \emptyset$\;
	\For{$k = 1$ \KwTo $K$}{
		$\Omega_{cum} \leftarrow \Omega_{cum} \cup i_k$\;
		\ForEach{$A \in {^\wedge S}$}{
			$B \leftarrow A \cup i_k$\;
			\If{$B \neq A$}{
				$X \leftarrow \displaystyle\argmin_{C \in B^{\uparrow{^\wedge S}}} \hspace{-0.cm}\left(|C| \right)$\;
				\If{$X \neq$ \text{NULL} \textbf{and} $X \subseteq A \cup \Omega_{cum}$}{
					$q(A) \leftarrow q(A) . q(X)$\;
				}
			}
		}
	}
	\caption{Computation of $\{q(B) ~/~ B \in {^\wedge S} \}$ based on $\{w(B) ~/~ B \in {^\wedge S} \}$.}\label{algo:w_dotF_rev}
\end{algorithm}

\begin{algorithm}
	\KwIn{$\{m(B) ~/~ B \in {S^{\downarrow ^\mathcal{L} S}} \}$, ${S^{\downarrow ^\mathcal{L} S}}$, $\overline{\iota}(S)$}
	\KwOut{$\{b(B) ~/~ B \in {S^{\downarrow ^\mathcal{L} S}} \}$}
	sort $\overline{\iota}(S)$ such that $\overline{\iota}(S) = \{i_1, i_2, \dots, i_{K}\},$ where $|i_1| \leq |i_2| \leq \dots \leq |i_{K}|$\;
	\ForEach{$A \in {S^{\downarrow ^\mathcal{L} S}}$}{
		$b(A) \leftarrow m(A)$\;
	}
	\For{$k = K ~\KwTo~ 1$}{
		\ForEach{$A \in {S^{\downarrow ^\mathcal{L} S}}$}{
			$B \leftarrow A \cap i_k$\;
			\If{$B \neq A$}{
				$b(A) \leftarrow b(A) + b(B)$\;
			}
		}
	}
	\caption{Computation of $\{b(B) ~/~ B \in {S^{\downarrow ^\mathcal{L} S}} \}$ based on $\{m(B) ~/~ B \in {S^{\downarrow ^\mathcal{L} S}} \}$.}\label{algo:b_dotL}
\end{algorithm}

\begin{algorithm}
	\KwIn{$\{b(B) ~/~ B \in {S^{\downarrow ^\mathcal{L} S}} \}$, ${S^{\downarrow ^\mathcal{L} S}}$, $\overline{\iota}(S)$}
	\KwOut{$\{m(B) ~/~ B \in {S^{\downarrow ^\mathcal{L} S}} \}$}
	sort $\overline{\iota}(S)$ such that $\overline{\iota}(S) = \{i_1, i_2, \dots, i_{K}\},$ where $|i_1| \leq |i_2| \leq \dots \leq |i_{K}|$\;
	\ForEach{$A \in {S^{\downarrow ^\mathcal{L} S}}$}{
		$m(A) \leftarrow b(A)$\;
	}
	\For{$k = 1 ~\KwTo~ K$}{
		\ForEach{$A \in {S^{\downarrow ^\mathcal{L} S}}$}{
			$B \leftarrow A \cap i_k$\;
			\If{$B \neq A$}{
				$m(A) \leftarrow m(A) - m(B)$\;
			}
		}
	}
	\caption{Computation of $\{m(B) ~/~ B \in {S^{\downarrow ^\mathcal{L} S}} \}$ based on $\{b(B) ~/~ B \in {S^{\downarrow ^\mathcal{L} S}} \}$.}\label{algo:b_dotL_rev}
\end{algorithm}

\begin{algorithm}
	\KwIn{$\{m(B) ~/~ B \in {S^{\uparrow ^\mathcal{L} S}} \}$, ${S^{\uparrow ^\mathcal{L} S}}$, $\iota(S)$}
	\KwOut{$\{q(B) ~/~ B \in {S^{\uparrow ^\mathcal{L} S}} \}$}
	sort $\iota(S)$ such that $\iota(S) = \{i_1, i_2, \dots, i_{K}\},$ where $|i_1| \leq |i_2| \leq \dots \leq |i_{K}|$\;
	\ForEach{$A \in {S^{\uparrow ^\mathcal{L} S}}$}{
		$q(A) \leftarrow m(A)$\;
	}
	\For{$k = 1 ~\KwTo~ K$}{
		\ForEach{$A \in {S^{\uparrow ^\mathcal{L} S}}$}{
			$B \leftarrow A \cup i_k$\;
			\If{$B \neq A$}{
				$q(A) \leftarrow q(A) + q(B)$\;
			}
		}
	}
	\caption{Computation of $\{q(B) ~/~ B \in {S^{\uparrow ^\mathcal{L} S}} \}$ based on $\{m(B) ~/~ B \in {S^{\uparrow ^\mathcal{L} S}} \}$.}\label{algo:q_dotL}
\end{algorithm}

\begin{algorithm}
	\KwIn{$\{q(B) ~/~ B \in {S^{\uparrow ^\mathcal{L} S}} \}$, ${S^{\uparrow ^\mathcal{L} S}}$, $\iota(S)$}
	\KwOut{$\{m(B) ~/~ B \in {S^{\uparrow ^\mathcal{L} S}} \}$}
	sort $\iota(S)$ such that $\iota(S) = \{i_1, i_2, \dots, i_{K}\},$ where $|i_1| \leq |i_2| \leq \dots \leq |i_{K}|$\;
	\ForEach{$A \in {S^{\uparrow ^\mathcal{L} S}}$}{
		$m(A) \leftarrow q(A)$\;
	}
	\For{$k = K ~\KwTo~ 1$}{
		\ForEach{$A \in {S^{\uparrow ^\mathcal{L} S}}$}{
			$B \leftarrow A \cup i_k$\;
			\If{$B \neq A$}{
				$m(A) \leftarrow m(A) - m(B)$\;
			}
		}
	}
	\caption{Computation of $\{m(B) ~/~ B \in {S^{\uparrow ^\mathcal{L} S}} \}$ based on $\{q(B) ~/~ B \in {S^{\uparrow ^\mathcal{L} S}} \}$.}\label{algo:q_dotL_rev}
\end{algorithm}

\begin{algorithm}
	\KwIn{$\{b(B) ~/~ B \in {S^{\downarrow ^\mathcal{L} S}} \}$, ${S^{\downarrow ^\mathcal{L} S}}$, $\overline{\iota}(S)$}
	\KwOut{$\{v(B) ~/~ B \in {S^{\downarrow ^\mathcal{L} S}} \}$}
	sort $\overline{\iota}(S)$ such that $\overline{\iota}(S) = \{i_1, i_2, \dots, i_{K}\},$ where $|i_1| \leq |i_2| \leq \dots \leq |i_{K}|$\;
	\ForEach{$A \in {S^{\downarrow ^\mathcal{L} S}}$}{
		$v(A) \leftarrow b(A)^{-1}$\;
	}
	\For{$k = 1 ~\KwTo~ K$}{
		\ForEach{$A \in {S^{\downarrow ^\mathcal{L} S}}$}{
			$B \leftarrow A \cap i_k$\;
			\If{$B \neq A$}{
				$v(A) \leftarrow v(A) / v(B)$\;
			}
		}
	}
	\caption{Computation of $\{v(B) ~/~ B \in {S^{\downarrow ^\mathcal{L} S}} \}$ based on $\{b(B) ~/~ B \in {S^{\downarrow ^\mathcal{L} S}} \}$.}\label{algo:v_dotL}
\end{algorithm}

\begin{algorithm}
	\KwIn{$\{v(B) ~/~ B \in {S^{\downarrow ^\mathcal{L} S}} \}$, ${S^{\downarrow ^\mathcal{L} S}}$, $\overline{\iota}(S)$}
	\KwOut{$\{b(B) ~/~ B \in {S^{\downarrow ^\mathcal{L} S}} \}$}
	sort $\overline{\iota}(S)$ such that $\overline{\iota}(S) = \{i_1, i_2, \dots, i_{K}\},$ where $|i_1| \leq |i_2| \leq \dots \leq |i_{K}|$\;
	\ForEach{$A \in {S^{\downarrow ^\mathcal{L} S}}$}{
		$b(A) \leftarrow v(A)^{-1}$\;
	}
	\For{$k = K ~\KwTo~ 1$}{
		\ForEach{$A \in {S^{\downarrow ^\mathcal{L} S}}$}{
			$B \leftarrow A \cap i_k$\;
			\If{$B \neq A$}{
				$b(A) \leftarrow b(A) . b(B)$\;
			}
		}
	}
	\caption{Computation of $\{b(B) ~/~ B \in {S^{\downarrow ^\mathcal{L} S}} \}$ based on $\{v(B) ~/~ B \in {S^{\downarrow ^\mathcal{L} S}} \}$.}\label{algo:v_dotL_rev}
\end{algorithm}

\begin{algorithm}
	\KwIn{$\{q(B) ~/~ B \in {S^{\uparrow ^\mathcal{L} S}} \}$, ${S^{\uparrow ^\mathcal{L} S}}$, $\iota(S)$}
	\KwOut{$\{w(B) ~/~ B \in {S^{\uparrow ^\mathcal{L} S}} \}$}
	sort $\iota(S)$ such that $\iota(S) = \{i_1, i_2, \dots, i_{K}\},$ where $|i_1| \leq |i_2| \leq \dots \leq |i_{K}|$\;
	\ForEach{$A \in {S^{\uparrow ^\mathcal{L} S}}$}{
		$w(A) \leftarrow q(A)^{-1}$\;
	}
	\For{$k = K ~\KwTo~ 1$}{
		\ForEach{$A \in {S^{\uparrow ^\mathcal{L} S}}$}{
			$B \leftarrow A \cup i_k$\;
			\If{$B \neq A$}{
				$w(A) \leftarrow w(A) / w(B)$\;
			}
		}
	}
	\caption{Computation of $\{w(B) ~/~ B \in {S^{\uparrow ^\mathcal{L} S}} \}$ based on $\{q(B) ~/~ B \in {S^{\uparrow ^\mathcal{L} S}} \}$.}\label{algo:w_dotL}
\end{algorithm}

\begin{algorithm}
	\KwIn{$\{w(B) ~/~ B \in {S^{\uparrow ^\mathcal{L} S}} \}$, ${S^{\uparrow ^\mathcal{L} S}}$, $\iota(S)$}
	\KwOut{$\{q(B) ~/~ B \in {S^{\uparrow ^\mathcal{L} S}} \}$}
	sort $\iota(S)$ such that $\iota(S) = \{i_1, i_2, \dots, i_{K}\},$ where $|i_1| \leq |i_2| \leq \dots \leq |i_{K}|$\;
	\ForEach{$A \in {S^{\uparrow ^\mathcal{L} S}}$}{
		$q(A) \leftarrow w(A)^{-1}$\;
	}
	\For{$k = 1 ~\KwTo~ K$}{
		\ForEach{$A \in {S^{\uparrow ^\mathcal{L} S}}$}{
			$B \leftarrow A \cup i_k$\;
			\If{$B \neq A$}{
				$q(A) \leftarrow q(A) . q(B)$\;
			}
		}
	}
	\caption{Computation of $\{q(B) ~/~ B \in {S^{\uparrow ^\mathcal{L} S}} \}$ based on $\{w(B) ~/~ B \in {S^{\uparrow ^\mathcal{L} S}} \}$.}\label{algo:w_dotL_rev}
\end{algorithm}

\begin{algorithm}
	\KwIn{$\{m(F) ~/~ F \in S \}$}
	\KwOut{$\{b(B) ~/~ B \in {^\vee S} \}$}
	\ForEach{$B \in {^\vee S}$}{
		$b(B) \leftarrow m(B)$\;
		\ForEach{$F \in S ~/~ F \subset B$}{
			$b(B) \leftarrow b(B) + m(F)$\;
		}
	}
	\caption{Computation of $\{b(B) ~/~ B \in {^\vee S} \}$ based on $\{m(B) ~/~ B \in S \}$ independently from $|\Omega|$.}\label{algo:b_dotF_noN}
\end{algorithm}

\begin{algorithm}
	\KwIn{$\{b(B) ~/~ B \in {^\vee S} \}$}
	\KwOut{$\{m(B) ~/~ B \in {^\vee S} \}$}
	sort ${^\vee S}$ such that ${^\vee S} = \{A_1, A_2, \dots, A_{K}\},$ where $|A_1| \leq |A_2| \leq \dots \leq |A_{K}|$\;
	\For{$i=1$ \KwTo $K$}{
		$m(A_i) \leftarrow b(A_i)$\;
		
		\ForEach{$B \in {^\vee S} ~/~ B \subset A_i$}{
			$m(A_i) \leftarrow m(A_i) - m(B)$\;
		}
	}
	\caption{Computation of $\{m(B) ~/~ B \in {^\vee S} \}$ based on $\{b(B) ~/~ B \in {^\vee S} \}$ independently from $|\Omega|$.}\label{algo:b_dotF_noN_rev}
\end{algorithm}

\begin{algorithm}
	\KwIn{$\{m(F) ~/~ F \in S \}$}
	\KwOut{$\{q(B) ~/~ B \in {^\wedge S} \}$}
	\ForEach{$B \in {^\wedge S}$}{
		$q(B) \leftarrow m(B)$\;
		\ForEach{$F \in S ~/~ F \supset B$}{
			$q(B) \leftarrow q(B) + m(F)$\;
		}
	}
	\caption{Computation of $\{q(B) ~/~ B \in {^\wedge S} \}$ based on $\{m(B) ~/~ B \in S \}$ independently from $|\Omega|$.}\label{algo:q_dotF_noN}
\end{algorithm}

\begin{algorithm}
	\KwIn{$\{q(B) ~/~ B \in {^\wedge S} \}$}
	\KwOut{$\{m(B) ~/~ B \in {^\wedge S} \}$}
	sort ${^\wedge S}$ such that ${^\wedge S} = \{A_1, A_2, \dots, A_{K}\},$ where $|A_1| \leq |A_2| \leq \dots \leq |A_{K}|$\;
	\For{$i=K$ \KwTo $1$}{
		$m(A_i) \leftarrow q(A_i)$\;
		
		\ForEach{$B \in {^\wedge S} ~/~ B \supset A_i$}{
			$m(A_i) \leftarrow m(A_i) - m(B)$\;
		}
	}
	\caption{Computation of $\{m(B) ~/~ B \in {^\wedge S} \}$ based on $\{q(B) ~/~ B \in {^\wedge S} \}$ independently from $|\Omega|$.}\label{algo:q_dotF_noN_rev}
\end{algorithm}

\begin{algorithm}
	\KwIn{$\{b(B) ~/~ B \in {^\vee S} \}$, ${^\vee S}$}
	\KwOut{$\{w(B) ~/~ B \in {^\vee S} \}$}
	sort ${^\vee S}$ such that ${^\vee S} = \{A_1, A_2, \dots, A_{K}\},$ where $|A_1| \leq |A_2| \leq \dots \leq |A_{K}|$\;
	
	\For{$i = 1$ \KwTo $K$}{
		$v(A_i) \leftarrow b(A_i)^{-1}$\;
		\ForEach{$B \in {^\vee S} ~/~ B \subset A_i$}{
			$v(A_i) \leftarrow v(A_i) / v(B)$\;
		}
	}
	\caption{Computation of $\{v(B) ~/~ B \in {^\vee S} \}$ based on $\{b(B) ~/~ B \in {^\vee S} \}$ independently from $|\Omega|$.}\label{algo:v_dotF_noN}
\end{algorithm}

\begin{algorithm}
	\KwIn{$\{v(B) ~/~ B \in {^\vee S} \}$, ${^\vee S}$}
	\KwOut{$\{b(B) ~/~ B \in {^\vee S} \}$}
	\ForEach{$B \in {^\vee S}$}{	
		$b(B) \leftarrow v(B)^{-1}$\;	
		\ForEach{$A \in {^\vee S} ~/~ A \subset B$}{
			$b(B) \leftarrow b(B) / v(A)$\;
		}
	}
	\caption{Computation of $\{b(B) ~/~ B \in {^\vee S} \}$ based on $\{v(B) ~/~ B \in {^\vee S} \}$ independently from $|\Omega|$.}\label{algo:v_dotF_noN_rev}
\end{algorithm}

\begin{algorithm}
	\KwIn{$\{q(B) ~/~ B \in {^\wedge S} \}$, ${^\wedge S}$}
	\KwOut{$\{w(B) ~/~ B \in {^\wedge S} \}$}
	sort ${^\wedge S}$ such that ${^\wedge S} = \{A_1, A_2, \dots, A_{K}\},$ where $|A_1| \leq |A_2| \leq \dots \leq |A_{K}|$\;
	
	\For{$i = K$ \KwTo $1$}{
		$w(A_i) \leftarrow q(A_i)^{-1}$\;
		\ForEach{$B \in {^\wedge S} ~/~ B \supset A_i$}{
			$w(A_i) \leftarrow w(A_i) / w(B)$\;
		}
	}
	\caption{Computation of $\{w(B) ~/~ B \in {^\wedge S} \}$ based on $\{q(B) ~/~ B \in {^\wedge S} \}$ independently from $|\Omega|$.}\label{algo:w_dotF_noN}
\end{algorithm}

\begin{algorithm}
	\KwIn{$\{w(B) ~/~ B \in {^\wedge S} \}$, ${^\wedge S}$}
	\KwOut{$\{q(B) ~/~ B \in {^\wedge S} \}$}
	\ForEach{$B \in {^\wedge S}$}{	
		$q(B) \leftarrow w(B)^{-1}$\;	
		\ForEach{$A \in {^\wedge S} ~/~ A \supset B$}{
			$q(B) \leftarrow q(B) / w(A)$\;
		}
	}
	\caption{Computation of $\{q(B) ~/~ B \in {^\wedge S} \}$ based on $\{w(B) ~/~ B \in {^\wedge S} \}$ independently from $|\Omega|$.}\label{algo:w_dotF_noN_rev}
\end{algorithm}
\end{appendices}

\bibliography{./complexity-reduction}

\end{document}